\documentclass[11pt]{article}
\usepackage{amssymb,amsmath}
\usepackage{bm} 
\usepackage{booktabs} 
\usepackage{array}
\usepackage{latexsym}
\usepackage{graphicx}
\usepackage{color}
\usepackage{datetime}
\usepackage[nosort]{cite}
\usepackage{verbatim}
\usepackage{enumerate}
\usepackage{chngpage} 
\usepackage{mathrsfs}
\usepackage{euscript}
\usepackage{psfrag}

\usepackage{float}
\usepackage{tikz}
\usetikzlibrary{decorations.text}
\usepackage{enumitem}
\setlist{itemsep=0pt}
\usetikzlibrary{math}

\usepackage{mciteplus}

\usepackage[colorlinks=true,      linkcolor=blue,      urlcolor=blue,      
            filecolor=blue,      citecolor=blue,       pdfstartview=FitH,     
						pdfpagemode=UseNone,      bookmarksopen=true]{hyperref}  
\usepackage[all]{hypcap}     

\newcommand{\comm}[1]{} 

\renewcommand{\arraystretch}{1.3}

\def\({\left(}
\def\){\right)}
\def\[{\left[}
\def\]{\right]}

\def\coeff#1#2{{\textstyle \frac{#1}{#2}}}

\def\One{{\hbox{ 1\kern-.8mm l}}}

\def\barray{\begin{array}}
\def\earray{\end{array}}
\def\be{\begin{equation}}
\def\ee{\end{equation}}
\def\bea{\begin{eqnarray}}
\def\eea{\end{eqnarray}}
\def\bal{\begin{align}}
\def\eal{\end{align}}

\def\-{\,-\,}
\def\={\,=\,}
\def\+{\,+\,}
\def\equi{\,\equiv\,}

\def\mfn{\mathfrak{n}}
\numberwithin{equation}{section} 

\definecolor{cardinal}{rgb}{0.6,0,0}
\definecolor{darkgreen}{rgb}{0,0.4,0}
\definecolor{golden}{rgb}{0.92, 0.7, 0}
\definecolor{midnight}{rgb}{0, 0, 0.5}
\definecolor{darkblue}{rgb}{0, 0, 0.7}
\definecolor{purple}{rgb}{0.5, 0, 0.5}

\newcommand{\subf}[2]{%
  {\small\begin{tabular}[t]{@{}c@{}}
  #1\\#2
  \end{tabular}}%
}

\def\IR{\mathbb{R}}

\def\ZZ{\mathbb{Z}}

\def\cF{{\cal F}}

\def\cL{{\cal L}}

\def\cN{{\cal N}}
\def\cP{{\cal P}}
\def\cO{{\cal O}}

\def\cV{{\cal V}}
\def\cW{{\cal W}}

\usetikzlibrary{positioning}

\tikzset{
 diffuse color/.initial = black,                       
}

\tikzset{
 linear opacity/.initial=0.5,                          
 linear stroke/.style = {                              
   preaction={                                         
     draw=\pgfkeysvalueof{/tikz/diffuse color},        
     line width = (2.0-#1)*\pgflinewidth,              
     opacity=\pgfkeysvalueof{/tikz/linear opacity},white}},  
 diffuse gradient/.style={                             
   draw = none,                                        
   linear opacity=#1,                                  
   linear stroke/.list={0.0,#1,...,1.0}},              
 diffuse gradient/.default=1,                          
}

\tikzset{
 non-linear stroke/.style = {                          
   preaction={                                         
     draw=\pgfkeysvalueof{/tikz/diffuse color},        
     line width = (2.0-#1)*\pgflinewidth,              
     opacity=#1,white}},                                     
 diffuse falloff/.style={                              
   draw = none,                                        
   non-linear stroke/.list={0.0,#1,...,1.0}},          
 diffuse falloff/.default=1,                           
}
 
\begin{document}

\phantom{AAA}
\vspace{-10mm}

\begin{flushright}
%
%
\end{flushright}

\vspace{1.4cm}

\begin{center}

{\huge {\bf The Great Escape: }} \\
{\huge {\bf \vspace*{.25cm}Tunneling out of Microstate Geometries }}

\vspace{1cm}

{\large{\bf {Iosif Bena$^1$,~Felicity Eperon$^{1}$,~Pierre Heidmann$^2$  and  Nicholas P. Warner$^{1,3,4}$}}}

\vspace{1cm}

$^1$Institut de Physique Th\'eorique, \\
Universit\'e Paris Saclay, CEA, CNRS,\\
Orme des Merisiers, Gif sur Yvette, 91191 CEDEX, France \\[12 pt]

$^2$Department of Physics and Astronomy,\\
Johns Hopkins University,\\
3400 North Charles Street, Baltimore, MD 21218, USA,\\[12 pt]

\centerline{$^3$Department of Physics and Astronomy}
\centerline{and $^4$Department of Mathematics,}
\centerline{University of Southern California,} 
\centerline{Los Angeles, CA 90089, USA}

\vspace{10mm} 
{\footnotesize\upshape\ttfamily iosif.bena @ ipht.fr, felicity.eperon @ gmail.com, pheidma1 @ jh.edu, warner @ usc.edu} \\

\end{center}
\vspace{2cm}
 

\begin{adjustwidth}{3mm}{3mm} 
 
\vspace{-1.2mm}
\noindent
We compute the quasi-normal frequencies of scalars in asymptotically-flat microstate geometries that have the same charge as a D1-D5-P black hole, but whose long BTZ-like throat ends in a smooth cap. In general the wave equation is not separable, but we find a class of geometries in which the non-separable term is negligible and we can compute the quasi-normal frequencies using WKB methods. We argue that our results are a universal property of all microstate geometries with deeply-capped BTZ throats. These throats generate large redshifts, which lead to exceptionally-low-energy states with extremely long decay times, set by the central charge of the dual CFT to the power of twice the dimension of the operator dual to the mode.  While these decay times are extremely long, we also argue that the energy decay is bounded, at large $t$, by $\big(\log(t)\big)^{-2} $ and is comparable with the behavior of ultracompact stars, as one should expect for microstate geometries. 

\end{adjustwidth}

\thispagestyle{empty}
\newpage


\baselineskip=13pt
\parskip=2.5pt

\setcounter{tocdepth}{2}

\baselineskip=15pt
\parskip=3pt


\tableofcontents

\newpage
\section{Introduction}
\label{sec:Intro}

One of the primary motivations for the construction of microstate geometries is that they approximate very closely  the behavior of black holes without leading to information loss. This happens because these geometries have a smooth cap at very high redshift but do not have a horizon.  The craft in constructing and analyzing such geometries lies in how well they approximate the black-hole behavior and this craft is becoming a well-developed science for BPS microstate geometries. In particular, we now have extensive families of BPS microstate geometries that look exactly like a BPS black hole, except that the infinite AdS$_2$ throat of the black hole is capped at some very large depth \cite{Bena:2006kb, Bena:2007qc, Bena:2016ypk, Bena:2017xbt, Heidmann:2017cxt, Bena:2017fvm, Avila:2017pwi, Ceplak:2018pws,Heidmann:2019zws, Heidmann:2019xrd}.  This cap affects an infalling observer less than a Planck time before crossing the would-be event horizon \cite{Tyukov:2017uig, Bena:2018mpb}. 

From a holographic perspective, the depth of the throat is one of the most important physical parameters of the solution, because it controls the energy gap of the excitations on top of the supersymmetric CFT ground state dual to the microstate geometry. For the states with the longest throat, this gap matches exactly the one expected from the typical CFT states that count the black hole entropy \cite{Bena:2018bbd, Heidmann:2019zws, Bena:2006kb, Tyukov:2017uig}.

Furthermore, if one computes, holographically, the two-point function in the heavy state dual to the microstate geometry \cite{Bena:2019azk}, this two-point function exhibits the same thermal decay as in the BTZ background, except that the information is not lost but is recovered after a return time of order the inverse of the energy gap. Hence microstate geometries look exactly like black holes on time-scales less than this return time, but after that they do indeed return the information about what was thrown into them and about the smooth cap at the bottom of the throat.

Intuitively, it is natural to expect that the cap region will be the repository of all the microstate structure and thus one should expect infalling matter to be trapped there for a very long time. Starting with \cite{Cardoso:2005gj}, there have now been several investigations  \cite{Chowdhury:2007jx,Chakrabarty:2015foa,Eperon:2016cdd,Chakrabarty:2019ujg} of trapping of matter in either BPS or non-BPS microstate geometries.  Furthermore, it was shown in \cite{Eperon:2016cdd} that there exist modes that decay extremely slowly, and this was confirmed by a matched-asymptotic-expansion calculation of the decay time \cite{Chakrabarty:2019ujg}. From a mathematical perspective, this extremely slow decay of a wave equation in a background was the slowest ever found, and this has created some interest in the mathematical community \cite{Keir:2016azt,Eperon:2017bwq,Keir:2018hnv}.

The  concern raised by the analysis of \cite{Eperon:2016cdd} was that such long-term trapping would lead to non-linear instabilities. This is because, in General Relativity alone, if matter accumulates in a region for a long period of time, it will tend to form black holes, or black extended objects.  Luckily, String Theory affords many other possibilities, and the guiding principle of fuzzballs and microstate geometries is that whenever GR predicts the formation of a black hole, the system should instead become a fuzzball, or, in its more coherent incarnations, it should transition into a new microstate geometry. This is because all the microstate geometries belong to a very large moduli space of solutions, whose dimension is of order the central charge of the CFT dual to the black hole ($6 n_1 n_5$ for the D1-D5-P black hole \cite{Rychkov:2005ji,Bena:2014qxa}). Hence, an excitation of these geometries has $\sim n_1 n_5$ available directions into which it can spread, and will generically explore this very large phase space \cite{Marolf:2016nwu,Bena:2018mpb} rather than form a black hole.

There is another problem with the decay-time analysis of  \cite{Eperon:2016cdd}, which can be best seen when analyzing the slowly-decaying modes using a matched asymptotic expansion \cite{Chakrabarty:2019ujg}. From a mathematical perspective, the allowed wave-numbers are unbounded. However, one cannot trust the supergravity approximation if the wavelength of these oscillations is smaller than the Planck scale. This puts an upper bound on the wave numbers, and the practical effect of this upper bound is to eliminate the slowest decaying modes. This in turn indicates that from a physics perspective, the non-linear instability found in  \cite{Eperon:2016cdd} is an artifact of considering sub-Planckian modes\footnote{Here we mean sub-Planckian wavelengths, which correspond to super-Planckian masses.}.

The purpose of this paper is to calculate the long-term trapping in, and tunneling from, a family of asymptotically-flat microstate geometries that have the same charges and the same long AdS$_2$ throat as a three-charge black hole with a large event horizon\footnote{The solutions considered in \cite{Eperon:2016cdd,Chakrabarty:2019ujg} have an angular momentum that exceeds the cosmic censorship bound \cite{Giusto:2004id,Giusto:2004ip,Giusto:2004kj,Giusto:2012yz}, and hence lack the long BTZ-like throat characteristic of typical black-hole microstate geometries.}.

There is a standard approach to this class of problems in which one uses matched asymptotic expansions \cite{Chowdhury:2007jx,Chakrabarty:2015foa,Eperon:2016cdd,Chakrabarty:2019ujg}.  Essentially, one constructs the modes in an asymptotically AdS space-time, and then matches the AdS asymptotics to the Bessel function asymptotics that are the staple of flat-space scattering problems.  This usually requires approximations in which the frequency is taken to be small or one considers the near-decoupling limit of the background, $Q_P \ll Q_1, Q_5$.  To date, this method has been applied successfully to computing the quasi-normal frequencies of atypical microstate geometries that do not have long black-hole-like throats \cite{Giusto:2004id,Giusto:2004ip,Giusto:2004kj,Jejjala:2005yu}.

The challenge in analyzing the known asymptotically-flat microstate geometries with a deeply-capped BTZ throat is that they usually depend non-trivially on several variables and the wave equation is not separable.  However, we have obtained a family of such microstate geometries in which the scalar wave equation is ``almost separable'': If one tries to make a separation of variables, one finds that it almost works except for one term. We then show that this term is  parametrically suppressed in the long-throat approximation, and  even more highly suppressed at low energies.  This means that the tunneling process is accurately governed by the separable pieces of the scalar wave equation, and all the interesting physics is encoded in the radial wave equation.

Rather than using matched asymptotic expansions, we use a technique similar to that of  \cite{Cardoso:2005gj,Bena:2019azk}: we reduce the radial equation to an equivalent Schr\"odinger equation in which the tunneling from the cap to the asymptotic region becomes a simple computation of a barrier penetration.  We then use WKB methods to compute the quasi-normal frequencies of the modes of this system.  This approach leads to a simple, more intuitive picture of the tunneling process, making the universality of our results for generic deep microstate geometries all the more apparent. 

At low energy, we show that the quasi-normal modes are mostly supported in the highly-redshifted AdS$_3$ cap. Thus, as in \cite{Chowdhury:2007jx,Chakrabarty:2015foa,Eperon:2016cdd,Chakrabarty:2019ujg}, the real parts of the frequencies correspond to the bound-state frequencies of the cap and their imaginary parts depend on the redshift between the cap and flat space.  We find that the decay times in our solutions are parametrically slower than the decay times found in \cite{Eperon:2016cdd,Chakrabarty:2019ujg}, and this comes from the fact that our solutions have a very long throat.

We also highlight another regime of energy that is absent in the geometries studied in \cite{Eperon:2016cdd,Chakrabarty:2019ujg}. At intermediate energy, the modes start to explore the BTZ throat. The rigidity of the AdS$_2$ region makes them leak very slowly into flat space. The end result is that the spectrum of quasi-normal modes is modulated by the BTZ response function in this regime of energy. 

We also show that the modes that can be described in supergravity, even if they decay very slowly, give a decay that is consistent with the trapping created by {\em extremely compact neutron stars}.  This result is in perfect accord with the physics that one would hope to see emerge from the microstate geometry programme.  By smoothly capping-off geometries just above the horizon scale of a black hole, one creates an extremely compact object whose states can still be seen and measured by distant observers.  It is therefore to be expected that the trapping of matter by microstate geometries should parallel the trapping of matter by extremely compact, ``normal'' objects. 

If one leaves physical considerations aside, and considers arbitrary sub-Planckian wavelengths, one finds that our geometry  also has modes that decay  slowly enough to give rise to non-linear instabilities.  As was shown in \cite{Eperon:2016cdd, Chakrabarty:2019ujg}, such modes are localized in the neighborhood of the evanescent ergosurface.  This was anticipated from the long-term trapping of exceptionally low-energy geodesic near such surfaces \cite{Eperon:2016cdd}.  We also find such modes in the superstratum geometries, but they are necessarily sub-Planckian and hence of no physical relevance.

Our analysis also reveals the existence of modes that are trapped forever.  Some of these modes have a very simple physical explanation: There are trapped modes with zero frequency, which correspond to BPS deformations of the  supersymmetric zero-energy superstratum into another zero-energy superstratum that is close in phase space.  In addition, there is an infinite family of trapped modes with negative momentum. These modes have positive energy but carry a (momentum) charge that is opposite to that of the background, so they will always be attracted to the bottom of the solution and will never be able to escape\footnote{The existence of modes that are trapped forever is in fact not only a feature of the superstratum geometry, but also of the overspinning Lunin-Mathur and GMS solutions \cite{Lunin:2001fv,Giusto:2004id,Giusto:2004ip,Giusto:2004kj} analyzed in \cite{Lunin:2001dt, Chakrabarty:2019ujg}. We thank Samir Mathur for explaining to us this physics.}.   The focus of this paper is on the quasi-normal modes but we will make some remarks about the non-trivial eternally-trapped modes in Section \ref{sec:Conclusions}.

In Section \ref{Sect:RadialEqn}, we present the general features of the six-dimensional microstate geometries whose quasinormal modes we compute, highlighting their key features and explaining how our WKB analysis works. In Section \ref{sec:SuperstratumPot}, we construct the asymptotically-flat $(2,1,n)$ supercharged superstrata, we discuss the separability of the minimally-coupled scalar-wave equation and the different limits of the potential that appears in the radial equation. In Section \ref{sec:Superstrata} we derive the spectrum of quasi-normal modes in two energy regimes using the WKB method and discuss the corresponding decay rates. In Section \ref{sec:OlderWork}, we review the analysis of \cite{Chakrabarty:2019ujg} for a family of atypical microstate geometries using similar convention as ours  and compare the results. In Section \ref{sec:Decay}, we discuss the decay timescales for the energy of the quasi-normal modes to leak to flat space and the potential instabilities. We make some final remarks in Section \ref{sec:Conclusions}.

\section{Tunneling and quasi-normal modes in microstate geometries}
\label{Sect:RadialEqn}

In this paper we compute the quasinormal modes for two classes of microstate geometries:  Superstrata \cite{Bena:2015bea,Bena:2016ypk,Bena:2017xbt,Ceplak:2018pws,Heidmann:2019zws}, and the Giusto-Mathur-Saxena (GMS), or the closely-related Giusto-Lunin-Mathur-Turton (GLMT) solutions \cite{Giusto:2004id,Giusto:2004ip,Giusto:2004kj,Giusto:2012yz}.   In this section, we give an overview of the geometry and the relevant class of Schr\"odinger problems and describe how to use WKB methods to analyze the decay of the states that are trapped deep within the microstate geometry.  

\begin{figure}
\begin{adjustwidth}{-2.0cm}{-2.0cm}
 \centering
\begin{tabular}{ccc}
\subf{\includegraphics[width=7.8cm]{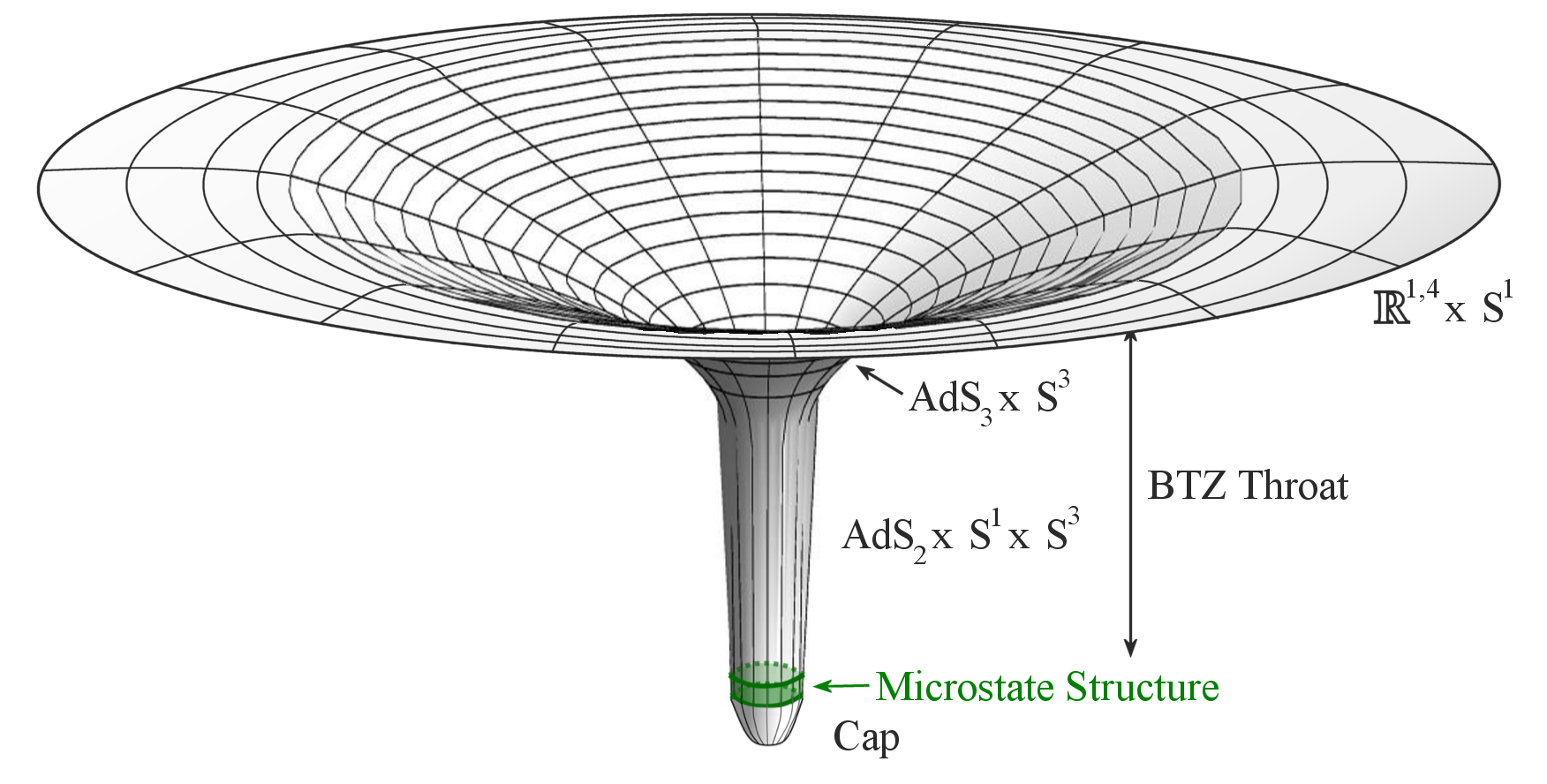}}
     {\begin{minipage}{20em}
{\footnotesize (a) A superstratum spacetime is asymptotically flat at infinity, then has an AdS$_3\times$S$^3$ region, then an AdS$_2\times$S$^1\times$S$^3$ throat and then the cap (which a redshifted global AdS$_3\times$S$^3$). Together the upper AdS$_3$ and the AdS$_2\times$S$^1$ regions form the BTZ throat of the geometry.}
\end{minipage}
     }
&
\subf{\includegraphics[width=7.8cm]{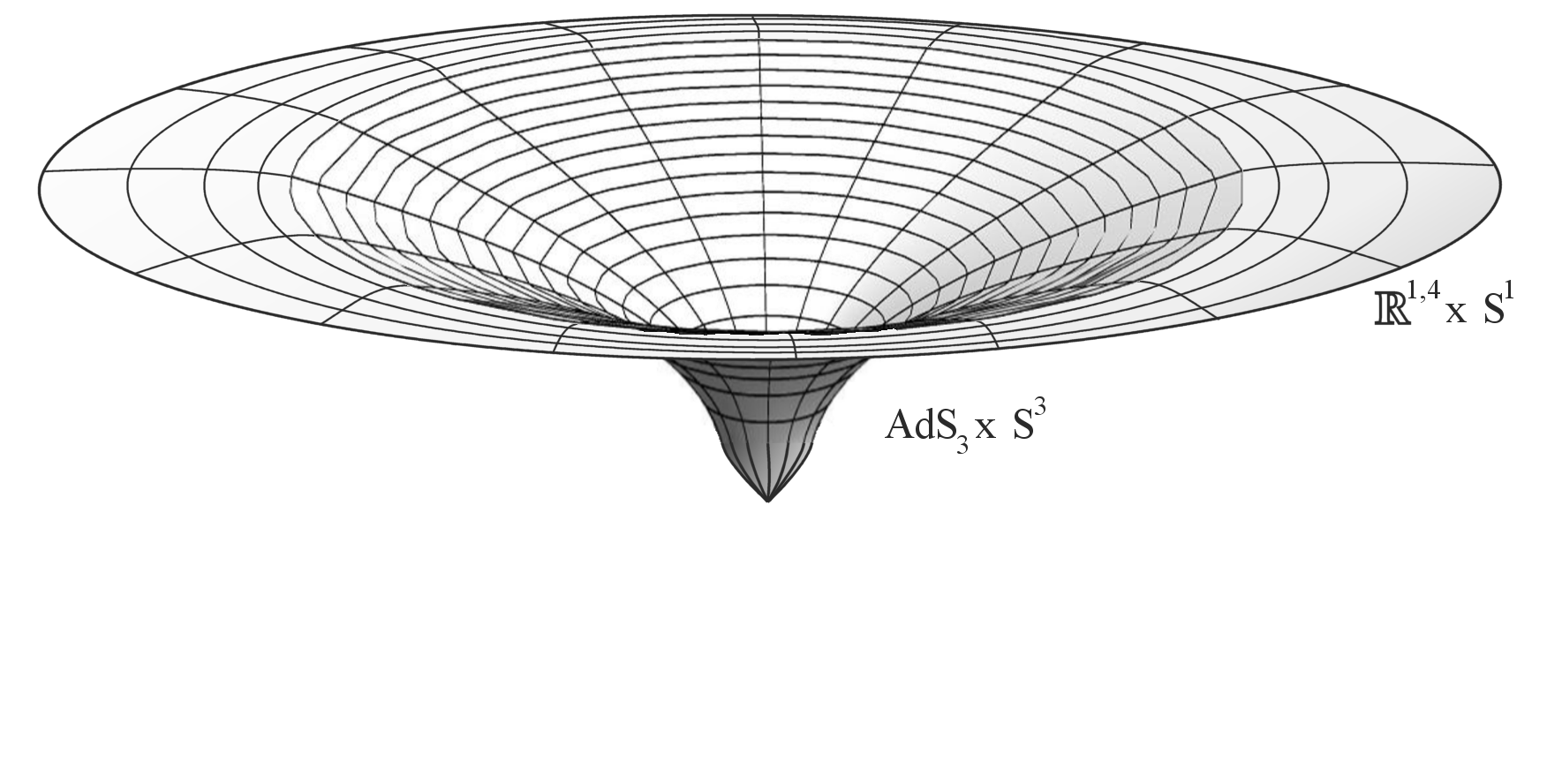}}
 {\begin{minipage}{20em} 
{\footnotesize    (b) A GMS  spacetime has flat space at infinity glued to a global AdS$_3\times$S$^3$ in the infrared. Unlike superstrata, the redshift between the cap and flat space is not controlled by the length of a BTZ throat but by the parameter, $k$, of a $\ZZ_k$ orbifold. To illustrate the presence of this orbifold we represent the cap as a cone.}
     \end{minipage}
   }
\end{tabular}
\end{adjustwidth}
\caption{Schematics of the two classes of microstate geometries considered in this paper.}
\label{fig:spacetime}
\end{figure}

At infinity the geometries are  asymptotic to $\IR^{1,4}\times$S$^1$.   Asymptotically-flat superstrata with a deeply-capped BTZ throat \cite{Bena:2017xbt,Ceplak:2018pws} have four regions: (i) The cap, (ii) The AdS$_2$ throat,  (iii) The AdS$_3$ region, and (iv) The flat region near infinity.  A schematic picture of this structure is shown in Figure \ref{fig:spacetime}(a).   The AdS$_3$ region and the AdS$_2$ throat together form a region of the geometry that is closely approximated by the BTZ metric.  In this paper, the superstrata  will always have a long AdS$_2$ throat, but the size of  AdS$_3$ region will depend upon the charges because the geometry may well transition rapidly from the AdS$_2$ throat to the asymptotically-flat region.   The geometry of the cap at the very bottom of the solution is closely approximated by a global AdS$_3\times$S$^3$ metric.   

The GMS solutions have two regions: (i) The cap and (ii) the flat region near infinity. A schematic picture of this structure is shown in Figure \ref{fig:spacetime}(b). The cap geometry is an S$^3$ fibration over a redshifted AdS$_3$ geometry. Since these solutions do not have a BTZ throat, the redshift is much smaller than for superstrata and yields to a larger energy gap.

\subsection{The parameters and charges of the solutions}
\label{sec:charges}

Superstrata are $\frac{1}{8}$-BPS solutions of type IIB supergravity on T$^4$ or $K3$ that have the same charges and mass as supersymmetric D1-D5-P black holes. From the perspective of the six-dimensional transverse space, they carry three charges, $Q_1$, $Q_5$ and $Q_P$, and two angular momenta, $J_L$ and $J_R$. 

We will consider a specific family of superstrata, denoted as the $(2,1,n)$ supercharged superstrata. They have five independent parameters, which we will denote by $Q_5$, $b$, $a$,  $R_y$ and an integer, $n$.   The parameter, $R_y$, is the radius of the common D1-D5 circle at infinity which we oftentimes  refer to as the ``$y$-circle,'' while $Q_5$ is the charge of the D5 branes.  The remaining parameters, $b$, $a$ and $n$, control the other 
supergravity charges via:
\begin{equation}
Q_1 Q_5 ~=~ R_y^2 \, \big(a^2 + \coeff{1}{2} b^2\big) \,, \qquad Q_P = \coeff{1}{4} \,(n+1) \,b^2 \,, \qquad   J_L ~=~ \coeff{1}{2} \,R_y \,a^2   \,, \qquad
J_R ~=~ \coeff{1}{2} \, R_y\,  \big(a^2 + \coeff{1}{2} b^2\big) \,.
\label{charges}
\end{equation}
The first of these relations is required by smoothness.
 
The quantized charges, $n_1, n_5, n_P, j_L$ and $j_R$ are related to the supergravity charges by:
\begin{equation}
Q_1 ~=~  \frac{(2\pi)^4\,n_1\,g_s\,\alpha'^3}{V_4}\,,\quad Q_5 = n_5\,g_s\,\alpha' \,, \quad n_P ~=~    \cN \, Q_P\,, \quad   j_{L,R} ~=~ \cN  \, R_y^{-1}\, J_{L,R}   \,,
\label{quantcharges}
\end{equation}
where $V_4$ is the volume of the internal manifold (T$^4$ or $K3$) of the Type IIB compactification to six dimensions and  $\cN$ is: 
\begin{equation}
\cN ~\equiv~ \frac{n_1 \, n_5\, R_y^2}{Q_1 \, Q_5} ~=~\frac{V_4\, R_y^2}{ (2\pi)^4 \,g_s^2 \,\alpha'^4}~=~\frac{V_4\, R_y^2}{(2\pi)^4 \, \ell_{10}^8} ~=~\frac{{\rm Vol} (T^4) \, R_y^2}{ \ell_{10}^8} \,,
\label{cNdefn}
\end{equation}
where $\ell_{10}$ is the ten-dimensional Planck length and  $(2 \pi)^7 g_s^2 \alpha'^4  = 16 \pi G_{10} ~\equiv~ (2 \pi)^7 \ell_{10}^8$.    The quantity, ${\rm Vol} (T^4)  \equiv (2\pi)^{-4} \, V_4$, is sometimes introduced \cite{Peet:2000hn} as a ``normalized volume'' that is equal to $1$ when the radii of the circles in the $T^4$ are equal to one in Planck units.

One should note that, unlike the superstrata with a long BTZ throat constructed in  \cite{Bena:2017fvm}, the right-moving angular momentum of our solutions is quite large
\begin{equation}
j_R ~=~ \coeff{1}{2}\, n_1 n_5 \,,
\label{maxspin}
\end{equation}
and remains finite as one makes the throat longer and longer by decreasing the parameter $a$. In contrast,  the left-moving angular momentum, $j_L$, becomes arbitrarily small in this limit. Hence, the microstate geometry we consider corresponds to a BMPV black hole with a finite five-dimensional angular momentum (similar to the microstate geometries constructed in \cite{Bena:2006kb, Heidmann:2017cxt})\footnote{If one were to compactify and dualize this solution to a microstate geometry of a four-dimensional D6-D2-D0 black hole, the right-moving angular momentum, $j_R$, (\ref{maxspin}) would be equal to the D0 charge, which would be locked to be the product of two of the D2 charges.}.

While one may wish to consider superstrata with lower values of $j_R$, this value of the charge was the accidental side-effect of a choosing a  ``nearly separable'' superstratum. 

It is also useful to note that 
\begin{equation}
\frac{n_P}{n_1 n_5} ~=~ \frac{ (n+1) \, b^2}{4\,  \big(a^2 + \coeff{1}{2} b^2\big) } ~\sim~  \coeff{1}{2}\, (n+1) \ \ {\rm for} \ \ b \gg a   \,,
\label{mom-c-ratio}
\end{equation}
which means that for $b \gg a$,  $n$ controls the momentum in units of the central charge of the system.

Generically we will take $b \gg a$ because this produces superstrata with deeply-capped BTZ throats that are likely to trap particles for the longest period of time.

The GMS solutions are also three-charge $\frac{1}{8}$-BPS solutions of type IIB supergravity on T$^4$ or $K3$. They also carry  three charges, $Q_1$, $Q_5$ and $Q_P$, and two angular momenta, $J_L$ and $J_R$. Unlike superstrata, the angular momenta are much larger than those of black holes, exceeding the black hole cosmic censorship bound. Therefore, these solutions cannot have a deeply-capped BTZ throat.
 They are determined by five parameters, $Q_5$, $R_y$, $a$, an integer-moded spectral-flow parameter $\mfn$ and an orbifold parameter $k$. They are related to the supergravity charges via
\be 
Q_1 Q_5 \= R_y^2 \, a^2 \,,\qquad Q_p \= \frac{\mfn (\mfn+1)}{k^2} a^2\,,\qquad J_L \= \coeff{1}{2} \,R_y \, \frac{a^2}{k} \,,\qquad J_R \= R_y\,(\mfn+\coeff{1}{2} \,)\, \frac{a^2}{k} \,,
\ee
The supergravity charges are in turn related to the quantized charges via the same compactification relations \eqref{quantcharges} with \eqref{cNdefn}.
%
\subsection{The Schr\"odinger problem}
\label{sec:Sch}

As explained in the Introduction, the fact that GMS solutions are composed only of two regions allows one to use easily matched asymptotic expansions to study their quasi-normal modes in certain limits. We will review this technique in detail in Section \ref{sec:OlderWork}. To cope with the more complex structure of superstrata, we will use the WKB approximation to derive the spectrum of quasi-normal modes. Since this technique has not been so widely used in analyzing supergravity solutions, we will review the key elements.

The family of superstrata we use are very similar to those analyzed in  \cite{Bena:2019azk}, except that the geometries considered here have an asymptotically flat region. The price of adding this region is that there is  no longer a simple recasting of the metric as an S$^3$ fibration over a three-dimensional space.  Moreover, the scalar wave equation is no longer separable.  However, the geometry still behaves as depicted in  Figure \ref{fig:spacetime}(a) and, as we will show in Section \ref{sec:SuperstratumPot}, while no longer separable, the failure of separability is extremely small for solutions with a deeply-capped BTZ throat, and hence we can still use a separated wave equation as an excellent approximation. 

\begin{figure}
\centering
\includegraphics[scale=0.65]{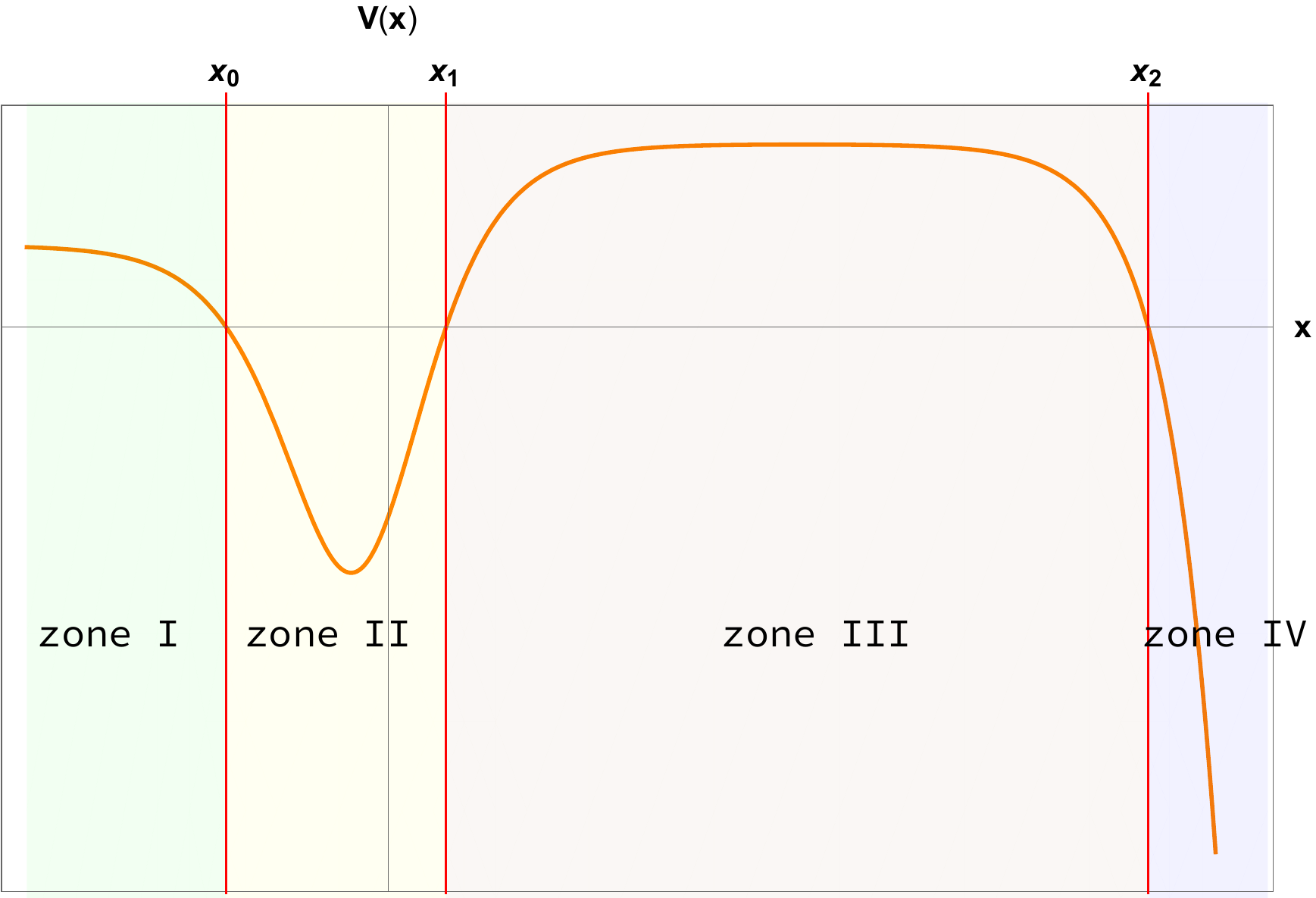}
\caption{Typical form of the potential $V(x)$.  It has four zones, corresponding to the centrifugal barrier, the cap, the BTZ throat and the asymptotically-flat region. Connecting these zones, there  are three classical turning points, $x_i$, where $V(x)$ vanishes.}
\label{fig:pictureofpotential}
\end{figure}

Just as in \cite{Bena:2019azk}, we find that the radial equation for the scalar modes can be recast in an equivalent Schr\"odinger form:
\begin{equation}
\frac{d^2}{dx^2} \,\Psi(x) ~-~ V(x) \, \Psi(x) ~=~  0 \,,
\label{Sch-form}
\end{equation}
for some potential, $V(x)$. The shape of the potential depends on several parameters, but, for the class of quasi-normal modes we wish to consider, the potential takes the form shown in Figure \ref{fig:pictureofpotential}.

There are four zones, delimited by the three classical turning points, $x_i$, defined by $V(x_i)=0$. Zone I is simply the centrifugal barrier at the center of the cap.  This barrier depends on the angular momenta of the mode and can be lowered to zero by considering $S$-waves.  Zone II is induced by the smooth cap geometry, and the lowest-energy states of the system are localized in this potential well.  Zone III corresponds to the barrier that the waves trapped in Zone II need to traverse in order to escape. It reflects the effects of the throat regions on the wave. Zone IV corresponds to the asymptotically flat region, and the potential decays without a lower bound because of the usual energy dilution in flat $\IR^{1,4}$.  

The fact that the potential drops arbitrarily low as $x$ becomes large, means that all the ``bound states'' in the cap are actually quasi-normal modes that will eventually escape to infinity by tunneling through the barrier in Zone III.  Our goal is to compute the quasi-normal excitations and estimate this barrier-penetration rate.

\subsection{The WKB analysis}
\label{sec:WKB}

The Schr\"odinger problem described above can be easily solved using a standard WKB analysis:  In each zone one uses a wave-function that is a superposition of functions of the form:
\begin{equation}
\Psi_\pm (x) ~=~ \left|V(x)\right|^{-\frac{1}{4}}\, \exp\left[ \pm \int^x  V(z)^{1/2} \,dz\right]\,.
\end{equation}
When $V(x)$ is negative these functions oscillate and when $V(x)$ is positive they grow or decay exponentially.  We consider modes that have a centrifugal barrier in Zone I and that therefore decay as $x \to -\infty$. Furthermore, the decay of our quasi-normal modes is captured by requiring outgoing modes as  $x \to +\infty$.  Thus we will need to correlate the boundary conditions at $+ \infty$  with the sign of the frequency.  

The matching at the classical turning points is then done using Airy functions, as in standard WKB problems. The only issue that can arise with this Airy-function matching is when two turning points (for example $x_1$ and $x_2$) are too close to each other; one then has to do a quadratic approximation through $x_1$ and $x_2$, using parabolic cylinder functions (see, for example, \cite{Iyer:1986np}).  Fortunately, for our problem, all the classical turning points are widely separated and we can apply the standard procedure.

We therefore take
\begin{equation}
\Psi(x)=
\left\{
\arraycolsep=0.1pt\def\arraystretch{1.8}
\begin{array}{rll}
& \dfrac{1}{|V(x)|^{\frac{1}{4}}} \left[D^\text{I}_+ \,\exp\left(\int^{x_0}_x |V(z)|^{\frac{1}{2}}dz \right) \,+\, D^\text{I}_- \,\exp\left(-\int^{x_0}_x |V(z)|^{\frac{1}{2}}dz \right) \right],  &\, x<x_0,\\
& \dfrac{1}{|V(x)|^{\frac{1}{4}}} \left[D^\text{II}_+ \,\exp\left(i \int_{x_0}^x |V(z)|^{\frac{1}{2}}dz \right) \,+\, D^\text{II}_- \,\exp\left(-i \int_{x_0}^x |V(z)|^{\frac{1}{2}}dz \right) \right] , &\, x_0<x<x_1,\\
&\dfrac{1}{|V(x)|^{\frac{1}{4}}} \left[D^\text{III}_+ \,\exp\left(\int_{x_1}^x |V(z)|^{\frac{1}{2}}dz  \right) \,+\, D^\text{III}_- \,\exp\left(-\int_{x_1}^x |V(z)|^{\frac{1}{2}}dz \right) \right], &\, x_1<x<x_2,\\
&\dfrac{1}{|V(x)|^{\frac{1}{4}}} \left[D^\text{IV}_+ \,\exp\left(i \int_{x_2}^x |V(z)|^{\frac{1}{2}}dz \right) \,+\, D^\text{IV}_- \,\exp\left(-i \int_{x_2}^x |V(z)|^{\frac{1}{2}}dz \right) \right] , &\,  x > x_2\,.
\end{array}
\right. 
\label{eq:wavefunction3turn}
\end{equation}
Around each turning point, $x\sim x_i$, the wave function behaves as:
\begin{equation}
d_+ \, \text{Bi}\left[\text{sign}(V'(x_i))|V'(x_i)|^{1/3}\,  (x-x_i)\right] \,+\, d_- \, \text{Ai}\left[\text{sign}(V'(x_i))|V'(x_i)|^{1/3}\,  (x-x_i)\right]\,.
\end{equation}

Matching the asymptotics of the Airy functions  to the WKB functions on both sides of each turning point, $x_i$, one can relate  $D^N_\pm$ to $D^I_\pm$ . This gives the \emph{connection formulae}:
\begin{align}
\label{connection3}
\begin{pmatrix} 
D^\text{II}_+ \\ 
D^\text{II}_-
\end{pmatrix} &~=~\begin{pmatrix} 
\frac{1}{2} \,e^{i\frac{\pi}{4}} & e^{-i\frac{\pi}{4}} \\ 
\frac{1}{2} \,e^{-i\frac{\pi}{4}} & e^{i\frac{\pi}{4}} 
\end{pmatrix}  \begin{pmatrix} 
D^\text{I}_+ \\ 
D^\text{I}_-
\end{pmatrix}, \qquad
  \begin{pmatrix} 
D^\text{III}_+ \\ 
D^\text{III}_-
\end{pmatrix} ~=~ \begin{pmatrix} 
-\sin \Theta & 2\, \cos \Theta\\ 
\frac{1}{2} \,\cos \Theta & \sin \Theta
\end{pmatrix}  \begin{pmatrix} 
D^\text{I}_+ \\ 
D^\text{I}_-
\end{pmatrix}, \nonumber\\
\begin{pmatrix} 
D^\text{IV}_+ \\ 
D^\text{IV}_-
\end{pmatrix} &~=~ \begin{pmatrix} 
\frac{1}{2} \,e^{i\frac{\pi}{4}} & e^{-i\frac{\pi}{4}} \\ 
\frac{1}{2} \,e^{-i\frac{\pi}{4}} & e^{i\frac{\pi}{4}} 
\end{pmatrix} \begin{pmatrix} 
\frac{1}{2} \,e^{-T}\cos \Theta & e^{-T}\sin \Theta \\
-e^{T}\sin \Theta & 2\, e^{T}\cos \Theta
\end{pmatrix} \begin{pmatrix} 
D^\text{I}_+ \\ 
D^\text{I}_-
\end{pmatrix},
\end{align}
where  
\begin{equation}
\Theta ~\equiv~  \int_{x_0}^{x_1} |V(z)|^{\frac{1}{2}}\,dz \,,\qquad T ~\equiv~  \int_{x_1}^{x_2} |V(z)|^{\frac{1}{2}}\,dz\,.
\label{Theta&Tdef1}
\end{equation}

The mode is required to decay in the centrifugal barrier and so one must take $D^\text{I}_+ = 0$. For a quasi-normal mode, one must have an outgoing wave for  large $x$.   If we assume that the wave depends on time as $e^{i \omega t}$, then equation (\ref{eq:wavefunction3turn}) implies that the wave function at large $x$ behaves as:
\begin{equation}
\Psi ~\sim~ D^\text{IV}_+ \, \left( \ldots \right) \, \exp \left[i \left( \omega t + f(r) \right) \right] ~+~ D^\text{IV}_- \, \left( \ldots \right) \, \exp \left[i \left( \omega t - f(r) \right) \right]\,,
\end{equation}
where $f(r)$ is a monotonically increasing function of  $r$.  This mode will be outgoing if  $D^\text{IV}_+ =0$ for $\text{Re}(\omega)>0$ and $D^\text{IV}_- =0$ for $\text{Re}(\omega)<0$.

These two boundary conditions lead to the following constraint on the matrix elements in the connection formula:
\begin{equation}
\cos \Theta ~+~ i\,\text{sign}\left( \text{Re}(\omega)\right)\,\frac{e^{-2 \,T}}{4}\, \sin \Theta ~=~ 0\,.
\label{eq:SpectrumWKBGen3TP}
\end{equation}

If we take the tunnelling barrier to be infinite, $e^{-2\,T} \to 0$, we find the standard WKB condition that leads to a tower of (real) bound-state spectrum labelled by a \emph{mode number} $N$:
\begin{equation}
\cos \Theta ~=~ 0 \qquad \Rightarrow \qquad \Theta ~=~ \frac{\pi}{2} ~+~ N\,\pi \,, \qquad N\in \mathbb{N}\,,
\label{eq:RealPartSpectrum}
\end{equation}
The quantity, $ \Theta$, defined in (\ref{Theta&Tdef1}),  depends upon $\omega$, and one uses (\ref{eq:RealPartSpectrum}) to determine the normal modes, $\omega^{(0)}_N$, of the bound states.

Since our superstrata have large but finite barriers, $0 < e^{-2\,T} \ll 1$, we can use perturbation theory to find the leading-order corrections to the spectrum.  
First, one expands  $\Theta$ around $\omega^{(0)}_N$ by taking $\omega ~=~ \omega^{(0)}_N ~+~ \delta \omega_N$:
\begin{equation}
\Theta (\omega) ~\approx~ \Theta(\omega^{(0)}_N)~+~ \delta \omega_N \,\frac{\partial \Theta}{\partial \omega}\bigg|_{\omega = \omega^{(0)}_N} ~=~ \frac{\pi}{2} ~+~ N\,\pi ~+~\delta \omega \, \frac{\partial \Theta}{\partial \omega}\bigg|_{\omega = \omega^{(0)}_N} \,.
\label{eq:Thetaexpansion}
\end{equation}
One also has:
\begin{equation}
\frac{\partial \Theta}{\partial \omega}\bigg|_{\omega = \omega^{(0)}_N} ~=~ \int_{x_0}^{x_1} \bigg( \frac{\partial}{\partial \omega}\,  \left| V(z)\right|^{1/2}\bigg) \bigg|_{\omega = \omega^{(0)}_N}  \, dz\,.
\end{equation}
The contribution from differentiating the endpoints of the integral with respect to $\omega$ vanishes by the fundamental theorem of integral calculus because $V$ vanishes at the end points.

Substituting (\ref{eq:Thetaexpansion}) into (\ref{eq:SpectrumWKBGen3TP}) leads to the leading-order correction:
\begin{equation}
 \delta \omega_N ~=~\coeff{1}{4} \,  i\, \text{sign}\left(\omega^{(0)}_N\right)\, \left(\frac{\partial \Theta}{\partial \omega} \right)^{-1}\,e^{-2 T} \bigg|_{\omega=\omega^{(0)}_N}\,.
\label{eq:ImPartSpectrum}
\end{equation}

There are several things to note.  First, this leading-order correction is purely imaginary.  There will also be shifts in the fundamental frequencies, $\omega_0$, but these arise at the next order in perturbations. Also note that for just about any physical system one has 
\begin{equation}
\text{sign}\left( \text{Re}(\omega)\right)\,\frac{\partial \Theta}{\partial \omega} ~>~ 0 \,.
\label{eq:DerThPos}
\end{equation}
This is because the fundamental frequencies of the system are given by solving  (\ref{eq:RealPartSpectrum}) for $\omega$ as a function of $N$.  The positivity condition (\ref{eq:DerThPos}) simply reflects the fact that the absolute values of the frequencies increase with the mode number.   As a result of (\ref{eq:DerThPos}), we see that the sign in (\ref{eq:ImPartSpectrum}) is precisely the correct one so that $e^{i \omega t}$ becomes a decaying mode, independent of the sign of $\omega$.

Taking this one step further, one can obtain a simple intuitive understanding of (\ref{eq:RealPartSpectrum}).  Recall that for a wave motion of frequency $\omega$ and wave number, $k$, the group velocity is given by $\frac{\partial \omega}{\partial k}$.  For a particle in a box of length, $L$, the wave number, $k$, is given by $k = \frac{2 N \pi}{L}$.  Thus, from (\ref{eq:RealPartSpectrum})  it follows that the group velocity  is given by $\frac{L}{2}\, (\frac{\partial \Theta}{\partial \omega})^{-1}$ and so the time for a round trip across the box (distance $2L$)  is  $4 (\frac{\partial \Theta}{\partial \omega})$.  Therefore the factor 
\begin{equation}
\frac{1}{4} \, \left(\frac{\partial \Theta}{\partial \omega} \bigg|_{\omega=\omega^{(0)}_N} \right)^{-1}\,
\label{eq:ImpactFrequency}
\end{equation}
in (\ref{eq:ImPartSpectrum}) represents the impact frequency of the bound-state wave against the potential barrier.  The factor  $e^{-2 T}$ in  (\ref{eq:ImPartSpectrum})  represents the transition probability per impact, and hence the complete expression represents the inverse time-scale for the decay.  

Finally, recall that  WKB methods work well if the potential is not ``too flat'' near its turning points, and provided that the turning points are widely separated.  In particular, this means that the ``plateau'' between $x_1$ and $x_2$ should be suitably high and wide. This guarantees that $e^{-2 T}$  will also be small and hence our perturbative computation of $\delta \omega$ will also be reliable.  As we will see, these conditions are satisfied by the quasi-normal modes of superstrata with a deeply-capped BTZ throat, as well as by the quasi-normal modes of GMS geometries in the near-decoupling limit that do not have a long capped BTZ throat\cite{Giusto:2004id,Giusto:2004ip,Giusto:2004kj,Giusto:2012yz}.

\section{The radial potential for asymptotically-flat $(2,1,n)$ superstrata}
\label{sec:SuperstratumPot}

Our ultimate goal is to compute the decay rate of deeply-bound states in asymptotically-flat superstrata.  One of the simplifying features of asymptotically-AdS superstrata is that the functions entering in their construction depend only on two variables \cite{Bena:2015bea,Bena:2016ypk}, and there are even simple families in which the massless scalar wave equation is separable \cite{Bena:2017upb,Walker:2019ntz}.   This was used to great effect in the study of bound states and Green functions in \cite{Raju:2018xue,Bena:2018bbd,Bena:2019azk}.  However, in more general superstrata, such as those constructed in \cite{Heidmann:2019xrd}, the functions that enter in the solution depend upon three or more variables and the wave equation fails to be separable.  The situation becomes even more complicated for asymptotically-flat superstrata \cite{Bena:2017xbt}, where even the flat-space analogues of the simplest asymptotically-AdS superstrata typically depend explicitly at least three, or more, variables and separability also fails.   

The key observation, that makes our entire analysis possible, is that there exist certain sufficiently simple asymptotically-flat superstrata in which the decay of perturbations can be computed.  First, if one uses the simplest ``supercharged superstrata'' \cite{Ceplak:2018pws,Heidmann:2019zws}, the geometry once again only  depends on two variables, even for asymptotically-flat superstrata.  Moreover,  there are families of such superstrata that have a ``nearly separable'' massless scalar wave equation.  This means that the wave-equation almost completely separates except for one term that spoils the separation.  Furthermore, we can show that this term can be made parametrically insignificant when the superstrata have a long capped BTZ throat.

We will therefore study the decay of bound states in  these simple ``supercharged superstrata.''  Specifically we will focus on what are known as the $(2,1,n)$ supercharged superstrata, whose asymptotically-AdS forms were constructed in  \cite{Ceplak:2018pws,Heidmann:2019zws}. It is relatively straightforward to generalize these results to obtain asymptotically-flat $(2,1,n)$ superstrata and we will give the solution in Section \ref{sec:SuperstratumGeo}.

The goal of this section is to reduce the problem of solving the massless wave equation in asymptotically-flat $(2,1,n)$ superstrata to solving a radial equation.  This equation comes with a  complicated potential function and we will examine, in considerable detail, its structure and elucidate the physics that emerges in various limits.  While the computational details depend upon the explicit form of this superstratum, we expect the physics that we extract to be a universal property of all superstrata with a deeply-capped BTZ throat.  

\subsection{The $(2,1,n)$ superstrata}
\label{sec:superstrata}

\subsubsection{The CFT states}

The asymptotically-AdS $(2,1,n)$ superstrata were constructed in \cite{Ceplak:2018pws,Heidmann:2019zws} and it is relatively straightforward to generalize these results to obtain asymptotically-flat $(2,1,n)$ superstrata.  

The asymptotically-AdS  $(2,1,n)$ superstrata are dual to coherent states of the D1-D5 CFT peaked around the Ramond-sector state:
 \begin{equation}
|+ \! + \rangle^{N_1} \otimes  |2,1,n,q=1\rangle^{N_2}  
\label{DualStates}
\end{equation}
In this expression,  $|+ \! + \rangle^{N_1}$ is the maximally spinning RR-ground state, and $|2,1,n,q=1\rangle$ is the spectral flow of the ``supercharged'' NS state:
 \begin{equation}
 |2,1,n,q=1\rangle_{NS} ~\equiv~ (L_{-1})^{n}  \left(G_{-\frac12}^{+1}G_{-\frac12}^{+2} + \frac{1}{2}J^+_0 L_{-1}\right)  |O^{--}\rangle_2\,.
 \label{SCstate}
\end{equation}
 The operators are all acting on the right-moving sector of the CFT. Note that we have followed the conventions of   \cite{Heidmann:2019zws}  in which we have re-labelled the states of \cite{Ceplak:2018pws}  by sending $n \to n+1$.  

The numbers, $N_1$ and $N_2$, of these states must  satisfy the constraint  
 \begin{equation}
 N_1 ~+~ 2\, N_2  ~\equiv~ n_1 n_5
\label{constraint1}
\end{equation}
where $n_1$ and $n_5$ are the numbers of D1 and D5 branes.

More details of the holographic dictionary can be found in \cite{Ceplak:2018pws}.  Here we will simply note  the RR states dual to the superstrata have quantum numbers:   
 \begin{equation}
 j_L   ~=~ \coeff{1}{2}\,N_1 \,,  \qquad  j_R   ~=~ \coeff{1}{2}\, (N_1 + 2\,  N_2)~=~  \coeff{1}{2}\,n_1 n_5 \,, \qquad  n_P  ~=~  (n+1)  \,N_2 \,
\label{AngMomP}
\end{equation}

In  the supergravity dual, the number of copies of each fundamental state, $N_1$ and $N_2$, are reflected in two Fourier coefficients, which will be denoted by $a$ and $b$.  The supergravity charges, $Q_1$ and $Q_5$ are proportional to $n_1$ and $n_5$  and the  numbers, $N_1$ and $N_2$,  are proportional to $a^2$ and $\frac{1}{4} b^2$. The supergravity analogue of  (\ref{constraint1}) becomes
 \begin{equation}
 \frac{Q_1 \, Q_5}{R_y^2}    ~=~ a^2 ~+~ \coeff{1}{2}\,b^2 \,, 
\label{smoothness}
\end{equation}
where $R_y$ is the radius of the common D1-D5 direction.  In supergravity this constraint emerges from requiring that the microstate geometry be smooth. The relationship between supergravity and quantized charges is given by \eqref{charges} and \eqref{quantcharges} and the precise details can be found in \cite{Bena:2015bea,Bena:2016ypk,Bena:2017xbt,Heidmann:2019zws}. One relation between supergravity charges and quantized charges that we will often use is
\be 
1 + \frac{b^2}{2\,a^2} \= \frac{n_1 n_5}{2\,j_L}\,.
\ee
This is the parameter that controls the depth of the BTZ throat: the redshift between the cap and infinity.

\subsubsection{The superstratum geometry}
\label{sec:SuperstratumGeo}

The construction technique for superstrata are well-documented in many places (see, for example, \cite{Bena:2015bea,Bena:2016ypk,Bena:2017xbt,Heidmann:2019zws,Shigemori:2020yuo}), and we are simply going to summarize the results of such an analysis. 

Superstrata are most simply described within the six-dimensional $(0,1)$ supergravity obtained by compactifying IIB supergravity  on T$^4$ (or $K3$) and then truncating the matter spectrum to tensor multiplets.     For supersymmetric solutions, the six-dimensional metric takes the form  \cite{Gutowski:2003rg,Cariglia:2004kk}:
\begin{equation}
d s^2_{6} ~=~-\frac{2}{\sqrt{\cP}}\,(d v+\beta)\,\Big[d u+\omega + \frac{\mathcal{F}}{2}(d v+\beta)\Big]+\sqrt{\cP}\,d s^2_4\,,
\label{sixmet} 
\end{equation} 
where
\begin{equation}
u ~\equiv~ \frac{1}{\sqrt{2}}(t-y)\,, \qquad \qquad v~\equiv~\frac{1}{\sqrt{2}}(t+y)
\label{uvty} 
\end{equation} 
are null coordinates and $y$ parametrizes the common $S^1$ of the D1 and the D5 branes.   

In the superstrata considered here, the metric, $ds^2_4$, is simply that of flat $\IR^4$ and it is most convenient to write it in terms of spherical bipolar coordinates:
\begin{equation}
 d s^2_4 =\Sigma \, \left(\frac{d r^2}{r^2+a^2}+ d\theta^2\right)+(r^2+a^2)\sin^2\theta\,d\phi^2+r^2 \cos^2\theta\,d\psi^2\,,
\end{equation} 
where 
\begin{equation}
\Sigma ~\equiv~ r^2+a^2 \cos^2\theta \,. 
\label{Sigdefn} 
\end{equation}

The vector, $\beta$, is chosen to be the potential for a self-dual magnetic field on  $\IR^4$ with a source along $r=0$, $\theta = \frac{\pi}{2}$: 
\begin{equation}
\beta ~=~  \frac{R_y\,a^2}{\sqrt{2}\,\Sigma}\,(\sin^2\theta\, d\phi - \cos^2\theta\,d\psi) \,.
\label{betadefn} 
\end{equation}

The remaining pieces of  (\ref{sixmet}), namely the vector,  $\omega$, that lies in $\IR^4$ and  the functions $\cP$ and $\cF$ are obtained by solving the BPS system following the techniques described in \cite{Bena:2015bea,Bena:2016ypk,Bena:2017xbt,Heidmann:2019zws}.  The data about the CFT states involve exciting particular Fourier modes in the three-form fluxes in the six-dimensional geometry.  However, the fluxes are not relevant to our problem and  so we will simply provide the metric quantities that emerge from solving the BPS system and refer the interested reader to \cite{Ceplak:2018pws} for the tensor fields.\footnote{The conventions for the tensor gauge fields are detailed in Appendix A.2 of \cite{Bena:2018bbd}. Then the pairs $(Z_I, \Theta^I)$ for our solutions can be found applying (6.16) of \cite{Ceplak:2018pws} taking $(k,m,n) \rightarrow (2,0,n-1)$.}
The metric is given by:
\begin{equation}
\begin{split}
\cP &~=~ \left( 1 ~+~ \frac{Q_1}{\Sigma} \right)\, \left(1 ~+~ \frac{Q_5}{\Sigma} \right)\,,\\
\cF &~=~ \frac{b^2 \, n(n+1)(n+2)(r^2+a^2)}{2 a^4} \,\bigg[ -\frac{4+2 n (n+2)\,(1-\Gamma)}{n^2 (n+1) (n+2)^2}~+~\frac{4 \sin ^2 \theta \, (\Gamma +1)}{n^2 (n+1) (n+2)^2}  \\
&\hspace{5.7cm} ~+~ \frac{\cos ^2\,\theta }{n^2} \,\Gamma ^n   ~-~ \frac{ 2 ~+~ \frac{(1-3 n)}{n} \sin ^2\theta}{n
   (n+1)} \, \Gamma ^{n+1}  \\
&\hspace{5.7cm} ~+~ \frac{n+3-(3 n+7) \sin ^2\theta}{(n+1) (n+2)^2} \, \Gamma ^{n+2} ~+~\frac{\sin ^2\theta }{(n+2)^2}\, \Gamma ^{n+3} \bigg] \,,\qquad 
\end{split} \nonumber
\end{equation}
\begin{equation}
\begin{split}
\omega &~=~\omega_0 ~+~ \frac{R_y\,b^2\, \left(a^2+r^2\right)^2}{2 \sqrt{2}\, a^4\, n (n+2) \,\Sigma } \Biggl[ \biggl(2n (n+2) \, (1-\Gamma)^2 - 4 \Gamma +(2 +n(1-\Gamma))^2\,  \Gamma ^{n+1}\biggr) \, \sin^2 \theta \,d\phi \\
& \hspace{5cm}~+~ \Gamma \,\biggl(4 - \left(2 +n(1-\Gamma) \right)^2 \,\Gamma^n \biggr) \,\cos^2 \theta \,d\psi\Biggr] \,,
\label{eq:PFbetaomega21n}
\end{split}
\end{equation}
where 
\begin{equation}
\omega_0  ~\equiv~    \frac{R_y \,a^2}{\sqrt{2}\,\Sigma}\,\left(\sin^2\theta\, d\phi + \cos^2\theta\,d\psi \right)\,, \qquad \Gamma ~\equiv~  \frac{r^2}{r^2 + a^2}\,.
\end{equation}

If one expands the metric (\ref{sixmet}) around infinity using  (\ref{eq:PFbetaomega21n}), one can extract  the angular momenta and momentum  given in (\ref{charges}).

Of particular importance in this paper will be the superstratum geometries that have very long capped BTZ throats, and hence cap off at very high redshift.  The hallmark of these geometries is that  $j_L$ is extremely small compared to the other charges.  From  (\ref{charges}) and (\ref{quantcharges})  it is evident that such solutions arise when:
\begin{equation}
 \qquad \frac{a^2 }{b^2}~\ll~  1 \qquad \Longleftrightarrow \qquad \frac{j_L}{n_1 n_5} \ll 1\,.
\label{eq:dslim}
\end{equation}
In this regime, the three-dimensional manifold parameterized by $(u,v,r)$ corresponds to a highly-redshifted  global AdS$_3$ cap region in the IR, $0 < r \lesssim \sqrt{n} \,a$. Then, as $\Gamma$ transitions from $0$ to $1$, the geometry resembles a BTZ throat.  In particular,  the geometry looks like AdS$_2\times$S$^1$ for $\sqrt{n}\, a \lesssim r \lesssim \sqrt{Q_P}$ and an ``upper'' AdS$_3$ region for $ \sqrt{Q_P}  \lesssim r \lesssim \sqrt{Q_{1,5}}$. 

It is also possible to have $Q_P \gtrsim Q_{1,5}$ (that is $b \gtrsim R_y$). For these charges,  the BTZ throat is reduced to a simple AdS$_2\times$S$^1$ throat that transitions to flat space without any intermediate AdS$_3$ region. As always with  brane configurations, the transition to the asymptotically-flat region occurs when the constants in the warp factors begin to dominate the terms that fall off with the radius.  This happens when $r \gtrsim  \sqrt{Q_I} $ and the metric becomes five-dimensional flat space times the $S^1$ common to the D1 and the D5 branes.

We therefore have three distinct sub-regions that are depicted in  Fig.\ref{fig:spacetime}(a):
\begin{itemize}
\item \underline{A global AdS$_3\times$S$^3$ cap region in the IR:}

The cap geometry is obtained by taking the limit $r\lesssim \sqrt{n} a$ (corresponding to $\Gamma^n \sim 0$) in \eqref{eq:PFbetaomega21n}. We decompose the six-dimensional cap metric as an S$^3$ fibration: 
\begin{equation}
\begin{aligned}
 ds_\text{cap}^2 ~=~   \sqrt{Q_1 Q_5} \,  \bigg[&  \;\! \frac{dr^2}{r^2 + a^2} - \frac{ (r^2 + a^2) }{a^2 \,R_y^2} \, d\tau^2  + \frac{r^2}{a^2 \,R_y^2}\left(dy +\frac{b^2}{2a^2} d\tau \right)^2  \+ \, d\theta^2  \,+\, {d\widehat{\Omega}_2}^2 \bigg] 
\,.
\end{aligned}
\label{eq:capmet}
\end{equation}
where $\tau \equiv \left(1+\frac{b^2}{2a^2}\right)^{-1}\,t \= \frac{2\,j_L}{n_1 n_5}\,t$ is the redshifted time and the  ${d\widehat{\Omega}_2}^2$ is the metric on S$^2$:
\begin{alignat}{1}
{d\widehat{\Omega}_2}^2 &\,=\, \sin^2 \theta \, \left(d\phi -\frac{d\tau}{R_y} + \frac{2 \,b^2\, r^4}{n(n+2)\,a^4\,(2 a^2 +b^2) R_y}\, (dt+dy) \right)^2 \\
&  \hspace{0.5cm}  + \,\cos^2 \theta \, \left(d\psi - \frac{dy}{R_y} -\frac{b^2}{2a^2}\frac{d\tau}{R_y} -  \frac{2 \,b^2\, r^4}{n(n+2)\,a^4\,(2 a^2 + b^2) R_y}\, (dt+dy)   \right)^2\,, \notag
\label{eq:metriccontrSup}
\end{alignat}
The $(r,\tau,y)$ manifold defines a (hugely red-shifted and boosted) global  AdS$_3$. The $d\widehat{\Omega}_2$ term in (\ref{eq:capmet}) give the metric on the $U(1) \times U(1)$ defined by $(\phi,\psi)$. The $\phi$-circles and $\psi$-circles universally pinch-off  at $\theta=0$ and $\theta=\frac{\pi}{2}$, respectively and so the $(d\theta , d\phi, d\psi)$ components  describe a round S$^3$ with non-trivial fibering over the three-dimensional space-time.

\item \underline{An intermediate S$^3$ fibration over a BTZ throat:}

For $ \sqrt{n} a \lesssim r \lesssim \sqrt{Q_{1,5}}$, we can approximate $\Gamma \approx 1$ and $\cP \approx \frac{Q_1 Q_5}{\Sigma^2}$. The metric reduces to:
\begin{equation}
\begin{aligned}
 ds_\text{BTZ}^2 ~=~   \sqrt{Q_1 Q_5} \,  \bigg[&  \;\! \frac{d\rho^2}{\rho^2} \,-\, \rho^2 \,\(dt^2-dy^2\)  \,+\,  \frac{2n+1}{4 \, R_y^2}\, (dy+dt)^2  \;\!  \\
  & + \, d\theta^2  \,+\,  \sin^2\theta \, \left(d\phi -\frac{dt+dy}{2 \,R_y}\right)^2 \,+\, \cos^2 \theta \, \left(d\psi - \frac{dt+dy}{2\,R_y} \right)^2  \;\! \bigg] \,,
\end{aligned}
\label{eq:BTZmetric21n}
\end{equation}
where $\rho \equiv \frac{r}{\sqrt{Q_1 Q_5}}$. This is simply a trivial S$^3$ fibration over a red-shifted extremal BTZ geometry. The left and right temperatures are 
\begin{equation}
T_L \,=\, \frac{\sqrt{2n+1}}{4\pi\, R_y}\,~, \qquad T_R \,=\,0\,.
\end{equation}

\item \underline{The product of   flat five-dimensional space-time and the common D1-D5 circle:}

For $\sqrt{Q_I} \lesssim r$, all quantities in the metric converge to a constant or to zero, giving  \be 
ds_\text{Flat}^2 \= -dt^2 \+ dy^2 \+ dr^2 \+ r^2 \left( d\theta^2 \+ \sin^2\theta \,d\phi^2 \+ \cos^2\theta \,d\psi^2 \right)\,.
\ee
\end{itemize}

Henceforth we will assume that $n$ is large.  This greatly simplifies the structure of the metric without losing the essential physics. 
This assumption means that the global AdS$_3$ cap will be large in units of the AdS radius, and hence will contain a large number of bound states. The bound states that localize in the cap only have small interactions with the rest of the geometry and, as we will see, their decay can be treated accurately in perturbation theory. 

\subsection{Scalar wave excitations}
\label{sec:Scalars}

We will look at the behavior of massless scalar modes satisfying 
\begin{equation}
\frac{1}{\sqrt{-\det g}} \: \partial_M \left( \sqrt{-\det g}\: g^{MN} \partial_N \, \Phi \right) \:=\: 0\,,
\label{eq:genericKleinGordon}
\end{equation}
where $g_{MN}$ is the six-dimensional metric defined in \eqref{eq:PFbetaomega21n}. Since the geometry is independent of $u$, $v$, $\phi$, and $\psi$, we can decompose the scalar into Fourier modes along these directions: 
\begin{equation}
\Phi = H(r,\theta)\,e^{i\left( \frac{\sqrt{2}\,\Omega}{R_y} \,u\,+\,\frac{\sqrt{2}\, P}{R_y} \, v \,+\,q_\phi \phi \,+\, q_\psi \psi \right)}\,.
\label{eq:modeprofile}
\end{equation}
The wave equation becomes  an expression of the form:
\begin{equation}
\bar{\cL}\, H(r,\theta) - \cV_r (r) \, H(r,\theta)  -  \cV_\theta (\theta) \, H(r,\theta)\, - \cW (r,\theta) \, H(r,\theta)\,=\, 0,
\label{eq:WEgenalpha}
\end{equation}
where we have defined the Laplacian operator, $\bar{\cL}$, via
\begin{equation}
\bar{\cL} \,\equiv\, \frac{1}{r}\, \partial_r \big( r (r^2 + a^2) \, \partial_r  \big)  ~+~  \frac{1}{\sin \theta \cos \theta}\partial_\theta \big( \sin \theta \cos \theta\, \partial_\theta   \big) .
\end{equation}
The angular and radial potentials, $V_r (r)$ and $V_\theta (\theta)$, and the non-separable term, $\cW(r,\theta)$, are given by:
\begin{equation}
\begin{split}
 \cV_\theta (\theta) \,\equiv \, &\frac{q_\phi^2}{\sin^2 \theta} +  \frac{q_\psi^2}{\cos^2 \theta} - \frac{4 a^2\, \cos^2\theta}{R_y^2} \,P\,\Omega\,, \\
\cV_r(r) \,\equiv\, &\cV_\text{asymp}(r) +  \frac{a^2 (q_\psi+P-\Omega)^2}{r^2} - \frac{a^2 \left(q_\phi+P+\left(1+\frac{b^2}{a^2}\right)\Omega\right)^2}{r^2+a^2}+ \frac{2\, a^2+b^2}{a^2 (n+2)+2 r^2} \frac{2b^2\,\Omega^2}{a^2} \,\\
& ~-~\frac{4 \,b^2 \Omega }{a^2 n (n+2)}\, F(r)\, \left[ q_\phi~+~P~+~\frac{(n+2) \left(a^2 n+\frac{b^2}{2} (n+1)\right)+\frac{b^2}{a^2}\, r^2}{a^2 (n+2)+2 r^2}\,\Omega   \right.\\
& \left. \hspace{3.7cm}-\frac{r^2+a^2}{a^2} \left(q_\phi ~-~q_\psi ~+~ \frac{b^2\,r^2}{a^4 \,n(n+2)} \,F(r) \,\Omega \right)\,\right]\,,\\
\cW(r,\theta) \,\equiv\, & \frac{4 b^2 \Omega^2 \cos^2\theta}{a^2\,R_y^2}\left[ \left(1-F(r) \right) \,\frac{r^2\, \left(2 (r^2+a^2)-a^2 n(n+1)\right)}{n (n+2) \,\left(a^2(n+2)+2 r^2 \right)} ~+~ \frac{a^2}{2} - \frac{r^2}{n(n+2)} \right.\\
& \left.\hspace{2.4cm} + \frac{2 r^2 \left( Q_1+Q_5+r^2+a^2 \cos^2\theta\right)}{a^2 n (n+2)} \right.\\
& \left.\hspace{2.4cm}  \times \left(1+\frac{a^2}{2r^2}-\left(1+\frac{a^2 (n+1)}{2 r^2}-\frac{a^2 n}{a^2(n+2)+2 r^2} \right) \left(1-F(r) \right)\right) \right]\,,
\label{eq:potentialsgen}
\end{split}
\end{equation}
where we have introduced a transition function, $F(r)$, between the cap and the outer region, and an asymptotic potential $\cV_\text{asymp}(r)$, which represents  the difference between the asymptotically-flat superstrata and the asymptotically-AdS$_3$ superstrata:
\begin{align}
\cV_\text{asymp}(r) ~\equiv~  &- \frac{4\,\Omega}{R_y^2}\left(Q_1+Q_5+r^2 \right)\label{eq:Vasymp} \\
&\qquad  \times \left[ P ~+~ \frac{b^2\,\Omega}{2a^4 n (n+2)} \frac{a^4 n(n+2)^2~+~ 2 r^2 \left( a^2 n (n+1) -2 (r^2+a^2)\right) \,F(r)}{a^2 (n+2) +2 r^2} \right],  \nonumber\\
F(r) ~\equiv~ & 1 ~-~ \left( \frac{a^2 (n+2)+2 r^2}{2\,(a^2+r^2)}\right)^2 \, \left(\frac{r^2}{a^2+r^2}\right)^n \label{eq:TransitionFunct}
\end{align}
Moreover, the wave profile \eqref{eq:modeprofile} must be $2\pi R_y$-periodic along $y$ which requires
\begin{equation}
q_y ~\equiv~ P-\Omega ~ \in~ \mathbb{Z} \,.
\label{eq:qyquant} 
\end{equation}
We now have three integer-moded quantum numbers related to the periodicities along $(y,\phi,\psi)$
\begin{equation}
\{ q_y \,, \, q_\phi \,,\, q_\psi \} ~ \in~  \mathbb{Z} \,.
\label{eq:allquant} 
\end{equation}

We conclude by noting that for quasi-normal modes, energy must be able to leak out at infinity and so  the potential must be negative at large $r$. This implies
\be 
\Omega P = \Omega \,(\Omega+q_y) >0\,.
\label{eq:condPositivityOmegaqy}
\ee
In contrast, the potential for the modes with  $\Omega P \leq 0$ is positive at infinity, and hence these modes are ``eternally trapped.''  We will discuss the modes further in Section \ref{sec:Conclusions}, and restrict our attention here to the quasi-normal modes.

\subsection{Particular limits of the scalar potential}
\label{ss:ScalarLims}

\subsubsection{Separability}

We begin by noting that the failure of separability of the wave equation is encapsulated entirely in the term  $\cW(r,\theta)$ in (\ref{eq:WEgenalpha}) and defined in (\ref{eq:potentialsgen}). 

First, we note that $\cW$ is proportional to $\Omega^2$ and, as we will show, it is extremely small for the lowest-energy quasi-normal modes.  However, independent of its coefficient, $\cW$ is also parametrically small when the BTZ throat is long.   

If one examines $\cW$ one can see that it contains  terms that could also be moved into the separable terms.  In fact, in defining $\cW$ we have been careful to adjust these terms so that $\cW$ is {\it parametrically smaller} than  $\cV_{\Omega^2}$, the coefficient of $\Omega^2$  in $\cV_r(r)$.  That is,
\begin{equation}
\left| \frac{\cW}{\cV_{\Omega^2}} \right| \xrightarrow[r\sim 0, +\infty]{} 0 \,,\qquad \left| \frac{\cW}{\cV_{\Omega^2}} \right| ~\sim~  \cO\left(\frac{a^2}{Q_{min}} \right)  \,, \qquad Q_{min} ~\equiv~ {\rm min}(Q_1, Q_5) \,.
\label{eq:potbound}
\end{equation}
Thus  $\cW$ is also negligible for all values of $\Omega$. 

For geometries with a deeply-capped BTZ throat with $Q_1 \sim Q_5$, Equation (\ref{smoothness}) implies $\frac{a^2}{Q_{min}}  \sim \frac{a^2}{b R_y}$, which is indeed small.  The negligibility of $\cW$ is then independent of the parameters of the mode $\{\Omega,\, q_y,\, q_\phi,\, q_\psi\}$ and only relies on having a solution with a long throat. 

We can therefore neglect $\cW$, and take $$H(r,\theta)  = K(r)\, S(\theta) \,\left(1+\cO\(\frac{a^2}{Q_{min}} \right) \right).$$ The wave equation  (\ref{eq:WEgenalpha}) then reduces to:
\begin{equation}
\begin{split}
\frac{1}{r}\, \partial_r \big( r (r^2 + a^2) \, \partial_r \,K(r) \big) ~-~ \cV_r (r) \, K(r)& ~=~ \lambda  \, K(r)\,,\\
\frac{1}{\sin \theta \cos \theta}\partial_\theta \big( \sin \theta \cos \theta\, \partial_\theta \,S(\theta) \big) ~-~  \cV_\theta (\theta) S(\theta) & ~=~  - \lambda \, S(\theta)\,.
\end{split}
\label{eq:separated}
\end{equation}
The second equation is almost, but not quite, the wave equation on a round S$^3$:
\begin{equation}
\frac{1}{\sin \theta \cos \theta}\partial_\theta \big( \sin \theta \cos \theta\, \partial_\theta \,S(\theta) \big) ~-~ \bigg(\frac{q_\phi^2}{\sin^2 \theta} +  \frac{q_\psi^2}{\cos^2 \theta} - \frac{4 a^2\, \cos^2\theta}{R_y^2} \,P\,\Omega\bigg) \,S(\theta) ~=~ - \lambda \, S(\theta)  \,.
\label{eq:S3modes}
\end{equation}
Without the last term (proportional to $P\Omega$), the smooth solutions of this equation are Jacobi polynomials and 
\begin{equation}
\lambda ~=~ \ell (\ell +2) \,, \qquad \ell \in \mathbb{N}  \,.
\label{eq:lamval}
\end{equation}
The $(P\Omega)$-term in (\ref{eq:S3modes}) comes directly from the coupling of the geometry to flat space and it arises in other investigations similar to ours (see for example, \cite{Chowdhury:2007jx,Chakrabarty:2015foa,Eperon:2016cdd, Chakrabarty:2019ujg}). We would like to argue that this term will only cause a small correction to the spectrum (\ref{eq:lamval}): 
\be 
\lambda ~=~ \ell ( \ell +2)  \,+\, \cO \left(\frac{a^2\,P \Omega}{R_y^2} \right) \,.
\ee
To make this correction parametrically small, we will take $a^2 \ll R_y^2$ and we will prove later that bound states have $P \Omega$ that scales at large $\ell$ as $\frac{a^4}{b^4} \, \ell^2$. Thus, once again, having a geometry with a long black-hole-like throat is enough to consider that the eigenvalue, $\lambda$, is given by (\ref{eq:lamval}) at leading order.

Without the term proportion to  $P\Omega$, the angular wave equation is exactly solvable and gives
\be 
S(\theta)\,\propto \, (\sin\theta)^{|q_\phi|}\, (\cos\theta)^{|q_\psi|}\,{}_2 F_1 \left(-\frac{1}{2}(\ell -|q_\phi|-|q_\psi|),\,\frac{1}{2}(\ell+2+|q_\phi|+|q_\psi|),\,|q_\psi|+1,\,\cos^2\theta \right)\,,
\label{eq:AngularWaveProfile}
\ee
which is regular at $\cos^2\theta =1$ if and only if one imposes the bound
\be 
|q_\psi| + |q_\phi| \leq \ell\,.
\label{eq:RegAngForm}
\ee

\subsubsection{Schr\"odinger form, the large-$n$ limit and regions of the geometry}

As in \cite{Bena:2019azk}, all the interesting physics is encoded in the potential function, $\cV_r(r)$.  The apparent complexity of its form \eqref{eq:potentialsgen} can be removed by dissecting it into various limits. Indeed, as explained in Section \ref{sec:SuperstratumGeo}, the superstratum geometry can be thought of being composed of a global-AdS$_3$ cap at small $r$, a BTZ throat in the middle, and an asymptotically-flat region at large $r$.  We will show that the scalar wave equation reflects this geometric structure.  We will also significantly simplify the discussion by taking the large-$n$ limit.  

We first convert the radial equation into an equivalent Schr\"odinger problem.  There is an infinite number of ways to do this, but we will use the approach of \cite{Bena:2019azk}, which was particularly effective and simple. Thus, we use:
\begin{equation}
K(r) ~\equiv~  \frac{\Psi(r)}{\sqrt{r^2+a^2}}\,,\qquad x ~\equiv~ \log \frac{r}{a}\,,\qquad x \in \mathbb{R}\,.
\end{equation}
The radial wave equation gives
\begin{equation}
\frac{d^2}{dx^2} \Psi(x) ~-~  V(x) \Psi(x) ~=~  0\,.
\end{equation}
where $V(x)$ is given by:
\begin{equation}
V(x) ~\equiv~  \frac{e^{2 x}}{e^{2 x}+1} \left[\,\(\ell+1 \)^2 + \frac{1}{e^{2 x}+1}~+~\cV_r(a\,e^x)\,\right ]\,.
\label{eq:potentialform21ngen}
\end{equation}

In terms of the geometry, the large-$n$ limit produces a large, highly-redshifted global AdS$_3$ cap region, with $0 < r \lesssim \sqrt{n} \,a$, which, at larger radius, transitions to the BTZ throat.  From the point of view of the scalar wave equation, this transition is driven by the behavior of the  transition function, $F(r)$, defined in (\ref{eq:TransitionFunct}), as it  goes from $1$ to $0$. The region with $F(r) \sim 1$ is the cap, while the bump in $F(r)$ that occurs at $ r \sim \sqrt{n} a$ corresponding to the beginning of the BTZ throat, as depicted Fig.\ref{fig:spacetime}(a).   

\begin{figure}
\begin{adjustwidth}{-1.5cm}{-1.5cm}
 \centering
\begin{tabular}{ccc}
\subf{\includegraphics[height=4.45cm]{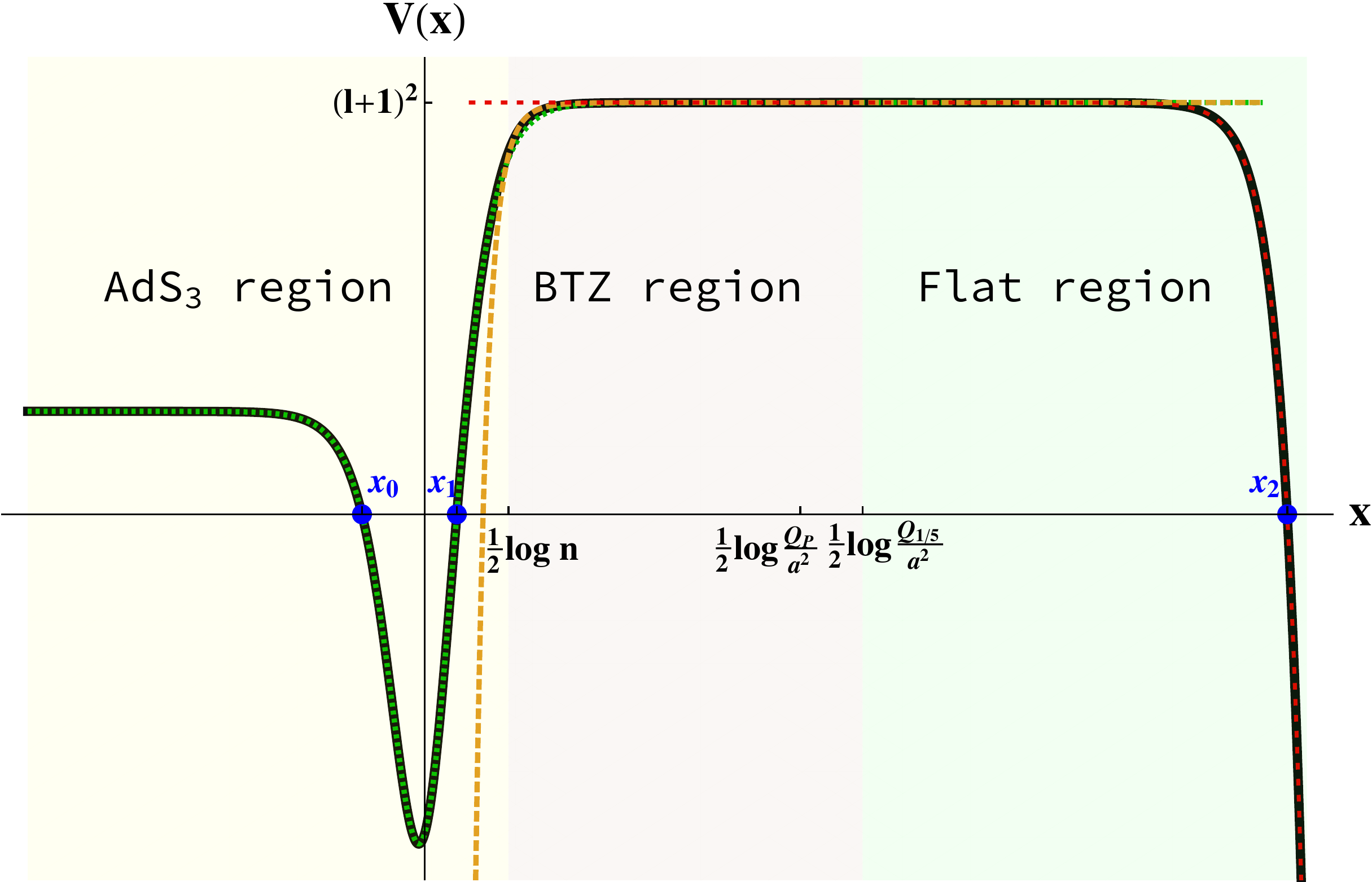}}
     {\\
     (a) When $N \lesssim \sqrt{n}\,\ell$.}
&
\subf{\includegraphics[height=4.45cm]{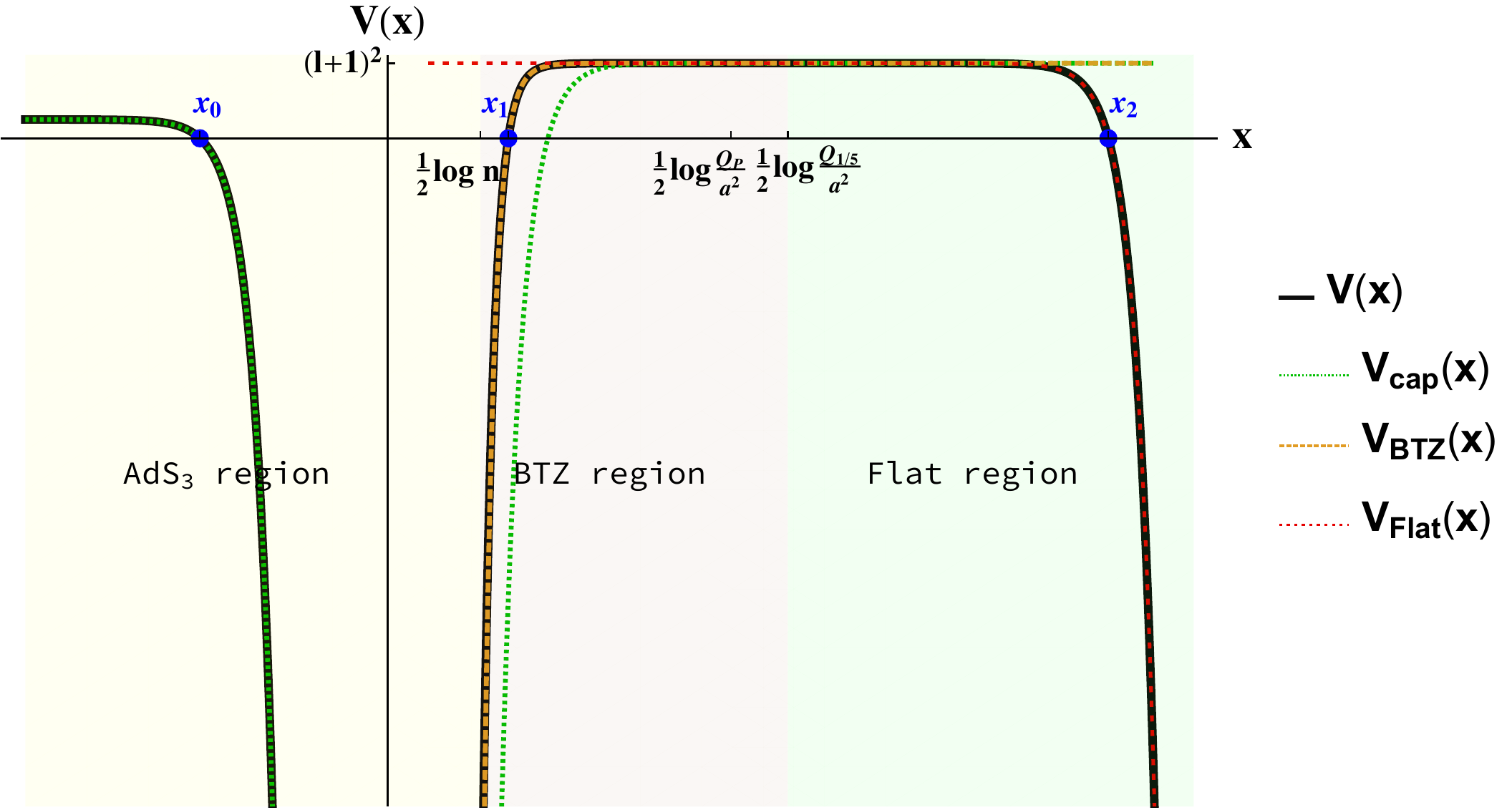}}
     {\\
     (b) When $N \gtrsim \sqrt{n}\,\ell$.}
\end{tabular}
\end{adjustwidth}
\caption{The potential $V(x)$ of a scalar field in an asymptotically-flat (2,1,n) superstratum with a deeply-capped BTZ throat and the three approximated potentials $\{V_\text{cap}(x),\,V_\text{BTZ}(x),\, V_\text{Flat}(x)\}$ \eqref{eq:ApproximatedPotentials}. The two figures are in different regimes of $(N,\ell)$ where $N$ is the mode number of the quasi-normal modes \eqref{eq:modeBehavior}.}
\label{fig:pictureofpotential21N}
\end{figure}

The transition to the asymptotically-flat region occurs when the $r^2$ term begins to dominate over $(Q_1 + Q_5)$ in the overall factor of 
$\cV_\text{asymp}(r)$ in (\ref{eq:Vasymp}). Thus, the asymptotically-flat region begins for $r \gtrsim  \sqrt{Q_{1,5}} $.  We therefore have three distinct sub-regions of the geometry in which we can simplify the potential.  These are the yellow, brown and green regions depicted in Fig.\ref{fig:pictureofpotential21N}.

\begin{itemize}
\item[-] \underline{\bf The global AdS$_3$ cap region:} when $0 < r \lesssim \sqrt{n} \,a$, or $x\lesssim \frac{1}{2}\log n$:  

\smallskip
The potential is well-approximated by the potential of a scalar field in a  AdS$_3$ background (green dotted curve in Fig.\ref{fig:pictureofpotential21N}):
\begin{equation}
\begin{split}
V(x) ~\sim~  V_\text{cap}(x) \equiv  \frac{e^{2 x}}{e^{2 x}+1} &\left[ \,\(\ell+1 \)^2 - 4 \,\frac{(Q_1+Q_5)\,P+Q_P\, \Omega}{R_y^2}\,\Omega ~+~ e^{-2 x}\, (q_\psi+P-\Omega)^2 \right. \\
&\,\, \left.  ~-~  \frac{1}{e^{2x}+1}\left( \left(q_\phi+P+\left(1+\frac{b^2}{a^2}\right)\Omega\right)^2 -1 \right)  \,\right]\,,
\end{split}
\label{eq:Vcap}
\end{equation}
As we will see, the factor $1+\frac{b^2}{a^2}$ make the cause the values of $\Omega$ where bound states occur to be highly redshifted compared to a simple global AdS geometry.

The form of this potential is simple because we have taken the large-$n$ limit:  At large $n$, the last term of the first line of $\cV(r)$, \eqref{eq:potentialsgen}, and all the terms in the second and third lines are negligible.  As explained in \cite{Heidmann:2019zws}, the cap structure is more complicated for small $n$.


\medskip
\item[-] \underline{\bf The BTZ region:}  when $ \sqrt{n} \,a \,\lesssim\, r \,\lesssim \,\sqrt{Q_{1,5}}$, or $\frac{1}{2}\log n\,\lesssim\, x\,\lesssim\, \frac{1}{2}\log \frac{Q_{1,5}}{a^2} $:

\smallskip
The potential is well-approximated by the potential of a scalar field in an extremal BTZ black-hole (orange dotted curve in Fig.\ref{fig:pictureofpotential21N}):
\begin{equation}
\begin{split}
V(x) ~\approx~ & V_\text{BTZ}(x) \\
 \equiv \,&\(\ell+1 \)^2 ~-~ 4 \,\frac{(Q_1+Q_5)\,P+Q_P\, \Omega}{R_y^2}\,\Omega \\ 
& - \frac{b^2}{a^2}\,\Omega \left(2 P+q_\phi +q_\psi+\frac{(1+n)(Q_1 +Q_5)}{R_y^2}\,\Omega \right)\, e^{-2 x}\
-\frac{b^4 (1+2n)\,\Omega^2}{4 a^4} \, e^{-4 x}\,.
\end{split}
\label{eq:VBTZ}
\end{equation}
The form of the potential is the same as one would have in a standard BTZ geometry  (see, for example, \cite{Bena:2019azk}) except that the parameters have been shifted by constants   proportional to $\frac{Q_I}{R_y^2}$.  These terms arise through the gluing to flat space.


\medskip
\item[-] \underline{\bf The flat region:} when  $ \sqrt{Q_{1,5}} \, \lesssim r $, or $ \frac{1}{2}\log \frac{Q_{1,5}}{a^2} \, \lesssim x $:

\smallskip
The potential is well-approximated by the potential of a scalar field in flat space, which is shown as a red dotted curve in Fig.\ref{fig:pictureofpotential21N}):
\begin{equation}
V(x) ~\approx~ V_\text{Flat}(x) \equiv\(\ell+1 \)^2~-~ 4 \,\frac{(Q_1+Q_5)\,P+Q_P\, \Omega}{R_y^2}\,\Omega ~-~ \frac{4 \,a^2\,\Omega\, P}{R_y^2}\,e^{2x} \,.
\label{eq:Vflat}
\end{equation}
This reflects the relative roles of the three-dimensional mass, $(\ell+1)$, and the asymptotic decay of the energy and momentum at large $x$. If $\Omega\, P$ is positive, the last term is negative and ``destabilizes'' the bound states at the cap to produce quasi-normal modes. However, if the last term is negative, the modes will be trapped forever in the geometry. 

\end{itemize}
 As we will show below, the bound states will have frequencies quantized in units of 
\begin{equation}
\frac{2 j_L}{n_1n_5} ~=~  \frac{a^2 R_y^2}{Q_1 Q_5} ~\sim~ \frac{2 a^2}{b^2} ~\ll 1~ \,.
\label{eq:quantunit}
\end{equation}
This is a consequence of the huge red-shift created by the long capped BTZ throat of the microstate geometry.

We label the modes by a  \emph{mode number} $N\in \mathbb{N}$ and their frequencies will behave as
\begin{equation}
\Omega_N ~\approx~  \frac{2 j_L}{n_1 n_5}  \times \cO\(N \)  ~\approx~  \frac{2 a^2}{b^2}  \times \cO\(N \)\,.
\label{eq:modeBehavior}
\end{equation}
For $N \lesssim\ell^2 \, \frac{Q_{1,5}}{a^2}$, we can simplify the constant term, 
$$\(\ell+1 \right)^2 - 4 \,\frac{(Q_1+Q_5)\,P+Q_P\, \Omega}{R_y^2}\,\Omega \,\approx \, \(\ell+1 \right)^2\,,$$ 
in each potential and work with\footnote{We have replaced $P$ by the quantized momentum charge $q_y$ in \eqref{eq:qyquant}.}:
\begin{equation}
\begin{split}
V_\text{cap}(x) & ~\equiv~  \frac{e^{2 x}}{e^{2 x}+1} \left[ \(\ell+1 \)^2 + e^{-2 x}\, (q_\psi+q_y)^2 -  \frac{1}{e^{2x}+1}\bigg(\bigg(q_\phi+q_y+2\bigg(1+\frac{b^2}{2a^2}\bigg)\Omega\bigg)^2 -1 \bigg) \right]\,, \\
V_\text{BTZ}(x) &~\equiv~   \,\(\ell+1 \)^2  - \,\frac{b^2\, \Omega \left(2q_y+q_\phi +q_\psi + \left(2+\frac{(1+n)(Q_1 +Q_5)}{R_y^2}\right)\,\Omega \right)}{a^2}  e^{-2 x} - \,\frac{b^4 (1+2n)\,\Omega^2}{4 a^4} e^{-4 x} \,,\\
V_\text{Flat}(x) &~\equiv~  \(\ell+1 \)^2 - \frac{4 \,a^2\,\Omega\, \(\Omega+q_y\)}{R_y^2}\,e^{2x} \,.
\label{eq:ApproximatedPotentials}
\end{split}
\end{equation}
For $N\gtrsim \ell^2 \, \frac{Q_{1,5}}{a^2} $, the constant term starts to be negative and quasi-normal modes no longer exist.

\subsubsection{Energy regimes,  mode numbers and mass }
\label{sec:EnergyRegimes}

One can arrive at an even simpler picture of the bound-state physics if one thinks about energetics in terms of the mode number, $N$, and the three-dimensional mass, $\ell$, of the six-dimensional massless mode.  Indeed, a closer study of Fig.\ref{fig:pictureofpotential21N}(a), suggests that the approximate potentials, $V_\text{cap}(x)$ and $V_\text{Flat}(x)$ actually match the full potential at low energy far outside the ranges in $r$ described above.  This means that we can think of the physics in this regime as being controlled by the highly red-shifted AdS cap transitioning directly to flat space.

As noted above, the cap potential is a good approximation for $ r \lesssim \sqrt{n} \,a$, or $x\lesssim  \frac{1}{2} \log n$.   For small $N$ and large $\ell$, the modes are strongly trapped by the gravitational potential and hence become localized in the cap and do not feel the other features of the full geometry.   

The potential barrier for tunneling is set by the barrier height, $(\ell +1)$, and so  the relevant question is when the potential starts to level off at this value. One possibility is that this transition is in the cap region and determined by the cap potential as in Fig.\ref{fig:pictureofpotential21N}(a); one sees that this occurs when the last two terms of $V_\text{cap}(x)$ in (\ref{eq:ApproximatedPotentials}) become smaller than the first term.  This happens when
\begin{equation}
\left( \frac{b^2 }{a^2} \,\Omega + q_\phi-q_\psi \right)\left( \frac{b^2 }{a^2} \,\Omega + q_\phi+q_\psi +2 q_y\right)\, e^{-2 x}~\lesssim~  (\ell+1)^2    \,.
\label{eq:barriertop}
\end{equation}
Using \eqref{eq:modeBehavior}, this is equivalent to
\begin{equation}
\left( 2 N  + \ldots \right)\left(2 N  + \ldots \right)\, e^{-2 x}~\lesssim~  (\ell+1)^2    \,.
\label{eq:barriertop2}
\end{equation}
Given that the cap region is approximately at $x\lesssim  \frac{1}{2} \log n$, we see that the transition is indeed in the cap region as soon as the mode numbers, $N$, is in the range 
\begin{equation}
N ~\lesssim~  \sqrt{n}\, \ell \,,
\label{eq:lowestmodes}
\end{equation}

Thus for the lowest modes, satisfying (\ref{eq:lowestmodes}),  the long BTZ throat plays a relatively minor role in interpolating between $V_\text{cap}(x)$ and $V_\text{Flat}(x)$.   One should note that the BTZ throat plays an essential role in the physics of superstata as it is enables the existence of an extremely highly-redshifted cap.  Moreover,  in the Green function computations of \cite{Bena:2019azk}, the BTZ throat led to thermal decay at intermediate times.   However, from the point of view of the lowest lying bound states and of the quasi-normal modes,  all that really matters is that the cap is there and that it transitions smoothly to flat space.

This leads to the following picture.

\begin{itemize}
\item[-] \underline{\bf The low-energy regime:}

When the mode number is bounded by (\ref{eq:lowestmodes}), the wave is essentially contained in the IR AdS$_3$ cap. Its potential will be well-approximated by the highly-redshifted AdS$_3$ potential glued to flat space and the BTZ part of the geometry has a negligible effect, as depicted in Fig.\ref{fig:pictureofpotential21N}(a):
\be
V(x)~\approx~\left\{\begin{array}{ll}
V_\text{cap}(x)\,,\qquad  & x \lesssim \frac{1}{2}\log \frac{Q_{1,5}}{a^2}\,,\\
V_\text{Flat}(x)\,,\qquad \qquad & x \gtrsim \frac{1}{2}\log \frac{Q_{1,5}}{a^2}\,.
\end{array}\right.
\label{eq:capregimepot}
\ee

In this sense, the physics here is similar to the analyses of quasi-normal modes  in other AdS$_3$ geometries that are glued to flat space in the UV \cite{Eperon:2016cdd,Chakrabarty:2019ujg}.  The important difference in our work is that  we have more parameters to control the depth of the throat and our AdS$_3$ region is highly redshifted, by a factor $\frac{n_1 n_5}{j_L}$.   One should therefore expect similar results to those of  \cite{Eperon:2016cdd,Chakrabarty:2019ujg} {\it except} that our frequencies are quantized in units of $\frac{2 j_L}{n_1 n_5}$ with arbitrarily low $j_L$.  This will  lead to a much slower decay rates for the modes trapped in superstrata with deeply-capped BTZ throats.

\item[-] \underline{\bf The intermediate-energy regime:}

When the mode number lies in the range $N \gtrsim \sqrt{n}\,\ell$, as depicted in Fig.\ref{fig:pictureofpotential21N}(b), the energy level is large enough for the wave to explore the BTZ throat of the geometry. In this regime, one necessarily has to make use of the details of the BTZ potential, $V_\text{BTZ}(x)$, to describe the transition from the cap potential to the flat potential.  One might therefore expect to find some effects of the BTZ throat on the spectrum of quasi-normal modes. 

We call this regime the ``intermediate-energy regime'' to differentiate between the high-energy modes, with $N\gtrsim \ell^2 \, \frac{Q_{1,5}}{a^2} $, that correspond to a potential where the barrier starts to be negative and where quasi-normal modes no longer exist.
\begin{figure}[ht]
\begin{center}
 \usetikzlibrary{snakes}
 \definecolor{myred}{rgb}{0.7,0.2,0.2}
 \definecolor{mygreen}{rgb}{0.1,0.6,0.3}
\definecolor{myblue}{rgb}{0.2,0.2,0.7}
\definecolor{myyellow}{rgb}{0.6,0.5,0.1}
\begin{tikzpicture}
\tikzmath{\xxm = -0.4; \xxM =12*3/4; \yym=0; \yyM = 7*3.5/3;\xxf=0.2*3/4;\xxs=6.55*3/4;\dashinter=2;\step=0.7*3/4;\npos=1*3/4;\Qpos=7.5*3/4;\sl=1;} 
\path[fill=darkgreen!10] (\yym,\xxm ) rectangle (\yyM,\xxM);
\path[fill=darkblue!10] (0,\yyM+\npos) rectangle (\yyM,\xxM);
\fill [darkblue!10, domain=\yyM+\npos:\npos, variable=\x]
      (0,\yyM+\npos) -- plot ({\sl*(\x-\npos)}, {\x}) --  (0,\npos) -- cycle;
\fill [myred!20,dashed, domain=\xxM:\Qpos, variable=\x]
     (0,\xxM)  -- plot ({2*(\x-\Qpos+1)^(1/2) -2},{\x}) --  (0,\Qpos)-- cycle;
\draw[dashed,thick=1] (\yym-1,\xxm) -- (\yym,\xxm);
\draw[line width=0.3cm,diffuse gradient=0.2,scale=1,domain=\yyM+\npos+1:\npos-1/2,smooth,variable=\x,white] plot ({\sl*(\x-\npos)},{\x});
\draw[scale=1,domain=\yyM+\npos+0.3:\npos,smooth,variable=\x,dotted,thick=2] plot ({\sl*(\x-\npos)},{\x});
\draw[black]	(\yyM+0.5,\xxM+0.6) node {{\footnotesize $N \sim \sqrt{n}\,\ell$}};
\draw[line width=0.3cm,diffuse gradient=0.2,scale=1,domain=\xxM+0.3:\Qpos-1/2,smooth,variable=\x] plot({2*(\x-\Qpos+1)^(1/2) -2},{\x});
\draw[dashed,myred,thick=10,scale=1,domain=\xxM+0.3:\Qpos,smooth,variable=\x] plot({2*(\x-\Qpos+1)^(1/2) -2},{\x});    
\draw[myred]	(2.6,\xxM+0.6) node {{\footnotesize $N \sim \frac{Q_{1,5}}{a^2}\,\ell^2$}};
\draw[darkgreen]	(1.8*\yyM/3,\npos+1.5) node[align=center]{{\small\underline{Low-energy Regime}}\\{\scriptsize (modes governed by the cap)}};
\draw[darkblue](3.7,\Qpos+1.5) node[align=center]{ {\small\underline{Intermediate-energy Regime}}\\  {\scriptsize (modes sensitive to the BTZ throat)}};
\draw[myred](0.85,\Qpos+2.8) node[align=center]{ {\small\underline{No Modes}}};
\draw[semithick] (\yym,\xxm) -- (\yym,\xxf);
\draw[dashed,thick=1] (\yym,\xxf) -- (\yym,\xxf+\step);
\draw[semithick] (\yym,\xxf+\step) -- (\yym,\xxs);
\draw[dashed,thick=1] (\yym,\xxs) -- (\yym,\xxs+\step);
\draw[->,semithick] (\yym,\xxs+\step) -- (\yym,\xxM+0.4) node[anchor=south east] {\textbf{$N$}};
\draw[->,semithick] (\yym,\xxm) -- (\yyM+0.4,\xxm) node[anchor=west] {\textbf{$\,\,\ell$  }};
\draw (0,\xxm)node[circle,fill,inner sep=1.0pt]{};
\draw (0,\xxm) node[anchor=north]{$1$};
\draw[shift={(0,\npos)},thick] (3pt,0pt) -- (-3pt,0pt) 
		(-0.1,0) node[anchor=east]{$\sqrt{n}$};
\draw[shift={(0,\Qpos)},thick] (3pt,0pt) -- (-3pt,0pt) 
		(-0.1,0) node[anchor=east]{$\frac{Q_{1,5}}{a^2}$};
\end{tikzpicture}
\caption{The different regimes of quasi-normal modes of $(2,1,n)$ asymptotically flat superstrata. For any mass quantum number $\ell$, there is a tower of quasi-normal modes labelled by $N$. At low energy, $N \lesssim \sqrt{n}\,\ell$ those modes live in the cap geometry. Their scalar potential is well-approximated by the cap potential glued to flat space \eqref{eq:capregimepot}. At intermediate energy, $ \sqrt{n}\,\ell\lesssim N \lesssim \frac{Q_{1,5}}{a^2}\,\ell^2$, the modes start to explore the BTZ throat of the geometry. At high-energy, $N \gtrsim \frac{Q_{1,5}}{a^2}\,\ell^2$, the barrier of the scalar potential starts to be negative and quasi-normal modes no longer exist.}
\label{fig:regimeEnergy}
\end{center}
\end{figure}
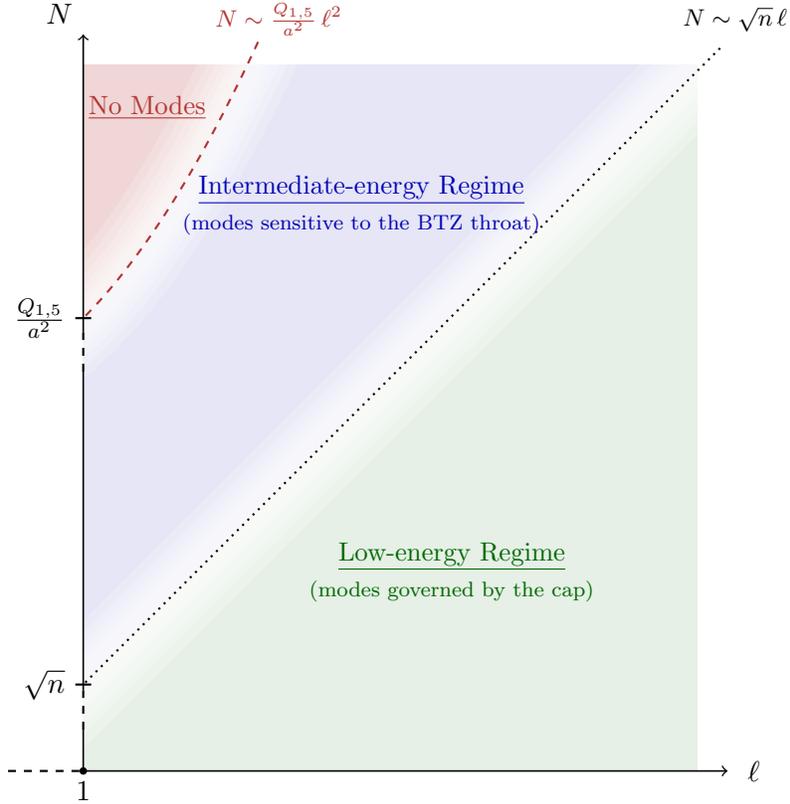
\end{itemize}

To summarize, we have  two energy regimes. We depict these regimes in Fig.\ref{fig:regimeEnergy}. They are separated by a boundary region around the line $N \sim \sqrt{n}\, \ell$. At low energy, $N \lesssim \sqrt{n}\,\ell$, the properties of the modes are determined by the red-shifted AdS$_3$ potential glued to flat space.    In the intermediate-energy regime, the spectrum of quasi-normal modes will be modified by the BTZ part of the geometry.   There are two parts of these regimes that will be important to us later.  The low-energy regime contains the large-$\ell$ limit at fixed $N$; the intermediate-energy regime contains large-$\ell$ region with $\frac{N}{\ell}$ fixed at a value larger than $\sqrt{n}$.

\section{Quasi-normal modes of asymptotically-flat superstrata}
\label{sec:Superstrata}

For superstrata, much of the essential physics is encoded in the radial components of the wave equation and so we have  examined various limits of the radial potential function.   In particular, in the last section we exhibited a deeply red-shifted, global AdS$_3$ cap that is connected to the asymptotically-flat region via a deep BTZ throat.  We also showed that, for low-energy modes, the effect of the BTZ region is negligible and such bound states are largely determined by the red-shifted cap.  We now use this structure to compute the quasi-normal modes.  As we remarked earlier, because the physical structure of superstrata with deeply-capped BTZ throats is universal, we expect our conclusions to be largely independent of the details. 

\subsection{Quasi-normal modes in the low-energy regime}
\label{ss:lowboundstates}

We apply the WKB techniques described in Section \ref{sec:WKB} to the superstratum. The quasi-normal modes are labelled by a mode number, $N \in \mathbb{N}$, defined by:
\be 
\Omega_N \= \Omega^{(0)}_N \+ \delta \Omega_N \,.
\ee
The zeroth-order ``normal'' frequencies, $\Omega^{(0)}_N$, are purely real and are given by the usual quantization relation 
\be 
\Theta^{(0)}_N \= \frac{\pi}{2} \left( 1 \+ 2 N \right)\,, \qquad N\in \mathbb{N}\,.
\label{eq:zerothOrder}
\ee
and the first-order correction, $\delta \Omega_N$, is purely imaginary and is given by:
\be 
\delta \Omega_N \= i\,\text{sign}\left(2\text{Re}(\Omega)+q_y\right)\, \left(\frac{\partial \Theta}{\partial \Omega} \right)^{-1}\,\frac{e^{-2 T}}{4}\,\,\Biggr|_{\Omega=\Omega^{(0)}_N }\,.
\label{eq:firstOrder}
\ee
Remember that $\Omega$ is the frequency along the $t-y$ direction and $P$ is the momentum along the $t+y$ direction, whereas, in the general formula \eqref{eq:ImPartSpectrum} of Section \ref{sec:WKB}, $\omega$ was the conjugate momentum of $t$. This means that   $\text{sign}(\text{Re}(\omega))$ is now replaced by $\text{sign}(\text{Re}(\Omega+P))=\text{sign}(2\text{Re}(\Omega)+q_y)$.

We now have to evaluate the integrals:
\begin{equation}
\Theta ~\equiv~  \int_{x_0}^{x_1} |V(z)|^{\frac{1}{2}}\,dz \,,\qquad T ~\equiv~  \int_{x_1}^{x_2} |V(z)|^{\frac{1}{2}}\,dz\,,
\label{eq:Theta&Tdef2}
\end{equation}
where $x_0$, $x_1$ and $x_2$ are the three turning points as depicted in Fig.\ref{fig:pictureofpotential21N}(a). For the three approximate potentials (\ref{eq:ApproximatedPotentials}) these integrals are elementary . 

\subsubsection{The normal frequencies, $\Omega^{(0)}_N$}

The potential $V_{\rm cap}(x)$ has the form
\begin{equation}
V_\text{cap}(x)  ~\equiv~\frac{1}{(e^{2 x}+1)^2} \,\Big[\, (\ell+1)^2 \, e^{4 x} - 2\, B\, e^{2 x} \,  + (q_\psi+q_y)^2  \,\Big],
\label{eq:potentialform}
\end{equation}
with
\begin{equation}
B ~\equiv~    \frac{1}{2}\, \left[\left(q_\phi+q_y+2\left(1+\frac{b^2}{2a^2}\right)\Omega\right)^2 -1-(\ell+1)^2 -\left(q_\psi+q_y \right)^2 \right] \, \,.
\label{eq:ABCdefn}
\end{equation}
The classical turning points that define the bound states are given by:
\begin{equation}
e^{2x_0}~=~   \frac{B - \sqrt{B^2-(\ell+1)^2\,(q_\psi+q_y)^2}}{(\ell+1)^2}\,,\qquad e^{2x_1}~=~   \frac{B + \sqrt{B^2-(\ell+1)^2\,(q_\psi+q_y)^2}}{(\ell+1)^2} 
\label{eq:TPx0x1}
\end{equation}
and the integral $\Theta$ in (\ref{eq:Theta&Tdef2}), yields:
\begin{equation}
\Theta ~=~  \frac{\pi}{2} \,\left[ \,-\ell-1 - \left| q_\psi+q_y \right| ~+~ \sqrt{ \left(q_\phi+q_y+2\left(1+\frac{b^2}{2a^2}\right)\Omega\right)^2 -1}\,\right] \,.
\label{eq:capTheta}
\end{equation}

The WKB approximation requires modes with a large number of oscillations between $x_0$ and $x_1$, and this means:
\begin{equation}
\left(q_\phi+q_y+2\left(1+\frac{b^2}{2a^2}\right)\Omega\right)^2 ~\gtrsim~ 10\,,
\end{equation}
and this leads to 
\begin{equation}
\Theta \approx  \frac{\pi}{2} \,\left[ \,-\ell-1 - \left| q_\psi+q_y \right| ~+~ \left| q_\phi+q_y+2\left(1+\frac{b^2}{2a^2}\right)\Omega\right|\,\right]\,.
\label{eq:ThetaRes}
\end{equation}
Therefore, using (\ref{eq:zerothOrder}), one finds that the ground-state frequencies are given by:
\begin{equation}
\begin{aligned}
\Omega^{(0)}_N & ~\approx~ \pm \,\frac{a^2}{2\left(a^2 +\frac{1}{2}\, b^2\right)} \, \left(2 N + \ell + 2 ~+~ |q_\psi +q_y| \mp (q_y+q_\phi) \right) \,, \\
&  ~=~ \pm \,\frac{j_L }{n_1 n_5} \, \left(2 N + \ell + 2 ~+~ |q_\psi +q_y| \mp (q_y+q_\phi) \right) \,,
\end{aligned}
\label{eq:zerothordercapregime}
\end{equation}
for  $ N \in \mathbb{N} $. As with any physical system, we have two branches of frequencies, one positive and one negative. Note that, as we anticipated in \eqref{eq:modeBehavior}, one has
\begin{equation}
\Omega_N ~=~ \pm  \frac{2 j_L}{n_1 n_5} \,(N ~+~ \ldots  \,)\,.
\end{equation}
In particular, the frequencies are quantized in units of  $\frac{2 j_L}{n_1 n_5} \ll 1$.  The fact that the frequencies are extremely small was essential in going from (\ref{eq:Vcap}) to the simpler form of $V_{\rm cap}$ in (\ref{eq:ApproximatedPotentials}).

Last but not least, the precision of the WKB approximation requires that $N\gtrsim 10$ to have a large number of oscillations between the turning points.
At the other extreme, to compute $\Theta$ using the potential, $V_{\rm cap}$, means that the classical turning point, $x_1$, must remain in the cap region, which means $x_1 < \frac{1}{2}\log n$, which is guaranteed if  $B <   \frac{1}{2} n  (\ell +1)^2$.  Using (\ref{eq:zerothordercapregime}) in (\ref{eq:ABCdefn}), the validity of the computation above leads to a bound on $N$: 
\begin{equation}
N ~\lesssim ~  \sqrt{n} \, \ell  \,,
\label{eq:Nbound}
\end{equation}
which is exactly the bound we have already established for the low-energy regime.

\subsubsection{The quasi-normal decay rates, $\delta \Omega_N$}
\label{ss:lowdecays}

We now apply (\ref{eq:ImPartSpectrum})  to obtain the perturbative imaginary corrections to the normal modes caused by the tunneling through the asymptotically flat region.

The first part is straightforward.  It follows from (\ref{eq:ThetaRes}) that 
\begin{equation}
\frac{\partial \Theta}{\partial \Omega}\Biggr|_{\Omega=\Omega^{(0)}_N } ~\approx~ \pi\,\left(1+\frac{b^2}{2a^2}\right) \,\text{sign}\left(\Omega_N^{(0)}\right)~=~ \frac{\pi\,n_1 n_5}{2\,j_L}\,\text{sign}\left(\Omega_N^{(0)}\right)\,.
\label{eq:dTheta}
\end{equation}

The evaluation of the integral,  (\ref{eq:Theta&Tdef2}), that defines $T$ is more of a challenge because it crosses between  regions in which we have made different approximations to the potential. Indeed, we first note that the endpoint,  $x_1$, of the integral is determined by $V_\text{cap}(x)$ and is given by (\ref{eq:TPx0x1}), while the other endpoint, $x_2$,  is determined by $V_\text{Flat}(x)$ and is given by:
\begin{equation}
e^{2x_2}  ~=~  \frac{(\ell+1)^2 \, R_y ^2 }{4 \, a^2 \, \Omega \, ( \Omega + q_y) }\,.
\label{eq:x2res}
\end{equation}

One can make a reasonably good estimate of the value of $T$ by approximating the entire integral by the area of a rectangular plateau of height $(\ell+1)$.  Since $V_\text{Flat}(x)$ is dropping exponentially fast, the  right end of the plateau is well approximated by $x_2$.  Locating the left end of the plateau, $\hat x_1$, is a little more difficult.  It turns out that $x_1$ is not a good estimate for this point because the ramp up to the plateau can be fairly gradual.  It is better to estimate the point at which $V_\text{cap}(x)$  is approaching $(\ell+1)$.  We claim that the following is a better estimate of $\hat x_1$:
\begin{equation}
0 ~<~ \hat x_1  ~<~ \coeff{1}{2} \, \log n \,.
\label{eq:hatx1}
\end{equation}
These bounds come from considering the competition between the first and the third term of $V_\text{cap}(x)$ in (\ref{eq:ApproximatedPotentials})  for the possible mode numbers with $N \lesssim \sqrt{n} \, \ell$.  The upper bound in (\ref{eq:hatx1}) arises from $N \sim  \sqrt{n} \, \ell$, and   corresponds to requiring that, in the low-energy regime, $\hat x_1$ lies in the cap.   The lower bound comes from small $N$. With those approximation, we have 
\begin{equation}
e^{-2 T} ~ \simeq~  e^{-2(\ell+1)(x_2 - \hat x_1)}~ \simeq~   \Bigg[ \frac{4 \, a^2 \,n^h\, \Omega \, ( \Omega + q_y)  }{(\ell+1)^2 \, R_y ^2  } \Bigg]^{(\ell+1)}  \,,
\label{eq:Testimate0}
\end{equation}
where $\hat x_1= \coeff{h}{2} \log n$ for   $0 < h < 1$.

\begin{figure}
\centering
\includegraphics[scale=0.70]{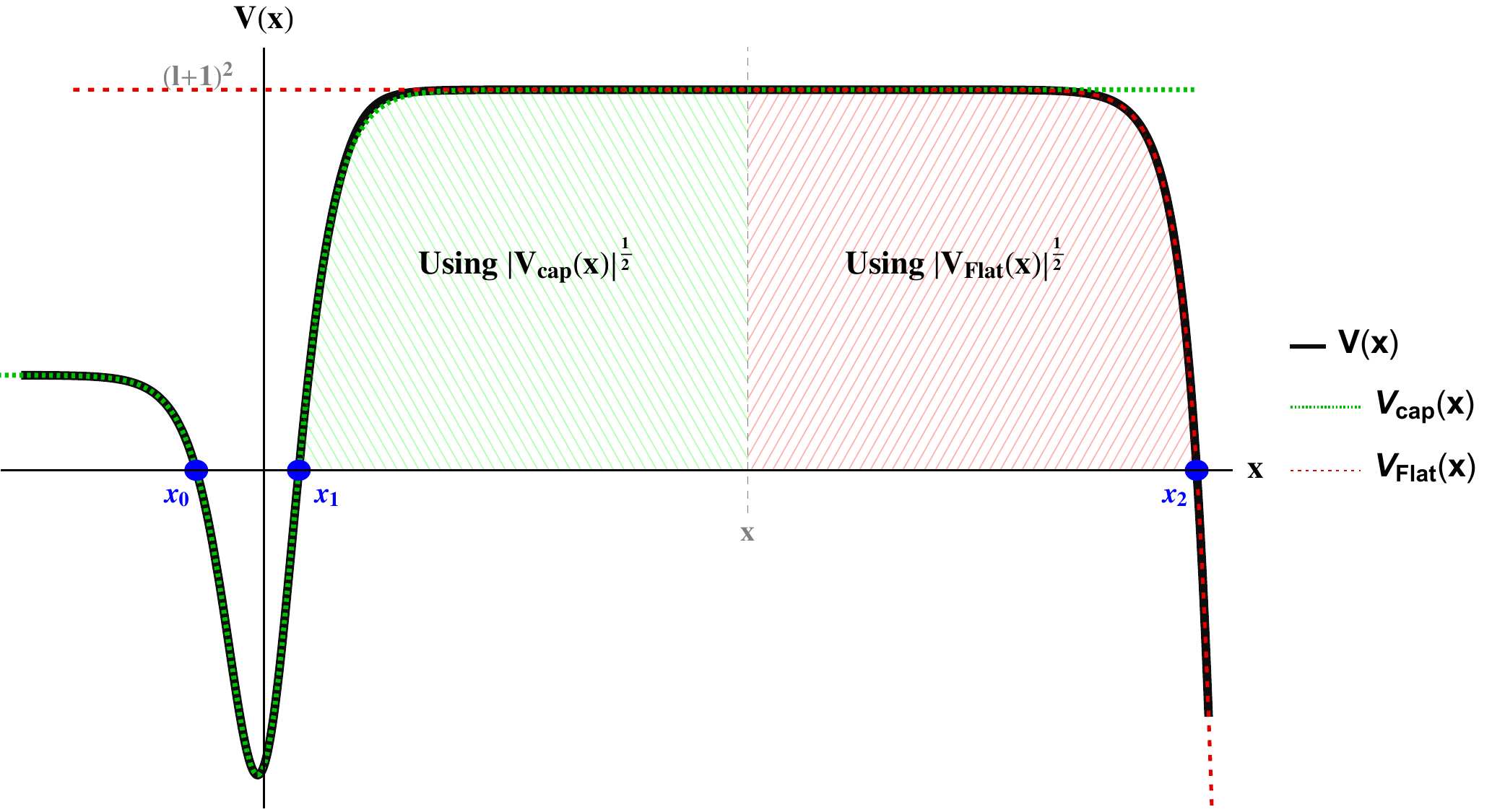}
\caption{The evaluation of $T$ using two domains: in the first we integrate $|V_\text{cap}(x)|^{\frac{1}{2}}$ and the second we integrate $|V_\text{Flat}(x)|^{\frac{1}{2}}$. The flatness of the potential ensures that the result is independent of the boundary between the two domains. } 
\label{fig:PotforT}
\end{figure}

We now make a much more precise evaluation  of $T$ by performing a calculation that may be viewed as the WKB analogue of a matched asymptotic expansion.  The strategy is extremely simple: we know that $V_\text{cap}(x)$ and $V_\text{Flat}(x)$  provide accurate approximations to the exact potential and that the domains of validity of these approximations overlap for a substantial interval at the top of the plateau where $V(x) \approx \ell +1$. We therefore know that, to a very good approximation, one has
\begin{equation}
T  ~\approx~    \int_{x_1}^x \,\left| V_\text{cap}(z)\right|^{\frac{1}{2}} \, dz ~+~    \int_x^{x_2} \, \left|V_\text{Flat}(z)\right|^{\frac{1}{2}} \, dz \,,
\label{eq:Testimate1}
\end{equation}
where $x_1 \ll x \ll x_2$ is chosen to lie in the overlap region at the top of the plateau as depicted in Figure \ref{fig:PotforT}

As we remarked earlier, both integrals in (\ref{eq:Testimate1}) are elementary and can be obtained in closed form. The detailed analysis may be found in Appendix \ref{App:QNMviaWKB}.  The general formulae are far from simple, however it is easy to make approximations that improve upon  (\ref{eq:Testimate0}).  Indeed, motivated by the results coming from matched asymptotic expansions like those of \cite{Lunin:2001dt,Eperon:2016cdd,Chakrabarty:2019ujg}, we have shown that the following result closely approximates the WKB expressions for $T$:
\begin{equation}
e^{-2 T} \,\,\bigl|_{\Omega = \Omega_N^{(0)}}~\simeq~ 4\pi^2 \,\left[ \Omega_N^{(0)}\,\( \Omega_N^{(0)}+q_y\)\,\frac{a^2}{R_y^2} \right]^{\ell+1}  \frac{1}{(\ell !)^2}\,\,\, {}   ^{\ell+1+N} C_{\ell+1}  \,\,\, {}^{\ell+1+N+|q_\psi+q_y|} C_{\ell+1}\,,
\label{eq:Testimate2}
\end{equation}
where ${}^pC_q$ is the standard binomial coefficient. Note that Stirling's approximation gives $\ell^\ell \sim \ell!$ to leading order, and so the first terms in  (\ref{eq:Testimate2}) coincide with the simple estimate, (\ref{eq:Testimate0}).

We have tested (\ref{eq:Testimate2})  against the WKB formula for $T$ in the Appendix \ref{App:WKBvsAsympMatch}. We found that they exactly match up to third order in the large-$N$ and large-$\ell$ expansions. Moreover, we used numerics to show that the mismatch is less than 1\% as soon as $(N,\ell)>10$.

Combining (\ref{eq:Testimate2}) with (\ref{eq:dTheta}) and (\ref{eq:firstOrder}), we arrive at the main result for the low-energy regime:
\begin{equation}
\delta \Omega_N \, \simeq \,2\pi i\, \bigg( \frac{j_L}{n_1 n_5}\bigg) \,\left[\Omega_N^{(0)}\,\(\Omega_N^{(0)}+q_y\)\,\frac{a^2}{R_y^2} \right]^{\ell+1}  \frac{1}{(\ell !)^2}\,\,\, {}   ^{\ell+1+N} C_{\ell+1}  \, {}^{\ell+1+N+|q_\psi+q_y|} C_{\ell+1}\,.
\label{eq:deltaOmegaN1}
\end{equation}
The sign-sensitive terms in  (\ref{eq:dTheta}) and (\ref{eq:firstOrder}) combine to the sign of $\Omega_N^{(0)}  (2\Omega_N^{(0)}+q_y) \ge \Omega_N^{(0)}  (\Omega_N^{(0)}+q_y) >0 $. The last inequality is simply  \eqref{eq:condPositivityOmegaqy}, which is required for having quasi-normal modes. 

Thus, the right-hand side of the expression is a {\it positive} purely imaginary  number. The time dependence of the modes is given by
\be 
e^{i\left( \frac{\sqrt{2}\,\Omega}{R_y} \,u\,+\,\frac{\sqrt{2}\, P}{R_y} \, v \right)}\= e^{i\left( \frac{2\,\Omega}{R_y} \,t\,+\,\frac{q_y}{R_y} \, (t+y)\right)} \= e^{i \, \frac{2 \,\delta\Omega_N}{R_y}\, t}\,e^{i\left( \frac{2\,\Omega_N^{(0)}}{R_y} \,t\,+\,\frac{q_y}{R_y} \, (t+y)\right)}\,,
\label{eq:decayinggrowing?}
\ee
which guarantees that the wave profile is decaying in time for both branches of frequencies \eqref{eq:zerothordercapregime}.  

One  important feature is that the decay time-scale is set by $ \frac{n_1 n_5}{j_L}   \sim   \frac{b^2}{a^2}   $, which is extremely long because of the very large red-shift between flat space and the cap.

We also note that the essential, leading-order physics of the quasi-normal decay is captured by the simple ``rectangle'' approximation that led to (\ref{eq:Testimate0}).  The more accurate computation leads to corrections that are sub-leading at large $\ell$.

Finally, for our analysis in Section \ref{sec:EikonalLowEnergy}, we note that  the low-energy regime contains the large $\ell$ limit of the spectrum of quasi-normal modes  for $\frac{N}{\ell} \lesssim \sqrt{n}$.

\subsection{Quasi-normal modes in the intermediate-energy regime}
\label{ss:quasi-normalBTZ}

We now consider mode numbers with $N \gtrsim \sqrt{n} \,\ell$.  These intermediate-energy states start exploring the BTZ throat of the geometry.   In particular, the middle classical turning point, $x_1$, is no longer in the cap region 
$$x_1 \gtrsim \frac{1}{2}\,\log n\,,$$
as depicted in Fig.\ref{fig:pictureofpotential21N}(b).  Once again to obtain the spectrum of quasi-normal modes via WKB, one needs to estimate the integrals $\Theta$ and $T$ \eqref{eq:Theta&Tdef2} using the approximate potentials.  The computation proceeds much as in Section \ref{ss:lowboundstates}

\subsubsection{The computation of $\Theta$} 

In the low-energy regime, the first two turning points are in the cap region.  This  facilitates the computation of $\Theta$ because it only involves $V_\text{cap}(x)$.   In the intermediate-energy regime, we simply  follow the approach of Section \ref{ss:lowdecays} and estimate $\Theta$ using $V_\text{cap}(x)$ from $x_0$ to $x\sim \frac{1}{2}\log n$ and $V_\text{BTZ}(x)$ from $x$ to $x_1$.  However, because of the depth of the potential well and the rapidity of the climb of the BTZ potential (see Fig.\ref{fig:pictureofpotential21N}(b)), almost all the support of the WKB integrals lies within the cap region.  

One can easily estimate the error in simply using $V_\text{cap}(x)$.  The crossover between the cap and the BTZ throat starts at $x \sim \frac{1}{2}\log n$, at which point the potential has some large, negative value, $V_c$. The potentials $V_\text{cap}(x)$ and $V_\text{BTZ}(x)$ lead to two different values,  $x_{1,\text{cap}}$  and $x_{1,\text{BTZ}}$, for the classical turning point (see Fig.\ref{fig:pictureofpotential21N}(b)). The difference of the WKB integrals for the two potentials is approximately the area of the triangle with base $x_{1,\text{BTZ}} -x_{1,\text{cap}}$ and height $\sqrt{|V_c|}$. This leads to an error estimate of order 
$$\frac{1}{\sqrt{n}}\,\sqrt{\frac{\sqrt{n}\,\ell}{N}}\,\log \frac{N}{\sqrt{n}\,\ell} ~<~ \frac{1}{\sqrt{n}} \,, $$
where the last inequality follows from  $N \gtrsim \sqrt{n} \,\ell$. 

Thus we find that $\Theta$ receives a small correction from the result for  the low-energy regime \eqref{eq:ThetaRes}:
\begin{equation}
\Theta \=  \frac{\pi}{2} \,\left[ \,-\ell-1 - \left| q_\psi+q_y \right| ~+~ \left| q_\phi+q_y+2\left(1+\frac{b^2}{2a^2}\right)\Omega\right|\,\right]\left( 1 \+ \cO\(\frac{1}{\sqrt{n}}\) \right)\,,
\label{eq:ThetaRes2}
\end{equation}
which, at zeroth order, leads to the two same branches of normal frequencies for $\Omega_N^{(0)}$, as in \eqref{eq:zerothordercapregime}.

\subsubsection{The computation of $T$ and the decay time} 

We compute $T$ just as in  Section \ref{ss:lowdecays}, but now we use $V_\text{BTZ}(x)$ to define the left side of the plateau.  In particular, the classical turning point is defined by the vanishing of $V_\text{BTZ}(x)$ in (\ref{eq:ApproximatedPotentials}).  This yields 
\begin{equation}
e^{2x_1}  ~=~  \frac{1}{2\, (\ell+1)^2}\,\bigg(\frac{b^2\, \Omega}{a^2}\bigg)\,\bigg( \,  \bar{p} ~+~ \sqrt{\bar{p}^2 + (n+\coeff{1}{2})^2 (\ell+1)^2}  \,\bigg)\,,
\label{eq:x1BTZ}
\end{equation}
where $\bar{p}$ is the effective BTZ momentum:
\begin{equation}
\bar{p} ~\equiv~ 2q_y+q_\phi +q_\psi + \left(2+\frac{(1+n)(Q_1 +Q_5)}{R_y^2}\right)\,\Omega  ~\approx~ 2q_y+q_\phi +q_\psi \,.
\end{equation}
Since $V_\text{BTZ}(x)$ rises to the plateau extremely fast, and $V_\text{Flat}(x)$ descends similarly fast, one expects that the WKB integral can be well approximated by a rectangular plateau of height $(\ell+1)$ and width  $x_2 -x_1$.    Using (\ref{eq:x2res}) and (\ref{eq:x1BTZ}), this leads to
\begin{equation}
\begin{aligned}
e^{-2 T} ~ \simeq~   e^{-2(\ell+1)(x_2 -  x_1)}   ~ \simeq~  & \Bigg[ \frac{2 \, a^2 \, \Omega \, ( \Omega + q_y)  }{(\ell+1)^4 \, R_y ^2  } \,\bigg(\frac{b^2\, \Omega}{a^2}\bigg)\, \bigg( \,  \bar{p} ~+~ \sqrt{\bar{p}^2 + (n+\coeff{1}{2})^2 (\ell+1)^2}  \,\bigg)\, \Bigg]^{(\ell+1)}\\
~ \simeq~  & \Bigg[ \frac{2 \, a^2 \, \Omega \, ( \Omega + q_y)  }{(\ell+1)^4 \, R_y ^2  } \,\bigg|\frac{n_1 n_5 \, \Omega}{j_L}\bigg|\, \bigg( \,  \bar{p} ~+~ \sqrt{\bar{p}^2 + (n+\coeff{1}{2})^2 (\ell+1)^2}  \,\bigg)\, \Bigg]^{(\ell+1)}  \,.
\end{aligned}
\label{eq:TestimateBTZ}
\end{equation}

A more precise computation in which one uses (\ref{eq:Testimate1}) with $V_\text{cap}$ replaced by $V_\text{BTZ}$ yields
\be 
e^{-2 \,T} ~\simeq~ \kappa\,\left[ \frac{\Omega\,\(\Omega+q_y\)}{(\ell+1)^4}\,\frac{a^2}{R_y^2} \right]^{\ell+1} \,\left| \frac{n_1 n_5\,\Omega}{j_L}\right|^{\ell+1} \left[\bar{p}^2 + \(n+\frac{1}{2}\)^2 \(\ell+1 \)^2 \right]^{\frac{\ell+1}{2}}\,,
\ee
where 
\be 
\kappa \equiv  \exp \left[ 2 \left(\ell+1 \+ \frac{\bar{p}}{1+2 n} \,\arctan \left[\frac{\(n+\frac{1}{2}\) \(\ell+1 \)}{\bar{p} + \sqrt{\bar{p}^2 + \(n+\frac{1}{2}\)^2 \(\ell+1 \)^2}} \right] \right)\right]\,.
\ee
The quantity $\kappa$ that is sub-leading in the large-$\ell$ expansion. Thus  (\ref{eq:TestimateBTZ}) does indeed yield a good  estimate of $T$.

Applying the WKB formula at zeroth order, using  \eqref{eq:firstOrder}  and (\ref{eq:zerothordercapregime}) we obtain the following results in the  intermediate-energy regime, $N\gtrsim \sqrt{n}\,\ell$:
\be 
\begin{split}
\Omega^{(0)}_N&\, \simeq \, \pm \,\frac{j_L }{n_1 n_5} \, \left(2 N + \ell + 2 ~+~ |q_\psi +q_y| \mp (q_y+q_\phi) \right) \,,\\
\delta \Omega_N &\, \simeq \, i\,\, \frac{j_L\, \kappa}{2\pi\,n_1 n_5} \,\left[ \Omega^{(0)}_N\,\(\Omega^{(0)}_N+q_y\)\,\frac{a^2}{R_y^2} \right]^{\ell+1} \,\left| \frac{n_1 n_5\,\Omega^{(0)}_N}{j_L\(\ell+1\)^4}\right|^{\ell+1} \left[\bar{p}^2 + \(n+\frac{1}{2}\)^2 \(\ell+1 \)^2 \right]^{\frac{\ell+1}{2}}\,.
\end{split}
\label{eq:modesBTZregime}
\ee
Once again, once can use the same arguments as in the low-energy regime to show that 
both branches of frequencies are decaying in time. Moreover, we also note that the essential, leading-order physics of the quasi-normal decay is captured by the simple ``rectangle'' approximation that led to (\ref{eq:TestimateBTZ}).  The more accurate computation leads to corrections that are sub-leading at large $\ell$.  Also, for our analysis in Section \ref{sec:EikonalLowEnergy}, we note that  the intermediate-energy regime contains the large $\ell$ limit of the spectrum of quasi-normal modes for $\frac{N}{\ell} \ge \sqrt{n}$.

\subsection{The eikonal limits}
\label{ss:quasi-normallargeL}

One of the original motivations for the stability analysis of \cite{Eperon:2016cdd} was the fact that microstate geometries with evanescent ergosurfaces will have time-like geodesics with extremely low energies and that are trapped for extremely long periods of time.   The link between this and the study of modes of the scalar wave equation arises through the standard geometric-optics limit in which the  phase function of the WKB solution provides a Hamilton-Jacobi function for geodesics.  In particular, the normals to the wave-fronts become the tangents of the geodesics.  Just as in WKB, the geometric optics limit, or eikonal limit, becomes more accurate at higher wave-numbers.  Moreover, by taking these limits in the right way, one can localize the wave in various geometric regions and use this to capture the physics of particular geodesics.   

For our geometries it is interesting to consider low-energy modes in the large-$\ell$ limit.  This is because the three-dimensional mass of the scalar modes is $(\ell+1)$, and massive modes are more strongly trapped in the three-dimensional geometry.  Equivalently, $(\ell+1)^2$ is the height of the potential barrier between the bound states  and the asymptotically flat region. As we have remarked, low-energy modes localize in the cap, near $r=0$. Moreover, at large-$\ell$  one can localize the scalar harmonics on the S$^3$, and especially near the evanescent ergosurface, $r=0$, $\theta = \frac{\pi}{2}$.  Such limits were a major focus of \cite{Eperon:2016cdd}.

There are several obvious limits to consider. First, one can take $\ell$ large while holding the other mode numbers, $q_\psi, q_\phi,  q_y$ and $N$ fixed, and small relative to $\ell$.  These are ``generic'' sphere modes in that they do not localize in any particular region.  Of more interest is to take $|q_\psi| = \ell$, $q_\phi =0$, or  $q_\psi  = 0$, $|q_\phi | = \ell$. (Remember that one must respect (\ref{eq:RegAngForm}).)  It is evident from (\ref{eq:AngularWaveProfile}) that these choices localize the wave at $\theta = 0$ or $\theta =  \frac{\pi}{2}$ respectively.   

From (\ref{sixmet}), one sees that the evanescent ergosurface is located where $\cP$ diverges.   From \eqref{eq:PFbetaomega21n}  and (\ref{Sigdefn}) one sees that this corresponds to  $r=0$  and $\theta = \frac{\pi}{2}$.   Thus we anticipate that stronger ``trapping'' of modes in the cap (localized near $r=0$) will arise if one takes $|q_\phi | = \ell$.

The physical difference between these limits, and the significance of $|q_\phi| =  \ell  $ become apparent in our results for the normal modes, (\ref{eq:zerothordercapregime}) and (\ref{eq:modesBTZregime}): 
\begin{equation}
\Omega^{(0)}_N  ~\approx~ \pm \,\frac{j_L }{n_1 n_5} \, \left(2 N + \ell + 2 ~+~ |q_\psi +q_y| \mp (q_y+q_\phi) \right) 
\end{equation}
One sees that a generic choice of mode numbers leads the $\Omega^{(0)}_N$ growing linearly with $\ell$.  However, this growth with $\ell$ can be cancelled to produce 
\begin{equation}
\Omega_N^{(0)} ~\approx~\pm \,\frac{j_L }{n_1 n_5} \, \left(2 N + 2 + |q_y| \mp q_y \right)\,.
\label{eq:lowgrowth2}
\end{equation}
if and only if we take 
\begin{equation}
q_\phi \= \pm\, \ell \,, \qquad q_\psi \= 0\,,
\label{eq:modesatevanescent}
\end{equation}
where the ``$\pm$'' depends on which branch of $\Omega_N^{(0)}$ is considered.

Thus generic modes have frequencies that grow linearly with $\ell$, and it is only the modes that localize near the evanescent ergosphere that have  frequencies that {\it do not} grow with $\ell$.  This is the wave-equation analogue of the statement that it is the geodesics that localize near the evanescent ergosphere that can have arbitrary low energy. 

From the results of Section \ref{ss:lowboundstates}, we found that the decay rates depend, to leading order, on the quantum numbers $\ell$ and $\Omega_N^{(0)}$ as:
\begin{equation}
\delta \Omega_N ~\sim~\frac{\Big(\Omega_N^{(0)}\Big)^{2(\ell +1)} }{(\ell !)^2} \,.
\label{eq:deltaOmegaleading}
\end{equation}
We therefore see the competition between mode energies and barrier height and length.  Observe that if $\Omega_N^{(0)}$ grows linearly with $\ell$, then the numerator and denominator grow with $\ell$ at the same leading-order rate.  If, however, $\Omega_N^{(0)}$ does not grow with $\ell$, $\delta \Omega_N$ becomes extremely small at large $\ell$.  These are the states that lie close to the ergosphere and are trapped for extremely long times.

We will therefore study the difference in decay times, at large $\ell$, for generic modes and for modes localized near the evanescent ergosphere.  It will be convenient to introduce the shorthand
\begin{equation}
\omega_N^{(0)} \equi 2 N + 2 + |q_y| \mp q_y \,\in \, \mathbb{N}\,.
\end{equation}
%

\subsubsection{The eikonal limits for low-energy modes}
\label{sec:EikonalLowEnergy}

The spectrum of low-energy quasi-normal modes, $N \lesssim \sqrt{n}\,\ell$, is given by (\ref{eq:zerothordercapregime}) and  (\ref{eq:deltaOmegaN1}). For modes localized around the evanescent ergosurface we have 
\begin{equation}
\Omega_N^{(0)} \= \pm \, \frac{j_L}{n_1 n_5} \,\omega_N^{(0)}\,. 
\label{eq:lowgrowth}
\end{equation}
Using Stirling's formula, we obtain the generic expressions 
\be 
\left(\ell\,! \right)^{-2} ~\approx~ \exp \left[- 2 \ell \log \ell +2 \ell -\log \ell \+ \cO(1) \right] \,,\qquad {}^{\ell+1+X} C_{\ell+1} ~\approx~ \exp \left[ X\,\log \ell \+ \cO(1) \right]\,, 
\label{eq:Stirling}
\ee
and then the imaginary part of the frequency \eqref{eq:deltaOmegaN1} behaves as
\begin{equation}
\begin{split}
\delta \Omega_N \= i\,\exp &\left[ -2 \ell \left(\log \(\frac{n_1 n_5}{j_L} \)+ \log \ell - \frac{1}{2} \log \left(\omega_N^{(0)} +\frac{n_1 n_5}{j_L}\,q_y \right) \right)\+ \left(2 + \log \frac{a^2\, \omega_N^{(0)}}{R_y^2} \right) \ell \right. \\
&  \+(2 N+|q_y| -1)\,\log \ell \- 3 \log \(\frac{n_1 n_5}{j_L}\)+ \log \left(\omega_N^{(0)} +\frac{n_1 n_5}{j_L}\,q_y \right) \+ \cO(1)\, \Biggr]\,.
\end{split}
\end{equation}
Thus, for the modes at the evanescent ergosurface, the decay rate at large $\ell$ is minimal when $q_y =0$ and the leading-order terms are
\begin{equation}
\delta \Omega_N \,\sim \, i\,\ell^{- 2 \ell}\,\left( \frac{j_L}{n_1 n_5} \right)^{2 \ell+3} \, \left( \frac{e \,a  (N+1) }{R_y} \right)^{2\ell}  \,\ell^{2N-1}\,\, \cO(1)\,,
\label{eq:dOmerg}
\end{equation}
where $e = \exp(1)$ and we have highlighted the factor of $\ell^{- 2 \ell}$.

For generic modes, we take $q_\phi$ and $q_\psi$ to be arbitrary but differing from \eqref{eq:modesatevanescent}. We will assume, for simplicity, that $N,\,q_y,\,q_\phi$ and $q_\psi$ are all fixed and small compared to $\ell$.  However, one will obtain a similar result, but with different coefficients,  if one allows some of the mode numbers to scale with $\ell$. Using the same procedure we obtain
\begin{equation}
\begin{split}
\delta \Omega_N \= i\,\exp &\left[ -2 \ell \log \(\frac{n_1 n_5}{j_L} \) \+ 2\left(1 + \log \frac{a}{R_y} +\frac{1}{2} \,\log \left(1+\frac{q_y}{\ell}\frac{n_1 n_5}{j_L} \right)\right) \ell \right. \\
&  \+(2 N+|q_y+q_\psi| +1)\,\log \ell \- 3 \log \(\frac{n_1 n_5}{j_L}\)\+ \cO(1)\, \Biggr]\,.
\end{split}
\end{equation}
The decay rate at large $\ell$ is minimal when $q_y =q_\psi =0$ and we have
\begin{equation}
\delta \Omega_N \,\sim \, i\,\left( \frac{j_L} {n_1 n_5}\right)^{2 \ell+3} \, \left( \frac{e \,a}{R_y} \right)^{2\ell} \, \ell^{2N+1}\,\, \cO(1)\,,
\label{eq:dOmgeneric}
\end{equation}
where $e = \exp(1)$.

Both expressions, (\ref{eq:dOmerg}) and  (\ref{eq:dOmgeneric}), for $\delta \Omega_N$ are products of $\ell^{\rm th}$ powers of small parameters.  Most notable is the factor  
\begin{equation}
\bigg( \frac{j_L} {n_1 n_5}\bigg)^{2 \ell+3} \,,
\label{eq:redshiftfactor}
\end{equation}
 which represents the effect of the large red-shift between flat space and the cap.  
 
 The primary, and most significant, difference between (\ref{eq:dOmerg}) and  (\ref{eq:dOmgeneric}) is the factor of $\ell^{- 2 \ell}$.  It is this factor that led to the suggestion that the evanescent ergosurfaces of microstate geometries give rise to exceptionally long-term trapping of matter.  We we will discuss this further, and explain why this conclusion is unwarranted, in Section \ref{sec:Decay}.  We will also discuss why the factor (\ref{eq:redshiftfactor}) carries the important physics of the quasi-normal decay of microstate excitations.

We note that the factor of $\ell^{- 2 \ell}$ in the decay rate is cancelled when the mode number, $N$,  scales with $\ell$. (A similar conclusion holds for  the mode number, $q_y$, so long as it has the proper sign.)  This means that the $\ell^{- 2 \ell}$ scaling is only a property of the lowest modes, whose frequencies and $y$-momenta remain small compared to $\ell$.  Since the intermediate-energy modes necessarily have frequencies that scale with $\ell$, one should also not expect the  $\ell^{- 2 \ell}$  factor in their decay rates, as we will now establish.   

\subsubsection{The eikonal limits for intermediate-energy modes}

The intermediate-energy modes are defined as the excitations with $N \gtrsim \sqrt{n}\,\ell$.  Their frequencies and decay rates are  given in (\ref{eq:modesBTZregime}). It is  evident from these expressions that even if one chooses  $q_\phi$ and $q_\psi$  as in (\ref{eq:modesatevanescent})  so as   to cancel the explicit $\ell$-dependence and arrive at (\ref{eq:lowgrowth2}), there is still the implicit $\ell$-dependence in $N$.  To take this into account, we define by $\alpha$ the fixed ratio
\be 
\alpha \= \frac{N}{\ell} \gtrsim \sqrt{n}\,.
\ee
For simplicity, we will also assume that $\{q_y, \,q_\phi,\,q_\psi \}$ are fixed (one obtains a similar result with different coefficients if they scale with $\ell$). By expanding $\delta \Omega_N$ in \eqref{eq:modesBTZregime}, we obtain
\be 
\begin{split}
\delta \Omega_N \= i\,\exp &\left[ -2 \ell \log \(\frac{n_1 n_5}{j_L} \) \+ \left(2 + \log \frac{\alpha^3 \, a^2 \(n+\frac{1}{2}\) }{R_y^2} +\log \left(1+\frac{q_y}{\alpha\,\ell}\frac{n_1 n_5}{j_L} \right)\right) \ell \right. \\
&  \- 3 \log \(\frac{n_1 n_5}{j_L}\)\+ \cO(1)\, \Biggr]\,.
\end{split}
\ee
The decay rate is then minimal when $q_y=0$ and we have
\be 
\delta \Omega_N \,\sim \, i\,\left(\frac{j_L} {n_1 n_5}\right)^{2 \ell + 3} \, \left( \frac{e \,a\,\alpha^{\frac{3}{2}} \,\(n+\frac{1}{2}\)^{\frac{1}{2}}}{R_y} \right)^{2\ell}\,\, \cO(1)\,.
\label{eq:dOmInter}
\ee

{\it A priori,} this decay is faster than that of (\ref{eq:dOmgeneric}) because of the factors of $\alpha^{3 \ell} n^\ell \gtrsim n^{\frac{5}{2}\ell}$.  This is because we are considering  intermediate-energy states that  have $N$ scaling with $\ell$. Hence, despite being highly-localized on the sphere, the high occupation numbers mean that these excitations are beginning to explore the BTZ throat and have more energy to tunnel through the barrier.   Such modes are no longer strongly localized near the evanescent ergosphere, located at $r=0$, $\theta = \frac{\pi}{2}$, and our analysis shows that these higher modes do not have the exceptionally low decay rates that result from the extra factor of  $\ell^{- 2 \ell}$ in  (\ref{eq:dOmerg}).

It is interesting to push (\ref{eq:dOmgeneric}) and (\ref{eq:dOmInter}) slightly outside their domains of validity and look at the crossover between these formulae at large $N$, as well as large $\ell$.  The ratio of these expressions is
\begin{equation}
\frac{\ell^{2N+1}}{\alpha^{3 \ell} (n+\coeff{1}{2})^\ell } \,.
\end{equation}
As $N$ becomes large, one sees that the numerator grows faster than the denominator.  This is because  (\ref{eq:dOmgeneric})  is based on  the AdS cap, which does not limit, or contain, the modes nearly as strongly as the BTZ throat. Indeed, (\ref{eq:dOmInter}) does not explicitly depend on $N$.  This is because the extremely steep BTZ throat strongly attenuates any mode that enters the throat and confines modes very strongly within the cap.  This attenuating effect of the BTZ throat was also very noticeable in the thermal decay of the Green functions studied in \cite{Bena:2019azk}.

\section{Quasi-normal modes of  other microstate geometries}
\label{sec:OlderWork}

One of the simpler families of three-charge microstate geometries, obtained by Giusto, Mathur and Saxena (GMS),  are those generated through a spectral flow of the Lunin-Mathur D1-D5 geometries \cite{Giusto:2004id,Giusto:2004ip,Giusto:2004kj}. These are closely related to the GLMT geometries, which are obtained by fractional spectral flow \cite{Giusto:2012yz}.

Because of their simple relationship with the two-charge D1-D5 system, the GMS and GLMT geometries and their scalar wave equations are relatively simple.  In fact, the wave equation is exactly separable.   It is for these reasons that GMS geometries were recently used \cite{Eperon:2016cdd,Chakrabarty:2019ujg} to study instabilities and compute quasi-normal modes. 

Both derivations have been done using an asymptotic expansion analysis but in different limits: in the large-$\ell$ limit (eikonal limit) for \cite{Eperon:2016cdd} and in the near decoupling limit for \cite{Chakrabarty:2019ujg}. Our purpose here is to re-examine the results of \cite{Eperon:2016cdd,Chakrabarty:2019ujg} and compare and contrast them with our WKB analysis of quasi-normal modes of superstrata.

Unlike superstrata, the GMS geometries do not have the same charges and angular momentum as a black hole with a macroscopically large horizon area, and hence are dual to a more restricted family of CFT states.  Because of this, GMS geometries do not develop a long black-hole-like throat. However, GMS geometries involve a $\ZZ_k$ orbifold and one can generate large red-shifts by taking the orbifold parameter, $k$,  to be large.  This leads to more stringent limits on the redshifts of GMS solutions when compared to superstrata because the supergravity approximation will break down for high levels of orbifolding.  Superstrata do not suffer from any such limitations.

\subsection{The GMS geometries}

Here we summarize the essential details of the GMS geometries, their charges and quantum numbers.  We refer the interested reader to the original papers \cite{Giusto:2004id,Giusto:2004ip,Giusto:2004kj} for more details about their construction and the holographically dual CFT states. 

As with superstrata, GMS solutions are most simply described within the six-dimensional $(0,1)$ supergravity obtained by compactifying and truncating IIB supergravity  on T$^4$ (or $K3$). The six-dimensional metric takes the form \cite{Giusto:2004id,Giusto:2004ip,Giusto:2004kj}\footnote{Note that we have reversed $y\rightarrow -y$ compared to \cite{Giusto:2004id,Giusto:2004ip,Giusto:2004kj} without restriction in order to have the same metric conventions for $u$, $v$ as the superstrata.}
\bea
ds_6^2 & = & -\frac{2}{\sqrt{\cP}} du \,dv + \frac{2Q_{p}}{\bar{\Sigma}\,\sqrt{\cP}}\,dv^2+ \bar{\Sigma}\,\sqrt{\cP} \left( \frac{dr^2}{r^2 +
(\gamma_1+\gamma_2)^2\eta} + d\theta^2
\right)\nonumber \\
         &+& \sqrt{\cP}\, \Bigl( r^2 + \gamma_1\,(\gamma_1+\gamma_2)\,\eta -
\frac{Q_1 Q_5\,(\gamma_1^2-\gamma_2^2)\,\eta\,\cos^2\theta}{\bar{\Sigma}^2\cP}
\Bigr)
\cos^2\theta \,d\psi^2  \nonumber \\
&+& \sqrt{\cP}\,\Bigl( r^2 + \gamma_2\,(\gamma_1+\gamma_2)\,\eta +
\frac{Q_1 Q_5\,(\gamma_1^2-\gamma_2^2)\,\eta\,\sin^2\theta}{\bar{\Sigma}^2\cP}
\Bigr) \sin^2\theta\,d\phi^2  \nonumber \\
&+& \frac{Q_P\,(\gamma_1+\gamma_2)^2\,\eta^2}{\bar{\Sigma}\,\sqrt{\cP}}
\left( \sin^2\theta\, d\phi + \cos^2\theta \,d\psi\right)^{2} \nonumber\\
&-& \frac{2 \sqrt{2 Q_{1}Q_{5}} }{\bar{\Sigma}\,\sqrt{\cP}}
\left(\gamma_2 \sin^2\theta \,d\phi+\gamma_1 \cos^2\theta \,d\psi \right)
\,dv
\nonumber \\
&+& \frac{\sqrt{2Q_1 Q_5}\,(\gamma_1+\gamma_2)\,\eta}{ \bar{\Sigma}\,\sqrt{\cP}}
\left( \sin^2\theta\, d\phi + \cos^2\theta \,d\psi\right) \(dv-du \), \label{metric6d}
\eea
where $u$ and $v$ are the null coordinates composed from the time coordinate and the common S$^1$ of the D1 and D5 branes in \eqref{uvty}. The functions $\bar{\Sigma} $ and $\cP$ are defined by
\begin{equation}
 \bar{\Sigma} ~\equiv~ r^2+ (\gamma_1+\gamma_2)\,\eta\,\bigl(\gamma_1\, \sin^2\theta + \gamma_2\,\cos^2\theta\bigr), \qquad \cP~\equiv~ \( 1+ \frac{Q_{1}}{\bar{\Sigma}}\)\,\( 1+ \frac{Q_{5}}{\bar{\Sigma}}\)\,.
\end{equation}
This metric is asymptotically flat and caps off in its center as an orbifold of global AdS$_3\times$S$^3$.   Once again we are not interested in the three-form fluxes of the solutions since scalar excitations are insensitive to them.  Explicit expressions can be found in the  references cited above. 

The solution depends on the parameters $Q_1$, $Q_5$, $a$, $\gamma_1$, $\gamma_2$ and $\eta$, which determine the charges of the system. As one would expect,  $Q_1$ and  $Q_5$ are the  D1- and D5-brane supergravity charges.  These are related to the parameter $a$ via the regularity condition: 
\begin{equation}
a=\frac{\sqrt{Q_1Q_5}}{R_y} \,,
\label{parameter_a}
\end{equation}
One should remember that this solution was constructed starting from a 16-supercharge asymptotically-AdS solution that only had D1 and D5 charges, and its momentum charge was added by performing a spectral flow\footnote{See also \cite{Lunin:2004uu} for a similar attempt.} rather than by adding an explicit momentum wave, as is done in superstrata. The parameters $\gamma_1$,$\gamma_2$ and $\eta$ are related to the momentum charge via:   
\begin{equation}
Q_P= -\gamma_1 \,\gamma_2\,, \qquad   \eta ~=~ {Q_1 Q_5\over Q_1 Q_5 + Q_1 Q_P + Q_5
Q_P} \,,
\label{eq:momch}
\end{equation}
By expanding the metric at infinity, one can also obtain the two angular momenta of the solution
\begin{equation}
J_L ~=~\frac{\gamma_1 +\gamma_2}{2} \,\sqrt{Q_1 Q_5}\,,\qquad J_R ~=~\frac{\gamma_2 -\gamma_1}{2} \,\sqrt{Q_1 Q_5}\,.
\label{eq:GMSangularmomenta}
\end{equation}
The parameters $\gamma_1$ and $\gamma_2$ are related to the spectral flow parameter, $\mfn$, and the orbifold parameter, $k \in \mathbb{Z}$, via \cite{Giusto:2012yz}: 
\begin{equation}
\gamma_1 ~=~  -{\sqrt{Q_1 Q_5}\over R_y}\,\mfn\, \gamma \,,\quad \gamma_2~=~ {\sqrt{Q_1 Q_5}\over R_y}\,(\mfn +1)\gamma\,, \qquad \gamma ~\equiv~  \frac{1}{k}\,.
\label{gamma12}
\end{equation}
Corresponding to the supergravity charges, $\left(Q_1,\, Q_5,\,Q_P,\,J_L,\,J_R\right)$, there are the quantized charges $(n_1, n_5, n_P, j_L, j_R)$ \eqref{quantcharges}.  These charges are related to the parameters via:
\begin{equation}
j_L \= \frac{n_1 n_5}{2}\, \gamma\,,\qquad j_R \= (\mfn+\coeff{1}{2})\, n_1 n_5\, \gamma \,, \qquad n_P =  \mfn\, (\mfn+ 1)\, n_1 n_5\, \gamma^2 \,.
\label{eq:GMSchgs}
\end{equation}
Finally, it will be convenient to define a scaled version of the $a$-parameter:
\begin{equation}
\bar{a} ~\equiv~  \sqrt{\eta}\gamma \,a\,.
\label{eq:baradefn}
\end{equation}

While the underlying CFT states and the geometry are different, it is convenient to relate the quantized charges to those of the superstratum in order to obtain an approximate correspondence. In particular, (\ref{eq:GMSchgs})  matches (\ref{AngMomP}) is we identify:
\begin{equation}
N_1 ~\leftrightarrow~   n_1 n_5\, \gamma \,, \qquad N_2 ~\leftrightarrow~   \mfn\, n_1 n_5\, \gamma \,, \qquad n+1 ~\leftrightarrow~ (\mfn+1) \,\gamma \,.
\label{eq:corr1}
\end{equation}
In this expression, the superstrata quantities are on the left and the GMS quantities are on the right.  The constraint, $N_1 + 2N_2 = n_1 n_5$ in (\ref{AngMomP}), leads to 
\begin{equation}
1 ~\leftrightarrow~ (2\, \mfn +1)\,\gamma  \,.
\label{eq:corr2}
\end{equation}
One can similarly match the supergravity charges to arrive at
\begin{equation}
 \frac{1}{4} \, \frac{b^2}{a^2}  ~\leftrightarrow~  \mfn   \,, \qquad \frac{2\,a^2}{b^2+ 2 a^2}   ~\leftrightarrow~    \gamma \,, \qquad a^2 + \coeff{1}{2} b^2 ~\leftrightarrow~ a_{\text{GMS}}^2   \,.
\label{eq:corr3}
\end{equation}
The GMS solution and the superstratum are not, of course, the same solution, and have different ranges of validity, but the charges in (\ref{eq:corr2}) and (\ref{eq:corr3}) correspond perfectly in the regions where the phase spaces overlap.   Therefore the superstratum can be compared, at the mathematical level,  to the GMS solution that satisfies the constraint  (\ref{eq:corr2}).  In particular, this correspondence provides a very useful comparison between the energy regimes of both geometries.

It is important to remember that geometric details are very different and that highly-redshifted GMS geometries have a pathology that superstrata do not share.  The redshift parameter between flat space and the core of the GMS geometry is  
\begin{equation}
\frac{2 \,j_L}{n_1 n_5} ~=~ \gamma~=~ \frac{1}{k}  \,.
\label{eq:GMSredshift}
\end{equation}
To obtain a highly-redshifted geometry one must take $k$ to be extremely large. Indeed, for $j_L \sim \cO(1)$ one must take $k \sim \cO(n_1 n_5)$. However, one should remember that the AdS$_3$ and the S$^3$ have  radii of order $(n_1 n_5)^{1/4}$ in Planck units.  This means that if the orbifold is to avoid breaking the geometry into sub-Planckian pieces one must require 
\begin{equation}
k  ~\lesssim~(n_1 n_5)^{1/4}  \,.
\label{eq:kbound}
\end{equation}
The explicit momentum wave of the superstratum enables the geometry to evade this pathology, and there are no such restrictions on their depth: The supergravity approximation remains valid all the way down to the deepest states with $j_L \sim \cO(1)$ and the cap redshift can be of order $n_1 n_5$ \cite{Bena:2006kb,Bena:2007qc,deBoer:2008zn,deBoer:2009un,Bena:2016ypk}.

\subsection{Scalar wave perturbations}

The massles Klein-Gordon equation \eqref{eq:genericKleinGordon} is directly separable in the GMS geometry. We consider the mode expansion\footnote{We use a different pair of mode momenta, $(\Omega,P)$, compared to \cite{Eperon:2016cdd,Chakrabarty:2019ujg}. There is a relative reversal of $y$ direction  and their momenta along $t$ and $y$ are related to ours as $\widetilde{\omega}^\text{theirs} = - ( \Omega+P) = - (2 \Omega +q_y)$ and $\widetilde{\lambda}^\text{theirs} =   \Omega-P = -q_y$.}:
\begin{equation}
\Phi = K(r)\,S(\theta)\,e^{i\left( \frac{\sqrt{2}\,\Omega}{R_y} \,u\,+\,\frac{\sqrt{2}\, P}{R_y} \, v \,+\,q_\phi \phi \,+\, q_\psi \psi \right)}\,.
\label{eq:modeprofile2}
\end{equation}
The radial and angular wave equations are then:
\be
\begin{split}
\frac{1}{r}\frac{d}{dr}\left(r(r^2+\bar{a}^2)\frac{d}{dr}\right)K(r) \- \mathscr{V}_r(r) \,K(r) &\= \lambda\,K(r)\,,\\
{1\over \sin 2\theta}{d\over
d\,\theta}\left(\sin 2\theta {d\over d\,\theta}
\right)\,S(\theta) \- \mathscr{V}_\theta(\theta) \,S(\theta)& \= - \lambda \, S(\theta)\,,
\end{split}
\ee
where $\bar{a}$ is defined in  (\ref{eq:baradefn}) and the potentials are defined by:
\be 
\begin{split}
\mathscr{V}_r(r) ~\equiv~ & \cV_\text{asymp}(r) ~+~ \cV_\text{cap}(r)   \\
\mathscr{V}_\theta(\theta) ~\equiv~ & \frac{q_\phi^2}{\sin^2 \theta} +  \frac{q_\psi^2}{\cos^2 \theta} - \frac{4 \bar{a}^2}{R_y^2} \,P\,\Omega\,\left(\(\mfn +1\) \,\cos^2\theta - \mfn  \, \sin^2\theta \right)\,, \\ 
 \cV_\text{asymp}(r)   ~\equiv~ & - \frac{4 \Omega P}{R_y^2}\,r^2 - 4 \,\frac{(Q_1+Q_5)\,P+Q_P\, \Omega}{R_y^2}\,\Omega \,,\\
\cV_\text{cap}(r)   ~\equiv~ &  \frac{\bar{a}^2}{r^2} \, \biggl(\mfn \,q_\phi  -  (\mfn+1) q_\psi+ k\,\(P-\Omega\) \biggr)^2  \\ 
& \qquad ~-~ \frac{\bar{a}^2 }{r^2+\bar{a}^2}\,\left(   (\mfn+1) q_\phi - \mfn \,q_\psi + k\, \( P+\left( \frac{2}{\eta}-1\right) \Omega \) \right)^2\,, 
\end{split}
\label{eq:GMSpotentials}
\ee

The asymptotic potential, $ \cV_\text{asymp}(r)$,  is identical to the one obtained in the superstratum solution at large-$n$ \eqref{eq:Vflat}. The only other part of the potential is $\cV_\text{cap}(r)$, which is purely of the form of a global AdS$_3$ potential.    There is no intermediate BTZ throat and no corresponding intermediate regime in the potential like that of  \eqref{eq:VBTZ}. 

The redshift factor can be extracted from the  coefficient of $\Omega^2$ in  $\cV_\text{cap}(r)$, and here we find a redshift of $\sim k$ (since $\eta \lesssim 1$) whereas  (\ref{eq:ApproximatedPotentials}) leads to a factor of $2(1+\frac{b^2}{2a^2})$.  This is in accord with the correspondence (\ref{eq:corr3}). The redshift is given universally by
\begin{equation}
 z ~\approx~  \frac{n_1 n_5}{j_L} \,.
\label{eq:redshift}
\end{equation}

The angular potential \eqref{eq:GMSpotentials} is almost identical to the superstratum angular potential \eqref{eq:S3modes}. The last term, proportional to $P\Omega$, generates a correction from the usual spherical harmonics on S$^3$.  Thus we have
\begin{equation}
\lambda ~=~ \ell (\ell +2) \,+\, \cO \left(\frac{\bar{a}^2\,P \Omega}{R_y^2}\right) \,, \qquad \ell \in \mathbb{N}  \,.
\label{eq:lamval2}
\end{equation}
In the near-decoupling limit, $a^2 \ll R_y^2$, we will show  that $P \Omega$ scales with $\frac{1}{k^2}\,\ell^2$ for bound states. Thus, for the low-energy excitations (when the mode number $N$ is smaller than $\frac{k R_y}{a}$), one can take $\lambda~=~ \ell (\ell +2)$ and  $S(\theta)$ is given in \eqref{eq:AngularWaveProfile} with 
\be 
|q_\psi| + |q_\phi| \leq \ell\,.
\label{eq:RegAngForm2}
\ee
Just as for the $(2,1,n)$-superstratum potential, we will use the integrally-moded quantum number, $q_y$,  and  replace $P=\Omega+q_y$.
 
\subsection{Quasi-normal modes via asymptotic matching}

Following  \cite{Chakrabarty:2019ujg}, we introduce the short-hand notation:
\be 
\begin{split}
\zeta & \equi  k\,q_y\+ \mfn \,q_\phi  -  (\mfn+1) q_\psi\,, \\
\xi & \equi k\, \left( q_y + \frac{2}{\eta}\, \Omega \right) \+  (\mfn+1) q_\phi - \mfn \,q_\psi\,,\\
\nu^2 & \equi \left(\ell+1 \right)^2 - 4 \,\frac{(Q_1+Q_5)\,q_y+(Q_1+Q_5+Q_P)\, \Omega}{R_y^2}\,\Omega \,.
\end{split}
\ee
The radial wave equation becomes
\be 
\frac{1}{r}\frac{d}{dr}\left(r(r^2+\bar{a}^2)\frac{d}{dr}\right)K(r)\- \Bigg[-\frac{4\,\Omega (\Omega+q_y)}{R_y^2}\, r^2\+ \nu^2-1 \+\frac{\bar{a}^2\, \zeta^2}{r^2} \-\frac{\bar{a}^2 \,\xi^2}{r^2+\bar{a}^2} \Bigg]\,K(r)=0\,, 
\ee

We will show that the frequencies of the normal modes have the form
\be 
\Omega \=\frac{1}{k} \,\cO(N)\,.
\ee
where  $N \in \mathbb{N}$ is the mode number. Since we will take $k$ large in order to have a solution with a long throat, this means that the terms involving $\Omega$ in the definition of $\nu$ can be taken to be small and so, just as for the superstratum, we have
$$\nu \sim \ell +1 \,.$$
If $N$ starts to become large, of the order $N\sim \frac{k R_y}{\sqrt{Q_{1,5}}} \,\ell$, then $\nu^2$ will become negative and there will be no quasi-normal modes. We therefore restrict our attention to modes with $N \ll \frac{k R_y}{\sqrt{Q_{1,5}}}\,\ell$

The standard approach to quasi-normal modes is to apply matched asymptotic expansions.  Indeed this was done in \cite{Chowdhury:2007jx,Chakrabarty:2015foa,Eperon:2016cdd,Chakrabarty:2019ujg}  and we briefly recap this computation.  The details can be found in Appendix \ref{App:QNMviaAsympMatch}. In doing this analysis, we will restore a sign restriction in \cite{Chowdhury:2007jx,Chakrabarty:2015foa,Chakrabarty:2019ujg} that make them suggest that one branch of quasi-normal modes is potentially unstable by growing with time. We will show that without this sign restriction, both branches correspond to quasinormal modes that decay with time for GMS backgrounds.

The wave equation in the inner region is the wave equation in the global-AdS$_3$ cap, 
\be 
\frac{1}{r}\frac{d}{dr}\left(r(r^2+\bar{a}^2)\frac{d}{dr}\right)K_\text{in}(r)\- \Bigg[ (\ell+1)^2-1 \+\frac{\bar{a}^2\, \zeta^2}{r^2} \-\frac{\bar{a}^2 \,\xi^2}{r^2+\bar{a}^2} \Bigg]\,K_\text{in}(r)=0 \,,
\label{eq:AdSpotGMS}
\ee
 whereas the wave equation in the outer region is the wave equation in flat space, 
\be 
\frac{1}{r}\frac{d}{dr}\left(r^3\frac{d}{dr}\right)K_\text{out}(r)\- \Bigg[-\frac{4\,\Omega (\Omega+q_y)}{R_y^2}\, r^2\+ (\ell+1)^2-1 \Bigg]\,K_\text{out}(r)=0\,.
\label{eq:flatpotGMS}
\ee
The inner equation is simply that of a global AdS$_3$ cap and the outer equation is solvable in terms of Bessel functions. 

In the near-decoupling limit  ($a^2 \ll R_y^2$), the overlapping region, where the radial potential is dominated by $(\ell+1)^2-1$, is large.   This means that the matching of $K_\text{in}$ and $K_\text{out}$ provides an accurate approximation. The wave profile of quasi-normal modes is constrained by imposing smoothness at the origin and an outgoing boundary condition at infinity. As for the $(2,1,n)$ superstrata, having an outgoing wave solution to \eqref{eq:flatpotGMS} necessarily requires
\be 
\Omega \,(\Omega+q_y) >0\,.
\label{eq:condoutgoing}
\ee
For more details of the method we refer the interested reader to Appendix \ref{App:QNMviaAsympMatch}. 

In a nutshell: one imposes the proper boundary conditions in each region and then matches the power-law behavior of the hypergeometrics of global AdS at large $r$, to the small-$r$ power-law behavior of the Bessel functions.  This leads to the following constraint\footnote{We have changed a sign restriction in \cite{Chakrabarty:2019ujg}. This paper derives the formula prematurely fixing the sign of $\text{Re}(\omega)$, where $\omega$ is the momentum of the modes along $t$ (corresponding to $\omega=2\Omega +q_y$ in our convention). However, we have two branches of frequencies, one positive and one negative \eqref{eq:GMSzerothorder}. Thus, their formula applied to the branch with opposite sign leads them to the conclusion that this branch might correspond to unphysical mode that grow in time. If we do not fix the sign convention prematurely, we can see from equation \eqref{matchingnew2} that we obtain a factor of $e^{i\, \text{sign}(2\text{Re}(\Omega)+q_y) \pi \ell} $ instead of $e^{i\, \pi \ell}$ and both branches lead to decaying modes in time.}:
\begin{equation}
\begin{split}
 \frac{\Gamma(\ell+1)}{\Gamma(-\ell-1)}
\frac{\Gamma\left(\frac{1}{2}(-\ell +  |\zeta| + \xi) \right)
\Gamma\left(\frac{1}{2}(-\ell +  |\zeta| - \xi) \right)}
{\Gamma\left(\frac{1}{2}(2+\ell +  |\zeta| + \xi) \right)
\Gamma\left(\frac{1}{2}(2+\ell +  |\zeta| - \xi) \right)} & \\
~=~ e^{i\, \text{sign}(2\text{Re}(\Omega)+q_y) \pi \ell} & \frac{\Gamma(-\ell)}{\Gamma(2+\ell)}  \left(\frac{\Omega\, (\Omega+q_y)\,\bar{a}^2 }{R_y^2}\right)^{\ell+1}\,.
\end{split}
\label{matchingnew2}
\end{equation}
This equation is not exactly solvable, but, because the right-hand side is small for the lowest-energy states, one can work perturbatively.  At zeroth order, the left-hand side must vanish. Thus the Gamma functions on the denominator must hit their poles.  This results in the spectrum of normalizable modes of the AdS$_3$ cap. For the two different Gamma functions, we have two branches of frequencies, one mostly positive and one mostly negative labelled by a mode number $N\in \mathbb{N}$:
\be 
\Omega_N ~\simeq~ \Omega_N^{(0)} \= \pm \frac{\eta}{2k} \biggl[2N+\ell+2 \+ \left| k\,q_y + \mfn \,q_\phi  -  (\mfn+1) q_\psi\right| \,\mp\, \left( k\, q_y +  (\mfn+1) q_\phi - \mfn \,q_\psi \right) \biggr]\,.
\label{eq:GMSzerothorder}
\ee
To find the first-order correction, $$\Omega = \Omega_N^{(0)} \+ \delta \Omega_N\,, $$
we expand the Gamma function around its pole and obtain a purely imaginary contribution
\be 
\delta \Omega_N \simeq
i\, \frac{ \pi\,\eta } {k\,(l!)^2} \left[\Omega^{(0)}_N \left( \Omega^{(0)}_N+q_y\right) \frac{\bar{a}^2}{R_y^2}\right]^{l+1}   \  {}   ^{l+1+N} C_{l+1}  \  {}^{l+1+N+|{\zeta}|} C_{l+1}\,,
\ee
where ${}^nC_m$ is the usual binomial coefficient. The time dependence of the modes is given by \eqref{eq:decayinggrowing?} which guarantees that the wave profile is decaying in time for both branches of frequencies \eqref{eq:GMSzerothorder}.

To summarize, the spectrum of quasi-normal modes of GMS solutions is given by two towers of frequencies labelled by $ N \in \mathbb{N}$, one positive and one negative,
\be 
\Omega_N \= \Omega^{(0)}_N \+ \delta \Omega_N \,.
\ee
With the condtion that $$\Omega^{(0)}_N\left( \Omega^{(0)}_N+q_y\right) >0\,, $$we have
\be 
\begin{split}
\Omega^{(0)}_N&\, \simeq \, \pm \, \frac{\eta}{2k} \biggl[2N+\ell+2 \+ \left| k\,q_y + \mfn \,q_\phi  -  (\mfn+1) q_\psi\right| \,\mp\, \left( k\, q_y +  (\mfn+1) q_\phi - \mfn \,q_\psi \right) \biggr] \,,\\
\delta \Omega_N &\, \simeq \,i\, \frac{ \pi\,\eta } {k\,(l!)^2} \left[\Omega^{(0)}_N \left( \Omega^{(0)}_N+q_y\right) \frac{\bar{a}^2}{R_y^2}\right]^{l+1}   \  {}   ^{l+1+N} C_{l+1}  \  {}^{l+1+N+|{\zeta}|} C_{l+1}\,.
\end{split}
\label{eq:modesGMS}
\ee

We could, equally well, have used the WKB approach that we used for superstrata. Indeed, the techniques are almost certainly equivalent in that we match two accurate but approximate solutions in an inner and outer region and this matching is achieved in the large overlap region where the potential is constant.   The advantage of the WKB method is that it is easily applicable to geometries with more than two regions, such as superstrata.

In Appendix \ref{App:QNMviaWKB}, we apply our WKB techniques to the GMS backgrounds. This allows us to check in Appendix \ref{App:WKBvsAsympMatch}  the precision of the WKB spectrum formulae \eqref{eq:RealPartSpectrum} and \eqref{eq:ImPartSpectrum} compared to the matched asymptotic expansion calculation. In a concrete example we show that the mismatch is below 5\% as soon as we take $N >10$ and $\ell >10$.

\subsection{The eikonal limits}

Once again, we are interested in the slowest possible decay rates and the discussion is directly parallel to our discussion for the superstrata.  

Slow decay means that  we look at the large-$\ell$ limit and arrange the mode numbers so that $\Omega_N^{(0)}$ remains as small as possible and, if possible, cancel the explicit growth with $\ell$.   This cancellation is slightly more tedious than for quasi-normal modes of the $(2,1,n)$ superstratum. This is caused by the nontrivial mixing of $q_\phi$ and $q_\psi$ with the parameter $\mfn$ of the background, as is evident in \eqref{eq:modesGMS}.   It is also related to the non-trivial form of the evanescent ergosphere. 
We will skip most of the details of the computation, which may be found in \cite{Chakrabarty:2019ujg,Eperon:2016cdd}. 

The important result is that for any value of $\mfn$, one can pick a pair of $(q_\phi, q_\psi)$ satisfying 
\be 
q_\phi \+ q_\psi \= \pm \ell\,,
\label{eq:modesatevanescent2}
\ee
where the $\pm$ depends on which branch of $\Omega_N^{(0)}$ is considered, and for which the ratio $\frac{q_\phi}{q_\psi}$ is bounded by $\frac{\mfn}{\mfn+1}$.
As for superstrata, these modes correspond to modes for which the wave profile is strongly localized at the evanescent ergosurface. One then finds
\be 
\Omega_N^{(0)} \= \pm  \,\frac{\eta}{2k}  \,\left( 2 N + 2 + k(|q_y| \mp q_y)\right) \equi \pm  \,\frac{\eta}{2k} \,\omega_N^{(0)}\,.
\ee
In addition to the generic formulas \eqref{eq:Stirling}, one will need
\be 
{}^{\ell+1+X+j\,\ell} C_{\ell+1} ~\approx~ \exp \left[ \left(j\, \log\(1+j^{-1}\)+\log(1+j) \right)\ell-\frac{1}{2}\,\log \ell \+ \cO(1) \right]\,,
\ee
and the imaginary part of the frequency \eqref{eq:modesGMS} behaves as
\begin{equation}
\begin{split}
\delta \Omega_N \simeq i\,\exp &\Biggl[ -2 \ell \left(2 \log k + \log \ell - \frac{1}{2} \log \left(\omega_N^{(0)} +k\,\frac{2q_y}{\eta} \right) \right)\+\left(N-\frac{3}{2}\right) \,\log \ell \\
&  \+ \left(2 + \log \frac{a^2\, \omega_N^{(0)}\eta^3(1+j)}{4R_y^2} +j\,\log(1+j^{-1})\right) \ell \- 5 \log k+ \log \left(\omega_N^{(0)} +k\,\frac{2q_y}{\eta}  \right) \Biggr]\,,
\end{split}
\end{equation}
where $j \equi \frac{|\zeta|}{\ell} \= \frac{\mfn q_\phi - (\mfn+1) q_\psi}{\ell} \= \cO(\ell^0)$. Thus, for the modes at the evanescent ergosurface, the decay rate at large $\ell$ is minimal when $q_y =0$ and the leading-order terms are
\begin{equation}
\delta \Omega_N \,\sim \, i\,\ell^{- 2 \ell}\,k^{-4 \ell -5}\, \left( \frac{e \,a\, \eta^{\frac{3}{2}} (N+1) \sqrt{1+j}}{R_y} \right)^{2\ell} \left(1+j^{-1} \right)^{j\,\ell} \,\ell^{N-\frac{3}{2}}\,\, \cO(1)\,,
\label{eq:dOmerg2}
\end{equation}
where $e = \exp(1)$ and we have highlighted the factor of $\ell^{- 2 \ell}$. 

For generic modes, we consider arbitrary $q_\phi$ and $q_\psi$ but differing from \eqref{eq:modesatevanescent2}. We will assume, for simplicity, that $N,\,q_y,\,q_\phi$ and $q_\psi$ are all fixed and small compared to $\ell$.  However, one will obtain a similar result, but with different coefficients,  if one allows some of the mode numbers to scale with $\ell$. Using the same procedure we obtain
\begin{equation}
\begin{split}
\delta \Omega_N \= i\,\exp &\left[ -4 \ell \log k \+ \left(2 + \log \frac{a^2\,\eta^3}{4\,R_y^2} + \,\log \left(1+\frac{k}{\ell}\frac{2\,q_y}{\eta} \right)\right) \ell \right. \\
&  \+(2 N+|\zeta| +1)\,\log \ell \- 5 \log k\+ \cO(1)\, \Biggr]\,.
\end{split}
\end{equation}
The decay rate at large $\ell$ is minimal when $q_y =\zeta=0$ and we have
\begin{equation}
\delta \Omega_N \,\sim \, i\,k^{-4 \ell-5} \, \left( \frac{e \,a\,\eta^{\frac{3}{2}}}{2\,R_y} \right)^{2\ell} \, \ell^{2N+1}\,\, \cO(1)\,.
\label{eq:dOmgeneric2}
\end{equation}
We can now compare this with the decay rate of low-energy quasi-normal modes of the superstrata, derived in \eqref{eq:dOmerg} for modes at the evanescent ergosurface and in \eqref{eq:dOmgeneric} for generic modes. Taking into account that $k = \frac{n_1 n_5}{2\,j_L}$, it may appear that the smallest decay rate for GMS backgrounds is smaller that the smallest decay rate for superstrata,  $$\ell^{-2\ell} \left(\frac{n_1 n_5}{j_L}\right)^{-4\ell-5} (\ldots) \qquad \text{vs}\qquad \ell^{-2\ell} \left(\frac{n_1 n_5}{j_L}\right)^{-2\ell-3} (\ldots).$$
However, as explained earlier, the GMS background has a reliable supergravity description if and only if $k \lesssim (n_1 n_5)^{\frac{1}{4}}$, which is equivalent to $j_L \gtrsim (n_1 n_5)^{\frac{3}{4}}$, whereas superstrata can have arbitrarily low angular momentum. Thus, the smallest decay rate for GMS background that can be accurately computed is actually at least $(n_1 n_5)^\ell$ times larger than the smallest decay rate for superstrata.

\section{Decay timescales}
\label{sec:Decay}

We now examine the leakage of energy from superstrata with a deeply-capped BTZ throat. Our discussion will closely follow that of \cite{Eperon:2016cdd}.   In particular, given the imaginary parts of the quasi-normal modes, one is looking for a uniform bounding function, $g(t)$, on a generic energy function, $E(t)$, that measures the energy of a scalar field in the microstate geometries.  The function, $E(t)$, is defined on space-like hypersurfaces, $\Sigma_t$, obtained by time slicing the microstate geometry. The goal is to find a bounding function, $g(t)$, that is independent of the details of the modes.

One should first recall the separated form of our wave-functions, (\ref{eq:decayinggrowing?}):
\begin{equation}
e^{i\left( \frac{\sqrt{2}\,\Omega}{R_y} \,u\,+\,\frac{\sqrt{2}\, P}{R_y} \, v \right)}\= e^{i\left( \frac{2\,\Omega}{R_y} \,t\,+\,\frac{q_y}{R_y} \, (t+y)\right)} \= e^{i \, \frac{2 \,\delta\Omega_N}{R_y}\, t}\,e^{i\left( \frac{2\,\Omega_N^{(0)}}{R_y} \,t\,+\,\frac{q_y}{R_y} \, (t+y)\right)}\,.
\label{eq:decay}
\end{equation}
We define 
\begin{equation}
 \omega_I  ~\equiv~ i \, \frac{2 \,\delta\Omega_N}{R_y}  \,,
\label{eq:decayrate}
\end{equation}
and note that it is always negative, and represents the inverse decay time of the quasi-normal mode.

Any basic energy function, $E_1(t)$ should be quadratic in the scalar and its first derivatives, and so should behave, for large quantum numbers, as
\begin{equation}
E_1(t) ~\sim~ \ell^2 e^{2 \omega_I  t} \,,
\label{eq:E1sim}
\end{equation}
where $\ell \gg 1$ is the dominant quantum number on the S$^3$.

As pointed out in  \cite{Eperon:2016cdd}, because microstate geometries have evanescent ergospheres,  and may involve trapping, the energy   function $E_1(t)$  might only be bounded by a second-order energy function, $E_2(t)$, which will be a quadratic in $\Phi$, $\partial \Phi$ and $\partial^2 \Phi$. Thus one seeks a ``universal'' function, $g(t)$, for which one has
\begin{equation}
E_1(t) ~<~  g(t) E_2(0) \,,
\label{eq:Ebound1}
\end{equation}
for $t >0$. The function, $E_2(t)$ will obey (\ref{eq:E1sim}), but with $\ell^2$ replaced by $\ell^4$.   Thus, we expect $E_2(0) ~<~ C \ell^4$ for some constant, $C$, that depends only on the energies of the waves and not on the details of the modes.  Thus, we are seeking a universal function, $g(t)$ that satisfies
\begin{equation}
 e^{2 \omega_I t}  ~<~ C\,g(t)\, \ell^2 \,,
\label{eq:Ebound2}
\end{equation}
in the large-$\ell$ limit, for some constant, $C$. 

For ultra-compact stars, at large $t$, the standard uniform bounding functions have the form \cite{Holzegel:2013kna,Keir:2014oka,Cardoso:2014sna,Moschidis:2015wya,Eperon:2016cdd}\footnote{It is also possible that higher-order energy functions can lead to higher negative powers of  $\log(t+2)$.  See, \cite{Keir:2014oka} for more details.  Such a possibility will not significantly modify our discussion here. For interesting further studies see \cite{Cardoso:2016rao,Cardoso:2019rvt}.}:
\begin{equation}
g(t)  ~=~ D\, \big(\log(t+2)  \big)^{-2} \,, \qquad {\rm as } \quad t \to \infty \,,
\label{eq:unibound}
\end{equation}
where $D$ is some constant that only depends on the background. One can then test this bounding function to see if it works for all modes at late times.  Indeed, consider  the time scale $t \sim e^{\tau \ell}$ for large $\ell$ and for some choice of $\tau$.  The condition (\ref{eq:Ebound2}) then becomes:
\begin{equation}
\log\bigg[\,\frac{\tau}{\sqrt{D\,C}}  \,\bigg]   ~\lesssim~ \left| \omega_I \right| \, e^{\tau \, \ell}  \,,
\label{eq:Ebound3}
\end{equation}
which must be satisfied at large $\ell$ for all values of $\tau >0$.  In particular, note that it is compatible with $\omega_I $ having the form:
\begin{equation}
\omega_I   ~\sim~  - \beta_0 \, e^{-\beta_1 \ell}  \,.
\label{eq:omIform}
\end{equation}
for any fixed  $\beta_0, \beta _1>0$, independent of $\ell$. Indeed, \eqref{eq:Ebound3} becomes
\be 
\log\bigg[\,\frac{\tau}{\sqrt{D\,C}}  \,\bigg]  ~\lesssim~ \beta_0 \, e^{(\tau - \beta_1) \,\ell}\,,
\ee
which is obviously satisfied at large $\ell$ by choosing $D$ appropriately.

The important point is that (\ref{eq:dOmgeneric}) has the form of (\ref{eq:omIform}) and so do most of the factors in (\ref{eq:dOmerg}).  The  problem \cite{Eperon:2016cdd}  is the factor of $\ell^{-2\ell}$, which means that for low-energy modes that localize near the evanescent ergosphere, $\omega_I$ contains a piece of the form 
\begin{equation}
\omega_I ~ \sim~ e^{-2 \ell \log \ell}    \,.
\label{eq:probelgrowth}
\end{equation}
This means that the right-hand side of (\ref{eq:Ebound3}) will always go to zero at large $\ell$, while the left-hand side of  (\ref{eq:Ebound3})  can be made arbitrarily large by taking $\tau$ large enough.  This led to the conclusion in  \cite{Eperon:2016cdd}  that bound states of microstate geometries decay more slowly than for ultra-compact stars.

However, if $\ell$ has a natural cut-off, $\ell \lesssim \Lambda$, then $e^{-2\ell \log\ell}$ will be bounded below by $e^{-2\Lambda \log \Lambda}$. Thus,
\be 
\left|\omega_I  \right| ~\gtrsim~ \beta_0 \, e^{-2\Lambda \log \Lambda}\, e^{-\beta_1 \,\ell}\,,
\ee
and we can show that for appropriately-chosen $D$ we have the bound 
\be 
\log\bigg[\,\frac{\tau}{\sqrt{D\,C}}  \,\bigg]  ~\lesssim~ \beta_0\, e^{-2\Lambda \log \Lambda} \, e^{(\tau - \beta_1) \,\ell} ~\lesssim~ \left| \omega_I \right| \, e^{\tau \, \ell}\,.
\ee
Thus, in the presence of a UV cut-off for $\ell$, the standard bounding functions for ultra-compact stars \eqref{eq:unibound} is also valid for microstate geometries.

The important point is that superstrata, and microstate geometries, have precisely such a cut-off imposed by the validity of the supergravity approximation. For superstrata, the radius of the S$^3$ is given by $(Q_1 \, Q_5 )^{\frac{1}{4}}$.  The $\ell^{\rm th}$ mode has a angular profile on S$^3$ that has necessarily $\ell$ zeroes between the North and South poles and so slices the sphere into sectors of size  $\ell^{-1} (Q_1 \, Q_5 )^{\frac{1}{4}}$.  For the six-dimensional supergravity approximation to remain valid, these modes must not slice the sphere into pieces that are smaller that the Planck scale or the compactification scale.  From  (\ref{cNdefn}) we see that 
\begin{equation}
\bigg(\frac{Q_1 \, Q_5}{n_1 \, n_5} \bigg)^{\frac{1}{4}}   ~=~\frac{ \ell_{10}^2}{\big({\rm Vol} (T^4)\big)^{\frac{1}{4}} } \,.
\label{wiggles}
\end{equation}
Thus, this means we need to limit $\ell$ by 
\begin{equation}
\ell  ~\lesssim ~(n_1 n_5)^{\frac{1}{4}}  \,,
\label{llimit}
\end{equation}
and $\ell$ has a UV bound given by $\Lambda \sim (n_1 n_5)^{\frac{1}{4}}$.

Thus the supergravity cut-off on the modes in $\ell$, means that all terms that decay slower than $e^{-\beta_1 \,\ell}$ are not an issue. Even more importantly, the primary effect on the  leakage of energy are the terms involving $ \frac{j_L}{n_1 n_5 }\approx \frac{1}{2} E_{\text{gap}} $ and  the energy decay is bounded by the standard expression, (\ref{eq:unibound}), as ultra-compact stars with a large value for $D$ that will depend on ${E_\text{gap}}^{-1}$.  This is precisely what one should expect for microstate geometries.  They look like black holes until very near the horizon scale.  They are thus as compact as an object can be, short of being a black hole.  It is also extremely natural that the time-scale for the decay is set by the inverse energy gap for the lowest-energy excitations of the system. 

We therefore find that, for modes below the supergravity cutoff, 
\be 
\left| \omega_I \right| ~ \gtrsim ~  \left(\frac{n_1 n_5}{j_L} \right)^{-2\ell-3} ~\sim~  \left( E_{\text{gap}} \right)^{2\,\ell+3} \,,
\ee
which leads to a minimal time-scale for the decay of order
\be 
t_\text{decay} ~\sim~ \big(E_\text{gap}\big)^{-2\,\ell -3}\,.
\ee

\section{Final comments}
\label{sec:Conclusions}

We have analyzed the decay rates of quasi-normal modes in superstrata.  While we only computed these decay rates using a WKB approximation in a particular family of superstrata in which the massless scalar wave equation is ``almost separable,'' we believe that our results are universal. We have shown that there are two regimes of energy. The low-energy modes are only sensitive to the highly-redshifted AdS$_3$ cap of the superstrata and the spectrum is the one we obtain in asymptotically-flat redshifted AdS$_3$ backgrounds, \eqref{eq:zerothordercapregime} and \eqref{eq:deltaOmegaN1}.   In particular, the time scale for the decay is set by the energy gap of the lowest-energy states in the microstate geometry:
\begin{equation}
E_{\rm gap} ~\approx~ \frac{2 j_L}{n_1 n_5} 
\end{equation}

At intermediate energy, the modes start exploring the BTZ throat of the geometry. We have shown that the real part of the frequencies have almost exactly the same frequencies as the low-energy modes, but the imaginary part is strongly attenuated in the BTZ throat, \eqref{eq:modesBTZregime}. 
This attenuatation has the effect of confining the modes for much longer in the cap, when compared to an AdS$_3$ cap glued directly to flat space. Intuitively, this effect can be thought of as coming from the strong rigidity against perturbation of the AdS$_2$ throat that interpolates between the cap in the IR and the flat space in the UV.

We have also shown that the extremely-long-duration trapping described in \cite{Eperon:2016cdd} is not an issue, neither for superstrata with deeply-capped BTZ throats nor for the shallow GMS geometries. The concern was that such trapping would lead to instabilities.  However, for superstrata with long throats, the modes that would be subject to such long-term trapping have extremely sub-Planckian wavelengths.  If one stays within the validity of the supergravity approximation, the trapping has the natural decay timescale determined by the energy gap. 


In addition, there exist families of modes that are trapped forever, and never decay. The non-trivial examples of such modes have a ``momentum charge'' opposite to the momentum of the solution, and the attraction between these opposite charges ensure that the force felt by these modes will always be attractive, and these modes will never be able to escape at infinity. Since these modes never decay, one might worry that if one creates them at the bottom of the solution they would give rise to non-linear instabilities and lead to black hole formation.

However, things are not so simple. First, the microstate geometries we consider have a moduli space whose dimension is $n_1 n_5$ \cite{Rychkov:2005ji, Bena:2014qxa}, and hence any energy one puts in them will excite the massless degrees of freedom corresponding to moving in this moduli space, and simply move the microstate geometry to another nearby one. This observation was also made in \cite{Marolf:2016nwu, Bena:2018mpb}.

Second, since the momentum charge of the eternally trapped modes is negative (compared  to the momentum charge of the background), we expect their physics to be similar to the physics of antibranes. In fact it is not hard to see that if one dualizes the anti-branes in bubbling solutions analyzed in \cite{Bena:2011fc, Bena:2012zi} to the D1-D5-P duality frame, one of the possible anti-brane charges corresponds exactly to the negative momentum of the eternally-trapped modes. 

Hence, we expect these modes to have other decay channels, similar to the brane-flux annihilation of anti-branes \cite{Kachru:2002gs}.  This process was studied in a dual frame where microstate geometries with multiple bubbles have charges corresponding to three M2 branes wrapping two-tori inside $T^6$ \cite{Bena:2011fc,Bena:2012zi}, and it was found that for microstate geometries with a very long throat this process can be very fast 
\cite{Bena:2015dpt}. It would be very interesting to work out the details of this non-perturbative process in the D1-D5-P duality frame, using superstrata instead of multi-bubble solutions, and calculate the decay times for the modes which perturbatively appear to be trapped forever.

It would also be interesting to try to construct the non-supersymmetric solutions sourced by these modes, especially in light of the recent observation that certain six-dimensional superstratum solutions can be described using a consistent truncation to three-dimensional supergravity \cite{Mayerson:2020tcl}. 

Returning to our study of the quasi-normal modes, we have shown that the WKB method can be used to extract the leading-order physics of trapping.  In particular, the decay rate is given by (\ref{eq:ImPartSpectrum}) and is determined by standard barrier penetration calculations.  Moreover, rough estimates of the area under the barrier provide the leading-order time-scales.  This leads us to believe that are results are universal for all deeply-capped BTZ geometries and not limited to the family of superstrata that we analyzed here.

The final result is that the decay time-scale for states in a superstratum with a deeply-capped BTZ throat is set by 
\begin{equation}
t_\text{decay} ~\sim~ \big(E_\text{gap}\big)^{-2\ell -3} \label{decay-gap}
\end{equation}
where $\ell$ is the ``three-dimensional mass,'' or the dominant wave-number on the S$^3$ that would represent the horizon of the corresponding black hole.  

It would be very interesting to compute this decay time using CFT methods. When $R_y^2 \gg Q_1 Q_5$ the solutions have an AdS$_3 \times$ S$^3$ throat in the intermediate region, between the AdS$_2$ throat and the flat space at infinity. As such, they are dual to certain states of the D1-D5 CFT \cite{Bena:2015bea, Bena:2016agb, Giusto:2015dfa}. This CFT has central charge $c_{\,\!_{CFT}} \equiv n_1 n_5$ and is unitary; thus it cannot capture all by itself the decay of our modes. To do this one should couple the CFT to flat space using the same technology as that used for computing the decay time of the JMaRT solution \cite{Chowdhury:2007jx,Chowdhury:2008bd,Chowdhury:2008uj,Avery:2009tu,Chakrabarty:2015foa}. This could be done by considering an operator of dimension $h$ and R-charge $j_L$ in the D1-D5 CFT coupled to flat space. In the language of this CFT, one expects the decay time (\ref{decay-gap}) to be
\begin{equation}
t_\text{decay} ~\sim~ \big(c_{\,\!_{CFT}}\big)^{2 h-1} \big(j_L\big)^{-2 h+1}\,,
\end{equation}
and one may envision doing a calculation of the type presented in \cite{Chowdhury:2007jx,Chowdhury:2008bd,Chowdhury:2008uj,Avery:2009tu,Chakrabarty:2015foa} in order to evaluate this decay time.

Moreover we have shown that the energy bounds on trapping in superstrata seem to be more consistent with the energy bounds of ultra-compact stars, rather than behaving as some exotic new class of objects.  This is precisely what one would hope for a microstate geometry: it is supposed to behave just like a black hole until close to the horizon region, where it caps off and looks just like an ultra-compact star.  In this framework, the information problem is resolved by having the state of the entire system encoded and accessible in precisely such an ultra-compact star created and supported by the microstate geometry.


\section*{Acknowledgments}
\vspace{-2mm}
We would like to thank Andrea Puhm and Harvey Reall for interesting discussions and Samir Mathur for sharing some of his insights. The work of IB was supported in part by the ANR grant Black-dS-String (ANR-16-CE31-0004-01), by the John Templeton Foundation grant 61149 and by the ERC Grant 772408-Stringlandscape. The work of PH was supported by the NSF grant PHY-1820784. The work of NW was supported in part by the ERC Grant 787320 - QBH Structure, and by the DOE grant DE-SC0011687. PH is very grateful to the IPhT, CEA-Saclay for hospitality while a substantial part of this work was done.

\vspace{2cm}

\appendix
\leftline{\LARGE \bf Appendices}

\section{Quasi-normal modes of asymptotically-flat AdS$_3$ backgrounds}
\label{App-GMS}

A careful study of quasi-normal modes of asymptotically-flat backgrounds is greatly facilitated if the scalar wave is separable, or approximately separable in the sense described in this paper.  If  no exact solutions can be found, then one must resort to approximate methods and two of the standard techniques are the WKB approximation and the asymptotic matching method. 
For black-hole physics it seems that WKB approximation is preferred (for example, see \cite{Zouros:1979iw,Iyer:1986np,Konoplya:2011qq,Konoplya:2019hlu}).  

On the other hand, the analysis of very simple microstate geometries has been performed via asymptotic matching. These include supersymmetric GMS solutions \cite{Eperon:2016cdd,Chakrabarty:2019ujg} and non-supersymmetric solutions \cite{Jejjala:2005yu,Chowdhury:2007jx,Chowdhury:2008bd,Chowdhury:2008uj}.   These three-charge solutions have an AdS$_3\times$S$^3$ cap that is directly glued to a flat five-dimensional space with an extra S$^1$.   This means that exact solutions can easily be constructed in separate, but overlapping, regions and  the quasi-normal modes can be obtained by asymptotic matching in the ovelap. 

Superstrata, and other microstate geometries with long, black-hole-like throats are far more complicated, and so require a more universally applicable approximation method and this is where WKB methods become more appropriate.  One of the goals of this Appendix is to assess the accuracy of WKB methods by making a detailed examination of  solutions with a global AdS$_3 \times$S$^3$ region glued in the UV to flat space. In particular, we derive the spectrum of quasi-normal modes in supersymmetric GMS solutions,  using both matched asymptotic expansions as in \cite{Chowdhury:2007jx}, and using  the WKB technique detailed in Section \ref{sec:WKB}.  We will see that these methods produce very similar results.

\subsection{The wave equation}

Using the same conventions as in the main sections of the paper, we  consider a three-dimensional background parameterized by a radial coordinate $r$, and two null coordinates 
\be 
u \equi \frac{t-y}{\sqrt{2}}\,,\qquad v \equi \frac{t+y}{\sqrt{2}}\,,
\ee
where $t$ is the time direction in the asymptotically-flat region and $y$ is the extra S$^1$. In the IR, the three-dimensional space is global AdS$_3$ whereas in the UV it is a S$^y$ fibration over a flat two-dimensional space. Moreover, $t$ and $y$ are isometries, so the scalar modes have an $r$-dependent profile and are decomposed into Fourier modes according to $u$ and $v$ by defining the conjugate momenta $\Omega$ and $P$:
\be 
\Phi (r,u,v) \= K(r) \, e^{i\left(\frac{\sqrt{2}\,\Omega}{R_y}\,u \+\frac{\sqrt{2}\,P}{R_y}\,v \right)}\,,
\label{eq:ModeExpApp}
\ee
where $R_y$ is the radius of the $y$-circle. 

\noindent The radial wave equation of a scalar of mass $\nu^2-1$ is
\be 
\frac{1}{r}\frac{d}{dr}\left(r(r^2+a^2)\frac{d}{dr}\right)K(r)\- \Bigg[-\frac{4\,\Omega P}{R_y^2}\, r^2\+ \nu^2-1 \+\frac{a^2\, \zeta^2}{r^2} \-\frac{a^2 \,\xi^2}{r^2+a^2} \Bigg]\,K(r)=0\,, 
\label{eq:WaveEqApp}
\ee
where $a$ is the curvature radius of the AdS$_3$ region and $\zeta$ and $\xi$ are function of the mode momenta. The parameter, $\zeta$, is the centripetal coefficient at the origin and regularity of the wave requires
\be 
\zeta ~ \in~  \ZZ\,.
\label{eq:centripetalinZ}
\ee

Here, the three-dimensional space is part of a bigger six-dimensional space via an S$^3$ fibration parameterized by three angles $(\theta,\phi,\psi)$. In this higher-dimensional space the scalar can be considered massless and the effective mass, $\nu$, arises from the eigenvalue problem of the angular wave equation. When this angular wave equation can be reduced to a spherical harmonic equation, $\nu$ is labelled by a positive integer, $\ell \in \mathbb{N}$, $$ \nu \equi \ell+1 \,.$$ Moreover, the coefficients $\zeta$ and $\xi$ will depend  on the momenta along $\phi$ and $\psi$ ($q_\phi$ and $q_\psi$). This dependence will be determined by the AdS$_3$ cap, essentially its redshift, and the details of the S$^3$ fibration. For the $(2,1,n)$ superstratum in the low energy regime, we have \eqref{eq:Vcap}
\be 
\zeta \= q_\psi \+ P \- \Omega\,, \qquad \xi \= q_\phi \+ P \+ \(1+\frac{b^2}{a^2} \) \,\Omega \,,
\label{eq:zetaxiSupApp}
\ee
whereas for GMS solutions, we have \eqref{eq:GMSpotentials}
\be
\zeta \= k\,\( P-\Omega\)\+ \mfn \,q_\phi  -  (\mfn+1) q_\psi\,, \qquad 
\xi  \= k\, \left( P + \(\frac{2}{\eta}-1\)\, \Omega \right) \+  (\mfn+1) q_\phi - \mfn \,q_\psi\,.
\label{eq:zetaxiGMSApp}
\ee
Because the computation does not require the details of the expression of $\zeta$ and $\xi$, we will keep them as arbitrary parameters.

By inspecting the various terms of the potential \eqref{eq:WaveEqApp}, we easily recognize  the potential of flat space, $-\frac{4\,\Omega\, P}{R_y^2} r^2 \+\nu^2 \- 1 $, as well as the potential of global AdS$_3$, $\nu^2-1 \+\frac{a^2\, \zeta^2}{r^2} \-\frac{a^2 \,\xi^2}{r^2+a^2} $. Thus, by requiring that the plateau given by $\nu^2-1$ is large, we expect that a WKB approximation or an asymptotic matching method are accurate. The size of the plateau requires us to impose a hierarchy of scales between the turning points of the flat-space potential and the turning points of the AdS$_3$ potential. This is guaranteed if
\be 
\frac{a}{R_y} \ll 1\,.
\label{eq:CondApp}
\ee 
In addition, the WKB approximation needs a large number of oscillations in the classical regions of the potential. This will require the classical turning points of the AdS$_3$ potential to be significantly separated; hence this method will not necessarily provide a good  approximation for the decay of the first few quasi-normal modes.

The quasi-normal modes are characterized by their oscillatory behavior at large distance. At infinity, their wave profiles are determined by the term $-\frac{4\,\Omega\, P}{R_y^2} r^2$. Having an oscillatory wave then requires 
\be 
\Omega \, P \,>\,0\,.
\label{eq:CondApp2}
\ee

We prefer to work with the conjugate momentum along the periodic direction $y$, $q_y$, which is integer-moded:
\be 
q_y \equi P - \Omega \in \ZZ\,.
\ee

\subsection{The spectrum of quasi-normal modes via asymptotic matching}
\label{App:QNMviaAsympMatch}

By assuming \eqref{eq:CondApp}, we have decomposed the scalar potential in two overlapping regions. The inner region is defined by
\be 
0 ~<~ \frac{r}{a} ~\lesssim ~ \epsilon \, \frac{R_y}{a} \,,
\label{eq:innerRegApp}
\ee
where we have introduced an arbitrary scale parameter, $\epsilon$, which can be chosen for convenience to be in the interval  $\frac{a}{R_y} \ll  \epsilon\ll 1$. One can check that in this range 
\be 
\frac{4\,\Omega\, P}{R_y^2} r^2 ~\lesssim ~ \cO(\epsilon^2)\,,
\ee
and the scalar potential is the one of the global AdS$_3$ cap. The outer region is defined by
\be 
 \frac{1}{\epsilon} \, \frac{R_y}{a}   ~\lesssim ~\frac{r}{a}\,,
 \label{eq:outerRegApp}
\ee
with the same $\epsilon$. Thus we have,
\be 
\frac{a^2\, \zeta^2}{r^2} \-\frac{a^2 \,\xi^2}{r^2+a^2} ~\lesssim ~ \cO(\epsilon^2)\,,
\ee
and the scalar potential is that of flat space. In the overlapping region,
\be 
 \frac{1}{\epsilon} \, \frac{R_y}{a}   ~\lesssim ~\frac{r}{a} ~\lesssim ~ \epsilon \, \frac{R_y}{a} \,,
 \label{eq:overlapRegApp}
\ee
both potentials are valid and their solutions can be matched. The philosophy of the matched asymptotic expansion is:
\begin{itemize}
\item[-] To solve the wave equation in the inner region by imposing the quasi-normal-mode boundary condition at $r \rightarrow 0$. For a smooth background as a global AdS$_3$ cap, this requires a smooth wave profile at $r\rightarrow 0$.
\item[-] To solve the wave equation in the outer region by imposing the quasi-normal-mode boundary condition at $r\rightarrow \infty$. This requires a purely outgoing wave at infinity.
\item[-] To match the asymptotic expansion of the wave profiles in the overlapping region. This matching will give an expression that will constrain the frequencies of the modes.
\item[-] To solve, perturbatively or exactly, the matching condition. This will give a tower of frequencies labelled by a mode number $N \in \mathbb{N}$. 
\end{itemize} 

\subsubsection{Solution in the inner region}

In the inner region, the scalar equation \eqref{eq:WaveEqApp} is approximated by the AdS$_3$ scalar equation
\be 
\frac{1}{r}\frac{d}{dr}\left(r(r^2+a^2)\frac{d}{dr}\right)K(r)\- \Bigg[ \nu^2-1 \+\frac{a^2\, \zeta^2}{r^2} \-\frac{a^2 \,\xi^2}{r^2+a^2} \Bigg]\,K(r)=0\,.
\ee
The solution regular at $r =0$ (satisfying $K_\text{in}(0)=0$) is
\be 
K_\text{in}(r) \=  C_\text{in} ~ r^{|\zeta|}  ~(r^2+a^2)^{\frac{\xi}{2}}~~ {}_2 F_1 \left[\frac{1}{2} \left(1-\nu +|\zeta| +\xi \right)\,,\, \frac{1}{2} \left(1+\nu +|\zeta| +\xi \right)\,,\,1+|\zeta|\,,\, -\frac{r^2}{a^2}\right]\,,
\ee
where $C_\text{in}$ is a normalization constant. 

\noindent In the overlapping region, $ \frac{r}{a} \sim \epsilon \, \frac{R_y}{a} \gg 1$, the radial wave profile behaves as
\be 
\begin{split}
K_\text{in}(r) \approx C_\text{in}\,\,a^{|\zeta|+\xi}\,\,\Gamma(1+|\zeta|) \, \Biggl[ &\frac{\Gamma(-\nu)}{\Gamma\left( \frac{1}{2}\left(1-\nu+|\zeta|+\xi \right)\right)\,\Gamma\left( \frac{1}{2}\left(1-\nu+|\zeta|-\xi \right)\right)} \, \left( \frac{r}{a}\right)^{-\nu-1}\\
& + \frac{\Gamma(\nu)}{\Gamma\left( \frac{1}{2}\left(1+\nu+|\zeta|+\xi \right)\right)\,\Gamma\left( \frac{1}{2}\left(1+\nu+|\zeta|-\xi \right)\right)} \, \left( \frac{r}{a}\right)^{\nu-1} \Biggr]\,.
\end{split}
\label{eq:waveoverlappingApp1}
\ee
Note that it we are implicitly considering that $\nu \not \in \mathbb{Z}$ which is contradiction with $\nu = \ell +1 \in \mathbb{N}$ for GMS backgrounds or superstrata. However, as in the usual holographic analysis, one has to consider $\nu\not \in \mathbb{Z}$ first, remove the divergences to obtain the quasi-normal modes, and then do an analytic continuation to integer $\nu$.

\subsubsection{Solution in the outer region}

In the outer region, the radial equation \eqref{eq:WaveEqApp} is approximated by the scalar equation in flat space
\be 
\frac{1}{r}\frac{d}{dr}\left(r^3\frac{d}{dr}\right)K(r)\- \Bigg[-\frac{4\,\Omega P}{R_y^2}\, r^2\+ \nu^2-1  \Bigg]\,K(r)=0\,.
\ee
The generic solutions are given by a linear combination of Bessel functions
\be 
K_\text{out}(r) \= \frac{1}{r} \, \left[ C_\text{out}^{(1)}~ J_\nu\left(\frac{2\sqrt{\Omega P}}{R_y} \,r\right)\+ C_\text{out}^{(2)}~ J_{-\nu}\left(\frac{2\sqrt{\Omega P}}{R_y} \,r\right)\right]\,.
\ee
In the overlapping region $\frac{r}{a} \sim \frac{1}{\epsilon} \, \frac{R_y}{a}$, that is $\frac{r}{R_y} \ll 1$, we have 
\be 
K_{\text{out}}(r) ~\approx~ \frac{C_\text{out}^{(1)}}{a\,\Gamma(1+\nu)}\,\left(\frac{\sqrt{\Omega P}\,a}{R_y}\right)^{\nu} \,\left( \frac{r}{a}\right)^{\nu-1}+\frac{C_\text{out}^{(2)}}{a\,\Gamma(1-\nu)}\,\left(\frac{\sqrt{\Omega P}\,a}{R_y}\right)^{-\nu} \,\left( \frac{r}{a}\right)^{-\nu-1}\,,
\label{eq:waveoverlappingApp2}
\ee
whereas in the asymptotic region, $r \gg R_y$, 
\be 
K_\text{out}(r) \,\propto\, \left(\frac{r}{a} \right)^{-\frac{3}{2}} \left[\,e^{2i \frac{\sqrt{\Omega P}}{a\,R_y}\, r -i \frac{\pi}{4}}\left(C_\text{out}^{(1)} e^{-i \nu \frac{\pi}{2}}+C_\text{out}^{(2)}e^{i \nu \frac{\pi}{2}}\right)\+ e^{-2i \frac{\sqrt{\Omega P}}{a\,R_y}\, r+ i \frac{\pi}{4}}\left(C_\text{out}^{(1)} e^{i \nu \frac{\pi}{2}}+C_\text{out}^{(2)} e^{-i \nu \frac{\pi}{2}}\right)\right]\,.
\ee
The time dependence of the mode is given by \eqref{eq:ModeExpApp}
\be 
\Phi(r,t,y) ~\sim~ K_\text{out}(r) \, e^{i\left(\frac{\Omega + P}{R_y}\,t \+\frac{P-\Omega}{R_y}\,y \right)}\,.
\ee
Thus, we see that if $\text{Re}(\Omega+P) >0$, the wave is outgoing when
 $$
 C_\text{out}^{(1)} \+C_\text{out}^{(2)}e^{i \nu \pi} \= 0\,. $$ 
However, if $\text{Re}(\Omega+P) <0$, the wave is outgoing when 
$$
C_\text{out}^{(1)} \+C_\text{out}^{(2)}e^{-i \nu \pi} \= 0\,. 
 $$ 
Consequently, the outgoing condition is
\be 
C_\text{out}^{(1)}\= -\, C_\text{out}^{(2)}e^{i\, \text{sign}(\text{Re}(\Omega+P))\,\nu \pi}\,.
\label{eq:outgoingcondApp}
\ee
References \cite{Chakrabarty:2019ujg} prematurely fix a convention for the sign of $\text{Re}(\Omega+P)$ at this point. However, the spectrum of quasi-normal modes gives two branches of frequencies, one with mostly-positive $\text{Re}(\Omega+P)$ and one with mostly-negative $\text{Re}(\Omega+P)$. 

The existence of branches with opposite signs led the authors of these references to the conclusion that the corresponding modes might grow with time, and, a posteriori, they show that they are not in the spectrum since they do not have the same sign as their convention. At a technical level, this is caused by fixing the sign of  $\text{Re}(\Omega+P)$ too early in the calculation. As we will see, if we carry the ``$\text{sign}(\text{Re}(\Omega+P))$" factors all along, both branches of frequencies will lead to quasi-stable modes that decay with time.

\subsubsection{Matching solutions in the overlapping region}

We match the asymptotic inner-wave profile \eqref{eq:waveoverlappingApp1} to the asymptotic outer-wave profile \eqref{eq:waveoverlappingApp2} in the overlapping region taking into account the outgoing condition \eqref{eq:outgoingcondApp}:
\be 
-e^{i \,\text{sign}(\text{Re}(\Omega+P))\,\pi \nu} \,\frac{\Gamma(1-\nu)}{\Gamma(1+\nu)}\left(\frac{\Omega\, P \,a^2}{R_y^2}\right)^{\nu}=\frac{\Gamma(\nu)}{\Gamma(-\nu)} \frac{\Gamma\left(\frac{1}{2}(1-\nu+|\zeta|+ \xi)\right) \Gamma\left(\frac{1}{2}(1-\nu+|\zeta|- \xi)\right)}{\Gamma\left(\frac{1}{2}(1+\nu+|\zeta|+ \xi)\right) \Gamma\left(\frac{1}{2}(1+\nu+|\zeta|- \xi)\right)}
\label{eq:MatchCondApp}
\ee
The quasi-normal-mode frequencies are obtained by solving this equation considering $\Omega$ as the variable. Because $\Omega$ may enter non trivially in $\xi$ and $\zeta$, this expression is not solvable analytically. 

However, we have assumed that $a \ll R_y$. Thus, as soon as we are considering low-energy modes (such that $\Omega P$ is not ``too large" which we will make precise later), the left hand-side of the equation is very small. This is a manifestation of the huge potential barrier that the wave has to go through in order to be able to leak to infinity. 

Under this assumption, one can solve the equation perturbatively. The zeroth-order solution is obtained by considering the left- handside to be zero, so the arguments in the Gamma functions on the denominator of the right hand-side must be at the poles. This will give two towers of normal frequencies, labelled by a mode number $N$, that correspond to the real part of the frequencies of the quasi-normal modes:

\be 
\Omega_N \= \Omega_N^{(0)} \+ \ldots\,, \qquad N \in \mathbb{N}\,, \qquad \text{with}\quad \Omega_N^{(0)} \in \IR\,.  
\ee
We obtain the first-order correction by perturbing the Gamma functions around their poles. This will give the imaginary leading-order correction to the normal frequencies: 
\be 
\Omega_N \= \Omega_N^{(0)} \+ \delta \Omega_N\,, \qquad N \in \mathbb{N}\,, \qquad \text{with}\quad \Omega_N^{(0)} \in \IR\,,\quad \delta \Omega_N \in i \, \IR\,.  
\ee

\subsubsection{The normal frequencies, $\Omega_N^{(0)}$}

As explained above, the zeroth-order expression is obtained when one of the two Gamma functions in the denominator of \eqref{eq:MatchCondApp} has a pole. This happens when 
\be 
\frac{1}{2} \left(1+\nu +|\zeta| \pm \xi \right) \= -N\,,\qquad N\in \mathbb{N}\,.
\label{eq:PoleGammasApp}
\ee
To find the final expression for $\Omega_N^{(0)}$, one needs to know the dependence of $\xi$ and $\zeta$ on the mode momenta $(\Omega,q_y,q_\phi,q_\psi)$. For the backgrounds considered in this paper, the superstrata \eqref{eq:zetaxiSupApp} or the GMS solutions \eqref{eq:zetaxiGMSApp}, the centripetal coefficient $\zeta$ does not depend on the frequencies $\Omega$ since $P-\Omega =q_y$ and $\Omega$ enters in $\xi$ as
\be 
\xi \=\frac{2}{E_\text{gap}} \, \Omega \+ \chi\,,
\label{eq:xigenApp}
\ee
where $\chi$ is independent of $\Omega$ and $E_\text{gap}$ is given by the background and will correspond to the gap of energy between two successive normal modes. For the $(2,1,n)$ superstratum at large $n$, $E_\text{gap} = \frac{2 j_L}{n_1 n_5}$ whereas for the GMS background $E_\text{gap} =\frac{\eta}{k} = \frac{j_L}{n_1 n_5} \, \eta$ where $\eta$ is defined in \eqref{eq:momch}. 

Thus, we have two branches of normal frequencies depending of the ``$\pm$'' choice:
\be 
\Omega_N^{(0)} \= \pm \, \frac{E_\text{gap}}{2}\, \biggl[2N\+\nu\+1 \+ \left|\xi\right| \,\mp\, \chi \biggr]\,,\qquad N\in \mathbb{N}\,.
\label{eq:AsympMatchzerothorder}
\ee
These normal frequencies are those of the bound states in the AdS$_3$ cap only. Indeed, the contribution from the gluing to flat space cannot be captured at zeroth order, because the left hand-side of \eqref{eq:MatchCondApp} is approximated to be zero. Moreover, as expected, we have two branches of normal frequencies, one positive and one negative. 

\subsubsection{The quasi-normal decay rates, $\delta \Omega_N$}

The computation of the first-order correction is slightly more involved. We proceed following the steps of \cite{Chowdhury:2007jx}. We change the argument of the divergent Gamma function from $-N$ to $-N - \delta N$ where $\delta N$ is small. We also replace $\xi$ by \eqref{eq:PoleGammasApp}, $\Omega$ by $\Omega_N^{(0)}$ and $P$ by $\Omega_N^{(0)}+q_y$ in the other arguments. The quasi-normal-mode equation \eqref{eq:MatchCondApp} gives
\be 
\frac{\Gamma\left(-N-\nu\right)\, \Gamma\left(N+|\zeta| +1\right)}{\Gamma\left(-N-\delta N\right) \,\Gamma\left(N+\nu+|\zeta|+1\right)}\= -e^{i \,\text{sign}(2\Omega_N^{(0)}+q_y)\,\pi \nu} \,\(\frac{\Gamma(-\nu)}{\Gamma(\nu)}\)^2\left[\frac{\Omega_N^{(0)}\, \left( \Omega_N^{(0)}+q_y\right) \,a^2}{R_y^2}\right]^{\nu}\,,
\label{eq:MatchCondApp2}
\ee
for both branches of normal frequencies \eqref{eq:PoleGammasApp}. 

We first simplify the non-divergent Gamma functions. We recall that $\zeta$ is an integer \eqref{eq:centripetalinZ} and, at this level, $\nu$ is considered to be a real number. Thus, using that 
$$\Gamma(X+i) \= (X)_i\, \,\Gamma(X) \,,$$ 
where  $\left(X \right)_i \= \prod_{j=0}^{i-1} (X+j) $ is the Pochhammer symbol, we have
\be 
\begin{split}
\frac{\Gamma\left(N+|\zeta| +1\right)}{\Gamma\left(N+\nu+|\zeta|+1\right)} & \= \frac{(N+|\zeta| )!}{\Gamma(\nu)\,\,(\nu)_{N+|\zeta|+1}}\,,\\
\Gamma\left(-N-\nu\right) &\= \frac{\nu\,\,\Gamma(-\nu)}{(-1)^{N}\,\,(\nu)_{N+1}}\,.
\end{split}
\ee
 Then, we expand the divergent Gamma function
\be 
\Gamma(-N-\delta N) ~ \sim~  \frac{1}{(-1)^N \, N!\,\, \delta N}\,.
\ee
We finally have
\be 
\delta N \= -\,e^{i \,\text{sign}(2\Omega_N^{(0)}+q_y)\,\pi \nu} \,\, \frac{\Gamma(-\nu)}{\Gamma(\nu)} \,\left[\frac{\Omega_N^{(0)}\,  \left( \Omega_N^{(0)}+q_y\right)\,a^2}{R_y^2} \right]^\nu \,\frac{(\nu)_{N+1}\,(\nu)_{N+|\zeta|+1}}{\nu\,N!\,(N+|\zeta| )!}\,.
\ee
When $\nu$ is not an integer, $\delta N$ has a real and an imaginary part and then the first-order correction also slightly changes   the normal frequencies. However, when we analytically continue $\nu$ to an integer, $\nu \= \ell+1 \in \mathbb{N}$, only the imaginary part gets a finite value. We use the relation
\be 
\sin\left(\pi \nu \right)\,\Gamma(-\nu) \= - \frac{\pi}{\nu\,\Gamma(\nu)}\,,
\ee
and obtain
\be 
\delta N \= i\,\, \frac{\pi}{(l!)^2}\,\, \text{sign}(2\Omega_N^{(0)}+q_y) \,\, \left[\frac{\Omega_N^{(0)}\,  \left( \Omega_N^{(0)}+q_y\right)\,a^2}{R_y^2} \right]^{\ell+1} \,\,{}^{\ell+1+N} C_{\ell+1}  \,\,{}^{\ell+1+N+|\zeta|} C_{\ell+1} \,,
\label{eq:delatNApp}
\ee
where we have introduced the usual binomial coefficients
\be 
{}^{p+q} C_p \= \frac{(p)_q}{p\,(q-1)!}\,.
\ee
The last step consists in replacing $\delta N$ by $\delta \Omega_N$. We differentiate \eqref{eq:PoleGammasApp}:
\be 
- \delta N \= \frac{1}{2} \, \partial_\Omega \(|\zeta| \pm \xi \) \, \delta \Omega_N \= -\frac{\text{sign}(\Omega_N^{(0)})}{E_\text{gap}} \, \delta \Omega_N\,.
\ee
By inserting the expression into \eqref{eq:delatNApp}, we see that we have a product of sign functions. In \cite{Chakrabarty:2019ujg}, because the sign convention for $2\text{Re}(\Omega)+q_y$ was fixed too early, the result obtained was dependent on $\text{sign}(\Omega_N^{(0)})$. As a result, one branch of normal frequencies had a positive decay rate $i\,\delta \Omega_N$ whereas the other appeared to have a negative decay rate. This led to the incorrect conclusion that some of the wave functions were growing with time signifying instabilities. 

By rectifying the sign restriction, we see that the sign of the decay rate does not depend on the sign of $\Omega_N^{(0)}$ but on the sign of $$\Omega_N^{(0)} (2 \Omega_N^{(0)} +q_y) ={\Omega_N^{(0)}}^2 +\Omega_N^{(0)}(\Omega_N^{(0)}+q_y) > 0\,,$$ where the last inequality has been obtained considering the condition of existence of quasi-normal modes \eqref{eq:CondApp2}. Thus, as soon as \eqref{eq:CondApp2} is satisfied, both branches of normal frequencies have the same sign for the decay rate. We have
\be 
\delta \Omega_N \= i\,\, \frac{\pi}{E_\text{gap}\,(l!)^2}\,\, \left[\frac{\Omega_N^{(0)}\,  \left( \Omega_N^{(0)}+q_y\right)\,a^2}{R_y^2} \right]^{\ell+1} \,\,{}^{\ell+1+N} C_{\ell+1}  \,\,{}^{\ell+1+N+|\zeta|} C_{\ell+1} \,.
\label{eq:firstorderAsympMatchApp}
\ee
The right-hand side of the expression is a {\it positive} purely imaginary  number. The time dependence of the modes is given by
\be 
e^{i\left( \frac{\sqrt{2}\,\Omega}{R_y} \,u\,+\,\frac{\sqrt{2}\, P}{R_y} \, v \right)}\= e^{i\left( \frac{2\,\Omega}{R_y} \,t\,+\,\frac{q_y}{R_y} \, (t+y)\right)} \= e^{i \, \frac{2 \,\delta\Omega_N}{R_y}\, t}\,e^{i\left( \frac{2\,\Omega_N^{(0)}}{R_y} \,t\,+\,\frac{q_y}{R_y} \, (t+y)\right)}\,,
\ee
which guarantees that the wave profile is decaying in time for both branches of frequencies \eqref{eq:AsympMatchzerothorder}.  

\subsection{The spectrum of quasi-normal modes via WKB}
\label{App:QNMviaWKB}

We now solve the  problem analyzed in Appendix \ref{App:QNMviaAsympMatch}  using the WKB approximation method detailed in Section \ref{sec:WKB}. We will see that it follows the same philosophy as the asymptotic matching method and it leads to a very similar result.

\begin{figure}
\centering
\includegraphics[scale=0.75]{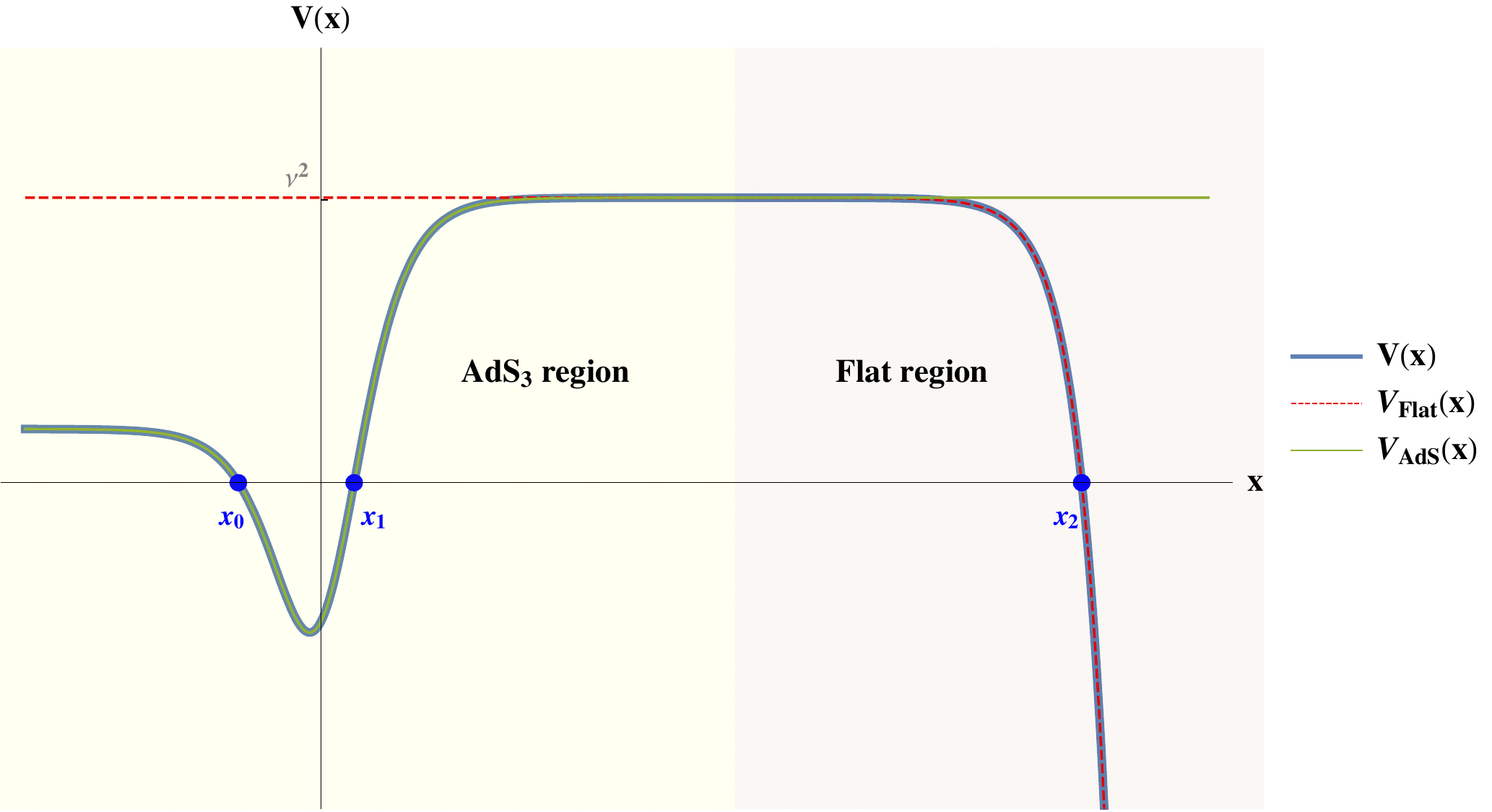}
\caption{The potential $V(x)$ and the two approximations $V_\text{Flat}(x)$ and $V_\text{AdS}(x)$.}
\label{fig:VirmaniPotential}
\end{figure}

As explained in Section \ref{sec:WKB}, we first need to transpose the wave equation \eqref{eq:WaveEqApp} into a Schr\"odinger problem. There are many ways to do so and we will use the one of \cite{Bena:2019azk} which gives a better accuracy for the mass term:
\begin{equation}
K(r) ~=~  \frac{\Psi(r)}{\sqrt{r^2+a^2}}\,,\qquad x ~=~ \log \frac{r}{a}\,,\qquad x \in \mathbb{R}\,.
\end{equation}
The wave equation gives
\begin{equation}
\frac{d^2}{dx^2} \Psi(x) ~-~  V(x) \Psi(x) ~=~  0\,.
\end{equation}
where $V(x)$ is given by:
\begin{equation}
V(x) ~\equiv~  \frac{e^{2 x}}{e^{2 x}+1} \left[\,- \frac{4 \,\Omega \,P\,a^2}{R_y^2} \,e^{2x} \+ \nu^2 \+  e^{-2 x} \,\zeta^2 \- \frac{\xi^2-1}{e^{2 x}+1}  \,\right ]\,.
\label{eq:potentialformApp}
\end{equation}
The form of the potential with the assumption \eqref{eq:CondApp} and \eqref{eq:CondApp2} is depicted in Fig.\ref{fig:VirmaniPotential}. In the inner region, $x\sim -\infty$ to the middle of the barrier $x \sim \log \epsilon \frac{R_y}{a}$, the potential is well-approximated by the AdS$_3$ potential, 
\be 
V_\text{AdS}(x) ~\equiv~  \frac{e^{2 x}}{e^{2 x}+1} \left[\, \nu^2 \+  e^{-2 x} \,\zeta^2 \- \frac{\xi^2-1}{e^{2 x}+1}  \,\right ]\,.
\ee
whereas from the middle of the barrier to the boundary, $x\sim + \infty$, the potential is given by the flat potential,
\be 
V_\text{Flat}(x) ~\equiv~ - \frac{4 \,\Omega \,P\,a^2}{R_y^2} \,e^{2x} \+ \nu^2 \,.
\ee

In Section \ref{sec:WKB}, we  showed that the WKB approximation gives a spectrum of quasi-normal modes as a tower of frequencies labelled by a mode number, $N \in \mathbb{N}$,
\be 
\Omega_N \= \Omega^{(0)}_N \+ \delta \Omega_N \,,
\ee
where the zeroth-order value, $\Omega^{(0)}_N$, is purely real and given by the usual quantization relation 
\be 
\Theta^{(0)}_N \= \frac{\pi}{2} \left( 1 \+ 2 N \right)\,, \qquad N\in \mathbb{N}\,.
\label{eq:zerothOrderApp}
\ee
The first-order correction, $\delta \Omega_N$, is purely imaginary and given by\footnote{Note a slight difference in the expression of the $\text{sign}(\ldots)$ compared to the general formula \eqref{eq:ImPartSpectrum}. In Section \ref{sec:WKB}, $\omega$ is the conjugate momentum of $t$ whereas here we are working with $\Omega$, the momentum along $u$ and with $P$, the momentum along $v$. The translation of the condition of having an outgoing wave  (which involves $t$) is:  $\text{sign}(\text{Re}(\omega))=\text{sign}(\text{Re}(\Omega+P))=\text{sign}(2\text{Re}(\Omega)+q_y)$.}
\be 
\delta \Omega_N \= i\,\text{sign}\left(2\,\Omega_N^{(0)}+q_y\right)\, \left(\frac{\partial \Theta}{\partial \Omega} \right)^{-1}\,\frac{e^{-2 T}}{4}\,\,\Biggr|_{\Omega=\Omega^{(0)}_N }\,.
\label{eq:firstOrderApp}
\ee
To find this correction one needs to evaluate the following integrals
\begin{equation}
\Theta ~\equiv~  \int_{x_0}^{x_1} |V(z)|^{\frac{1}{2}}\,dz \,,\qquad T ~\equiv~  \int_{x_1}^{x_2} |V(z)|^{\frac{1}{2}}\,dz\,,
\label{eq:Theta&TdefApp}
\end{equation}
where $x_0$, $x_1$ and $x_2$ are the three turning points as depicted in Fig.~\ref{fig:VirmaniPotential}. Unfortunately, the square root of the potential $|V(x)|^{\frac{1}{2}}$ is not integrable in a closed form and one will need to use the approximate potentials to estimate $\Theta$ and $T$. They will strongly depend on the values of $x_0$, $x_1$ and $x_2$. The two first can be obtained using $V_\text{AdS}(x)$ whereas $x_2$ is given by $V_\text{flat}(x)$:
\be 
\begin{split}
e^{2 x_2} \=\frac{\nu^2\,R_y^2} {4 \,\Omega P \,a^2}\,,\qquad e^{2x_1} \=  \frac{A + \sqrt{A^2-\nu^2\,\zeta^2}}{\nu^2}\,,\qquad e^{2x_0} \=  \frac{A - \sqrt{A^2-\nu^2\,\zeta^2}}{\nu^2}\,,
\label{eq:turningpointsApp}
\end{split}
\ee
where we have defined
\be 
A \equiv \frac{1}{2}\, \left(\xi^2-1-\nu^2 -\zeta^2 \right)\,.
\label{eq:AApp}
\ee

\subsubsection{The zeroth order, $\Omega_N^{(0)}$}

To obtain the real part of the quasi-normal-mode frequencies, $\Omega_N^{(0)}$, we need to compute $\Theta$. The integral is supported in a region where the potential is given by the AdS$_3$ potential, and one can simply use $\left|V_\text{AdS}(x)\right|^{\frac{1}{2}}$ which is integrable. We obtain
\be
\begin{split}
\Theta \= \frac{\pi}{2} \,\left[ \,-\nu \-\left|\zeta\right|\+ \sqrt{\xi^2 -1}\,\right]\,.
\end{split}
\ee
We know that the WKB approximation is precise when there are significantly many oscillations between $x_0$ and $x_1$, which happens when $\xi^2 \gg 1 $. This excludes the first few modes. For the higher modes we then have
\be
\begin{split}
\Theta \approx \frac{\pi}{2} \,\left[ \,-\nu \-\left|\zeta\right|\+ \left|\xi\right|\,\right]\,,
\label{eq:ThetaExprApp}
\end{split}
\ee
Thus, the zeroth-order expression obtained via WKB is exactly identical to the zeroth-order expression of the matched asymptotic method reviewed in the previous section \eqref{eq:PoleGammasApp}:
\be 
\begin{split}
\frac{1}{2} \left(-1-\nu -\left|\zeta\right| \+ \left|\xi\right| \right) \= N \,, \qquad N\in \mathbb{N}\,.
\end{split}
\label{eq:zerothspectrumWKBApp}
\ee
Using again the fact that, in the backgrounds we consider, $\zeta$ is independent of $\Omega$ and $\xi$ behaves as in equation \eqref{eq:xigenApp}, we obtain  the same branches of normal frequencies
\be 
\Omega_N^{(0)} \= \pm \, \frac{E_\text{gap}}{2}\, \biggl[2N\+\nu\+1 \+ \left|\xi\right| \,\mp\, \chi \biggr]\,,\qquad N\in \mathbb{N}\,.
\label{eq:WKBzerothorder}
\ee
Returning to the assumption $\xi^2 \gg 1 $, we can check that it indeed requires that $N \gtrsim 10$ and therefore the WKB method loses accuracy for the first few quasi-normal frequencies.

\subsubsection{The first order, $\delta \Omega_N$}

We aim to apply \eqref{eq:firstOrderApp} to obtain the expression of $\delta\Omega_N$. Using \eqref{eq:ThetaExprApp} with \eqref{eq:xigenApp}, we have
\be 
\frac{\partial \Theta}{\partial \Omega} \approx \frac{\pi}{E_\text{gap}}\,\text{sign}\(\text{Re}(\xi)\) \= \frac{\pi}{E_\text{gap}}\,\text{sign}\(\Omega_N^{(0)}\)
\ee
Thus, 
\be 
\text{sign}\left(2\,\Omega_N^{(0)}+q_y\right)\, \left(\frac{\partial \Theta}{\partial \Omega} \right)^{-1} \approx \frac{E_\text{gap}}{\pi}\,,
\ee
where we have used that $\text{sign}\left(2\,\Omega_N^{(0)}+q_y\right)\text{sign}\(\Omega_N^{(0)}\right) \= 1$ thanks to \eqref{eq:CondApp2}. 

\medskip

Now, we need to estimate the integral $T$ in equation \eqref{eq:Theta&TdefApp} which is a bit of a challenge since it is supported in the overlapping region between $V_\text{AdS}(x)$ and $V_\text{Flat}(x)$. We will use an intuitive procedure that is equivalent to translating the asymptotic matching method in WKB language. This will consist in computing the integral of $|V(z)|^{\frac{1}{2}}$ in two pieces. The first piece from $x_1$ to the middle of the barrier will use $V_\text{AdS}(x)$ and the second piece from the middle of the barrier to $x_2$ will use $V_\text{Flat}(x)$. More concretely, we take
\be 
T \,\simeq\, \int_{x_1}^{x} \, |V_\text{AdS}(z)|^{\frac{1}{2}} \,dz \+  \int_{x}^{x_2} \, |V_\text{Flat}(z)|^{\frac{1}{2}} \,dz\,,
\ee
where $x_1 \ll x \ll x_2$ which is allowed assuming \eqref{eq:CondApp}. The two integrals give
\be 
\begin{split}
\exp \int_{x_1}^x  |V_\text{AdS}(z)|^{\frac{1}{2}}\, dz ~\approx~ & e^{\nu \left(x-x_1 \right)} \, \left(\frac{4\,e^{2 x_{1}}}{e^{2 x_{1}}-e^{2 x_{0}}}\right)^{\frac{\nu}{2}}\left(\frac{e^{x_{1}}+e^{x_{0}}}{e^{x_{1}}-e^{x_{0}}}\right)^{\frac{\nu}{2} \,e^{x_1+x_0}}\\
& \,\times\, \left(\frac{\sqrt{e^{2x_1}+1}-\sqrt{e^{2x_0}+1}}{\sqrt{e^{2x_1}+1}+\sqrt{e^{2x_0}+1}}\right)^{\frac{\nu}{2}\sqrt{\left(e^{2x_1}+1\right)\left(e^{2x_0}+1\right)}} \,,\\
\exp \int_{x}^{x_2}  |V_\text{Flat}(z)|^{\frac{1}{2}} \,dz ~\approx~ &  \left(\frac{2}{e} \right)^\nu e^{\nu (x_2 -x)}\,,
\label{eq:asymptoticExpWKB}
\end{split}
\ee
where $e=\exp(1)$. We will introduce $\mu$ as
\be 
\mu \equi  \left(\frac{2}{e} \right)^\nu \,\left(\frac{4\,e^{2 x_{1}}}{e^{2 x_{1}}-e^{2 x_{0}}}\right)^{\frac{\nu}{2}}\left(\frac{e^{x_{1}}+e^{x_{0}}}{e^{x_{1}}-e^{x_{0}}}\right)^{\frac{\nu}{2} \,e^{x_1+x_0}} \,\left(\frac{\sqrt{e^{2x_1}+1}-\sqrt{e^{2x_0}+1}}{\sqrt{e^{2x_1}+1}+\sqrt{e^{2x_0}+1}}\right)^{\frac{\nu}{2}\sqrt{\left(e^{2x_1}+1\right)\left(e^{2x_0}+1\right)}}\,.
\ee
Then, we have
\be 
e^{-2 T } \,\simeq\, \mu^{-2}\,e^{2\nu(x_1-x_2)}\,,
\ee

Finally, we need to calculate the location of the turning points in  \eqref{eq:turningpointsApp} when $\Omega = \Omega_N^{(0)}$. For this we simplify the expression of $A$ \eqref{eq:AApp} using the quantization relation \eqref{eq:zerothspectrumWKBApp}:
\be 
\begin{split}
A \,\bigr|_{\Omega =\Omega_N^{(0)}} \=\nu  \(2N + |\zeta|+1\)\+  \frac{1}{2}\,\(2N +1\)^2 +|\zeta|\,.
\end{split}
\ee
Hence, from equation \eqref{eq:firstOrderApp} we find the WKB first-order correction to the frequency:
\be 
\delta \Omega_N \= i \,\frac{E_\text{gap}}{4\pi\,\mu^2} \,\left[4\,\Omega_N^{(0)}\,\left( \Omega_N^{(0)} +q_y\right)\,\frac{a^2}{R_y^2} \right]^{\nu} \,  \left( \frac{A+\sqrt{A^2-\nu^2\zeta^2}}{\nu^4}\right)^\nu\,.
\ee
We see that some factors are identical to the correction obtained via asymptotic matching \eqref{eq:firstorderAsympMatchApp}. Thus, the expressions agree if and only if
\be 
\frac{1}{2 \pi \,\mu^2}  \left( \frac{A+\sqrt{A^2-\nu^2\zeta^2}}{\nu^4}\right)^\nu \, \simeq \, \frac{2 \pi}{4^\nu\,((\nu-1)!)^2}\,\,\,  {}   ^{\nu+N} C_{\nu}  \  {}^{\nu+N+|{\zeta}|} C_{\nu}\,.
\label{eq:compAsympMatchWKB}
\ee
Given the completely different functions that appear on the left and on the right, this approximate equality is far from obvious. We will show in the next section that this approximate equality is satisfied with a difference less than 1\%  as soon as $(N,\ell) \gtrsim 10$ and that the large-$\ell$ or the large-$N$ expansions are miraculously identical until the third order!

\subsection{Comparison}
\label{App:WKBvsAsympMatch}

As stated above, the difference in the spectrum obtained via WKB or via asymptotic matching is determined by the difference between two quantities that we denote $L$ and $R$:
\be 
L \equi \frac{1}{2 \pi \,\mu^2}  \left( \frac{A+\sqrt{A^2-\nu^2\zeta^2}}{\nu^4}\right)^\nu \,,\qquad R \equi \frac{2 \pi}{4^\nu\,((\nu-1)!)^2}\,\,\,  {}   ^{\nu+N} C_{\nu}  \  {}^{\nu+N+|{\zeta}|} C_{\nu}\,.
\ee
The functions depend on three variables: the mode number $N$, the mass $\nu$ and the centripetal coefficient $\zeta$. Because of the very different forms of $L$ and $R$, it appears complicated to do a direct comparison. This is why we will show that the large-$\ell$ expansions for arbitrary $(N,\zeta)$ are strictly identical until the third order and similarly at large $N$ for arbitrary $(\ell,\zeta)$. For values in between, we will simply give a numerical plot.
\begin{itemize}
\item[-] \underline{At large $\nu$:}
\end{itemize}
We have
\be 
A \= \frac{2N +|\zeta|+1}{\nu}\,\left(1 \+ \cO\(\nu^{-1} \right)\right)\,,\quad e^{2 x_1}\= \frac{\alpha_+}{\nu}\,\left(1 \+ \cO\(\nu^{-1} \right)\right)\,,\quad e^{2 x_0}\= \frac{\alpha_-}{\nu}\,\left(1 \+ \cO\(\nu^{-1} \right)\right)\,,
\ee
where we have defined
\be 
\alpha_\pm \= \(2N + |\zeta|+1\) \pm \sqrt{\(2N+1 \)\(2N +2 |\zeta|+1\)}\,,
\ee
which gives
\be 
 \left( \frac{A+\sqrt{A^2-\nu^2\zeta^2}}{\nu^4}\right)^\nu \= e^{-3\,\nu\, \log \nu \+ \nu \log \alpha_+\+ \cO(1)}\,,\quad \mu^2 \= e^{-\nu \log \nu \+ \nu ( \log \alpha_+ + \log 4 - 2) \- \frac{\alpha_+ + \alpha_-}{2} \, \log \nu \+ \cO(1)}
\ee
Thus,
\be 
L \= \exp\biggl[ -2\nu \log \nu \+ 2\nu (1-\log 2) \+ \(2N +|\zeta|+1\)\,\log \nu\+ \cO(1)\biggr]\,.
\ee
On the other hand, using Stirling's formula \eqref{eq:Stirling}, $R$ behaves as
\be 
R\= \frac{1}{\Gamma(N+1)\,\Gamma(N+|\zeta|+1)}\exp\biggl[ -2\nu \log \nu \+ 2\nu (1-\log 2) \+ \(2N +|\zeta|+1\)\,\log \nu\+ \cO(1)\biggr]\,.
\ee
The three first leading orders then match exactly. One can also push to the fourth order, $e^{\cO(1)}$, and we can see that even if they do not match exactly, they are very close to each other when $N \gtrsim 10$.

\begin{itemize}
\item[-] \underline{At large $N$:}
\end{itemize}
We have
\be 
A\= 2\,N^2 \,\left(1 \+ \cO\(N^{-1}\) \right) \,,\quad e^{2 x_1}\= \frac{4 \,N^2}{\nu^2}\,\left(1 \+ \cO\(N^{-1}\) \right)\,,\quad e^{2 x_0}\= \frac{\zeta^2}{4 \,N^2}\,\left(1 \+ \cO\(N^{-1}\) \right)\,,
\ee
which gives
\be 
 \left( \frac{A+\sqrt{A^2-\nu^2\zeta^2}}{\nu^4}\right)^\nu \= \left(\frac{4 N^2}{\nu^4} \right)^\nu \left(1 \+ \frac{\nu \(\nu + |\zeta|+1 \) }{N}\+\cO(N^{-2}) \right)\,,\quad \mu^2 \= \( \frac{2}{e}\)^{4 \nu} \left(1 \+ \cO( N^{-2}) \right)
\ee
Thus,
\be
\begin{split} 
L &\= \frac{1}{2\pi} \,  \( \frac{e}{\sqrt{2}\,\nu}\)^{4 \nu}\,N^{2\nu} \,\left(1 \+ \frac{\nu \(\nu + |\zeta| +1 \) }{N}\+\cO(N^{-2}) \right)\,, \\
R&\=\frac{2\pi}{4^\nu \,\nu^2\, \Gamma(\nu)^4}  \,N^{2\nu} \,\left(1 \+ \frac{\nu \(\nu +  |\zeta|+1 \) }{N}\+\cO(N^{-2}) \right)\,.
\end{split}
\ee
The two first leading orders then match exactly. As for the coefficient in front, one can actually show as soon as $\nu \gtrsim 10$, the expansion of the Gamma function gives, 
$$ \frac{2\pi}{4^\nu \,\nu^2\, \Gamma(\nu)^4} \,\approx\, \frac{1}{2\pi} \,  \left( \frac{e}{\sqrt{2}\,\nu}\right)^{4 \nu}\,.$$

\begin{figure}
\centering
\includegraphics[scale=0.55]{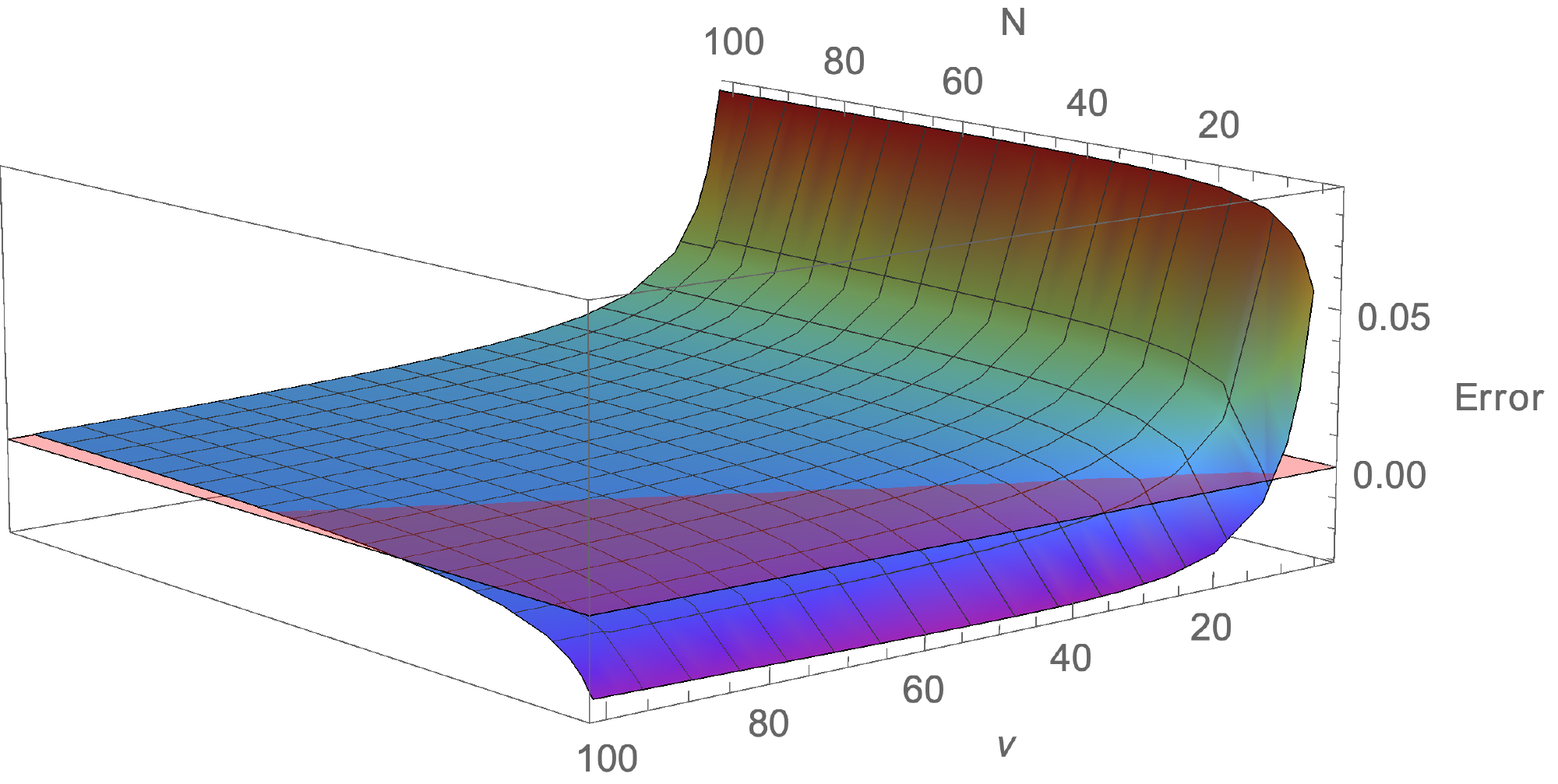}
\caption{Plot of the error function \eqref{eq:ErrorFunc} that compares the relative difference of the spectra obtained via WKB and via asymptotic matching, for $|\zeta|=1$.}
\label{fig:ErrorFunc}
\end{figure}

We have thus proven that, despite their very different analytical expressions, the results obtained via WKB and via asymptotic matching agree incredibly well at large $\ell$ and large $N$. To corroborate this, we can plot the error function, \be \text{Error} \= \frac{L \- R}{R}\,, \label{eq:ErrorFunc}\ee
as a function of $N$, $\nu$ and $\zeta$. It is not hard to observed that the value of $\zeta$ does not modify significantly this error function. Thus in Fig.\ref{fig:ErrorFunc} we show the dependence of this error function on $N$ and $\nu$, and we can see that as soon as $N\gtrsim 10$ and $\nu \gtrsim 10$ the difference between the WKB and the asymptotic-matching result is less than 1\%.

Our WKB techniques are therefore as accurate as asymptotic matching, and can be used for backgrounds that have more than two overlapping regions, such as superstrata.


\begin{adjustwidth}{-1mm}{-1mm} 

\bibliographystyle{utphys}      

\bibliography{microstates}       

\end{adjustwidth}

\end{document}